\documentclass{mn2e}
\usepackage{natbib2,natbibmnfix}
\usepackage{astrojournals}
\usepackage{epsfig}
\usepackage{amssymb}
\usepackage{multirow}
\usepackage{multicol}
\usepackage{color}
\usepackage{url}
\usepackage{amsmath}
\usepackage{ctable}
\usepackage{color}
\usepackage{fixltx2e} 
\usepackage[utf8]{inputenc}
\usepackage{hyperref}

\newcommand{\be}{\begin{equation}}
\newcommand{\ee}{\end{equation}}

\newcommand{\degree}{^{\circ}}

\def\mdot{\dot{M}}
\def\msun{{\rm M_{\odot}}}
\def\mbh{M_{\rm BH}}
\def\mcore{M_{\rm core}}
\def\rcore{r_{\rm core}}
\def\racc{r_{\rm acc}}

\def\tilt{\theta_{\rm tilt}}
\def\Jdisc{{\mathbf J_{\rm disc}}}

\def\Ldisc{{\mathbf L_{\rm disc}}}
\def\Lshell{{\mathbf L_{\rm shell}}}

\def\vrot{{v}_{\rm rot}}

\def\avisc{{\alpha_{\rm visc}}}
\def\npart{N_{\rm part}}
\def\ndisc{N_{\rm disc}}
\def\nshell{N_{\rm shell}}

\def\kms{\rm \, km \, s^{-1}}
\def\myr{\rm \,Myr}

\def\simlt{\mathrel{\rlap{\lower 3pt\hbox{$\sim$}}\raise 2.0pt\hbox{$<$}}}
\def\simgt{\mathrel{\rlap{\lower 3pt\hbox{$\sim$}} \raise 2.0pt\hbox{$>$}}}
\def\lsim{\mathrel{\rlap{\lower 3pt\hbox{$\sim$}}\raise 2.0pt\hbox{$<$}}}
\def\gsim{\mathrel{\rlap{\lower 3pt\hbox{$\sim$}} \raise 2.0pt\hbox{$>$}}}

\def\msunpc3{\,\msun~{\rm {pc^{-3}}}}

\def\kms{{\rm\,km\,s^{-1}}}
\def\pc{\mbox{\, pc}}
\def\kpc{\mbox{\, kpc}}

\newcommand{\acknowledgments}{\begin{small}\section*{Acknowledgments}\end{small}}
\newcommand\altaffilmark[1]{$^{#1}$}
\newcommand\altaffiltext[1]{$^{#1}$}
\title[Overlapping flows and SMBH growth]{Overlapping Inflow Events as Catalysts for Supermassive Black Hole Growth}

\author[Carmona-Loaiza et al.]{
\parbox[t]{\textwidth}{ 
Juan M. Carmona-Loaiza\thanks{E-mail:jcarmona@sissa.it}\altaffilmark{1},
Monica Colpi\altaffilmark{2,3},
Massimo Dotti\altaffilmark{2,3}
\&\ Riccardo Valdarnini\altaffilmark{1,4} 
} 
\vspace*{6pt} \\
\altaffiltext{1}{Scuola Internazionale Superiore di Studi Avanzati, 
   Via Bonomea 265, 34136 Trieste, Italy.} \\
\altaffiltext{2}{Dipartimento di Fisica G. Occhialini, Universit\`a
  degli Studi di Milano Bicocca, Piazza della Scienza 3, 20126 Milano,
  Italy.} \\
\altaffiltext{3}{INFN, Sezione di Milano-Bicocca, Piazza della Scienza 3, 20126 Milano, Italy. }\\
\altaffiltext{4}{INFN, Trieste - Iniziativa Specifica QGSKY, Italy. } \\
}

\date{Submitted to MNRAS, in the near future.\vspace{-0.6cm}}
\begin{document}
\maketitle
\label{firstpage}

\begin{abstract}
One of the greatest issues in modelling black hole fuelling is our lack of understanding of the processes by which gas loses angular momentum and falls from galactic scales down to the nuclear region where an accretion disc forms, subsequently guiding the inflow of gas down to the black hole horizon. It is feared that gas at larger scales might still retain enough angular momentum and settle into a larger scale disc with very low or no inflow to form or replenish the inner accretion disc (on $\sim 0.01$ pc scales). In this paper we report on hydrodynamical simulations of rotating infalling gas shells impacting at different angles onto a pre-existing, primitive large scale ($\sim 10 \pc$) disc around a super-massive black hole. The aim is to explore how the interaction between the shell and the disc redistributes the angular momentum on scales close  to the black hole's sphere of influence. Angular momentum redistribution via hydrodynamical shocks leads to inflows of gas across the inner boundary, enhancing the inflow rate by more than 2-3 orders of magnitude. In all cases, the gas inflow rate across the inner parsec is higher than in the absence of the interaction, and the orientation of the angular momentum of the flow in the region changes with time  due to gas mixing. Warped discs or nested misaligned rings form depending on the angular momentum content of the infalling shell relative to the disc. In the cases in which the shell falls in near counter-rotation, part of the resulting flows settle into an inner dense disc which becomes more susceptible to mass transfer.
\end{abstract}


\section{Introduction}
\label{sec:intro}

According to the black hole paradigm, Active Galactic Nuclei
(AGN) host super massive black holes accreting gas from their
surroundings. Observations indicate that these black holes have grown over cosmic time
mostly by radiative efficient
accretion around $z \sim 3$ (\citealt{Marconi04,
Merloni04}),  and that some of them were able to grow even faster, as
revealed by the recent observations of $\gsim 10^9 \msun$ black holes at $z
\sim 6-7$ (see, e.g. \citealt{Mortlock11}).
The tight correlation between the black hole mass $\mbh$ 
  and the stellar velocity dispersion $\sigma$ of the galaxy's host, observed
in the local universe, further indicates that 
 the accretion history of black holes is closely linked with the evolution of the galaxies which they inhabit 
(\citealt{Ferrarese05}). Understanding the growth mechanisms of
super massive black holes is therefore central, not only for explaining the AGN phenomenon but also
the evolution of galaxies across cosmic ages. A great deal of progress on
this field has been made in recent years, and a state of the art review is in \cite{Kormendy13}.

Identifying the processes which drive gas from the galactic scales ($R_{\rm gal} \sim 10$ kpc) down to the black hole's
innermost stable circular orbit ($R_{\rm isco}=3 \times 10^{-5}  (M_{\rm BH}/10^8\,\msun)$ pc for a non-rotating black hole) is still a challenging problem as an ab-initio treatment of all key 
processes is complex over such wide interval of physical scales (e.g. \citealt{Jogee06}).
In many studies, it has been recognised that the fuelling of QSOs and of the less luminous AGN requires not only a large reservoir of gas, but most importantly 
a mechanism for efficient transport of angular momentum that can bring gas from galactic scales 
down to the scale of the black hole horizon, or as a minimal condition, down
to the scale where an accretion disc forms (\citealt {Shakura73}), of size $R_{\rm disc}\simlt 1000\, R_{\rm isco} \simeq 3 \times 10^{-2}$ pc, as at larger radii discs become unstable to due to self-gravity  (\citealt{Pringle81, Goodman03}).
It is only below this scale that 
internal viscous stresses (\citealt{Balbus98}) guide the radial drift of gas toward the black hole on a
time scale that is short enough 
compared with the cosmic time (\citealt{Pringle81,King07}).
The specific
angular momentum of a test-particle in a circular orbit at $R_{\rm disc}$, $j_{\rm disc},$  is far smaller than that at $R_{\rm gal}$ (typically $j_{\rm gal}/j_{\rm disc}\simgt 10^3$; \citealt{Jogee06}).
Thus, unless inflows are almost radial already from the large scales, a sizeable fraction of 
the angular
momentum of the self-gravitating gas on galactic scales needs to be transported 
outwards via global  gravitational instabilities, such as a cascade of nested bars 
(\citealt{Shlosman89,Shlosman90}) or cancelled out via shocks and turbulence 
(\citealt{Wada04,Hobbs11}).

As AGN activity occurs in isolated galaxies of different morphologies and gas content, as 
well as in
interacting/merging galaxies, several mechanisms have been explored to overcome
this angular momentum problem  (e.g. \citealt{Jogee06} and references therein).
External triggers such as galaxy  major or minor mergers have been proposed, 
as they effectively funnel  large amounts of gas into the innermost central regions ($\sim100\pc$) 
of the remnant galaxy via tidal torques,
making it possible to build up a reservoir of gas that ultimately feeds the central black hole
(\citealt{Toomre72,Hernquist89, Barnes91, DiMatteo05, Mihos94, Mihos96, Saitoh04, Saitoh09,Kawakatu09, Debuhr10, Hopkins10sim, Gabor13}). 
With use of multi-scale SPH simulations, \cite{Hopkins10sim} 
 explore both major galaxy mergers and isolated bar-(un)stable disc galaxies, varying
the gas fraction and the bulge to total mass ratio. 
 They  show  that in gas-rich, disc-dominated galaxies the gas displays, on scales between 100 parsecs
and 10 parsecs, a rich array of morphologies, including bars, spirals, nuclear rings, single or three armed systems, and clumps that can transport angular momentum via a range of unstable modes. They further find that, 
on scales of about 1 pc, precessing eccentric discs develop that drive inflows of gas toward the  very vicinity of the black hole.
Numerical simulations also indicate that, in gas-rich galaxies, dense massive clouds form and interact with each other producing torques that lead to global gas inflows toward the nuclear regions, inducing high fluctuations in the accretion rate
onto the central black hole (\citealt{Bournaud11,Gabor13}).


\cite{Hobbs11} (Hobbs11 hereafter) take a simplified alternative to study the angular momentum problem, performing a set of idealised hydrodynamical SPH simulations of collapsing spherical gas clouds 
with different initial degrees of rotational support and supersonic turbulence. 
In this way, Hobbs11 correlate the level of angular momentum mixing in the flow to
the degree of turbulence, and  discuss  the role of turbulence in the feeding of the central black hole. 
In the purely rotating case (no turbulence) gas is found to settle into a nearly circular, rotationally supported ring-disc structure, 
as efficient mixing causes the initially broad angular momentum spectrum to collapse around its averaged value.
The inclusion of supersonic turbulence in the collapsing cloud leads instead to a large increase of the inflow rate: dense filaments,
created by the shocked converging turbulent flow, travel along nearly undisturbed, ballistic paths of low
angular momentum allowing a fraction of the gas to impact the innermost boundary of the simulation.

In this paper we consider as a potential trigger of black hole accretion in a galaxy, the interaction of 
a misaligned rotating gas cloud  with a pre-existing ``stalled'' disc-ring of gas present on scales $\simlt 10$ parsec that failed to 
reach the inner parsec to be accreted by the central black hole. 
 We aim at exploring how nested, misaligned inflows of interacting gas clouds redistribute their angular momentum, triggering inflows that are able to impact the region around a massive black hole.
We start  deliberately from the idealised case of overlapping non-equilibrium spherical shells, and follow
Hobbs11 for generating the gaseous shells in our simulations (not including turbulence). 
A primitive disc is created by a first inflow event, mimicked as an infalling non-equilibrium rotating shell.  
A subsequent event is represented again by an initially rotating 
spherical shell around the black hole with its rotation
axis tilted with respect to the disc rotation axis. We will show that the impact of the two fluids
can trigger large inflows across the inner boundary, in particular when the shell
counter-rotates relative to the primitive, relic disc, and that a warped complex disc structure
forms when the flow settles into dynamical equilibrium. 
A  sizeable fraction of the angular momentum (around 70\%)is lost in a single event.

The paper is organized as follows:
We describe the computational method 
and the initial conditions in sections \ref{sec:meth} and \ref{sec: ic}. Section \ref{sec: the simulations}
describes the outcome of the SPH simulations, while 
sections \ref{sec: ang mom evol} and \ref{sec: ang mom cons} focus  on the angular
momentum evolution. Inflow rates are given in Section \ref{sec: inflow}, 
and the conclusions of this study are presented in section \ref{sec:conclusions}.

\section{Computational Method}\label{sec:meth}
The physical system under study consists of a gaseous disc rotating around a central black hole and a
gaseous shell initially surrounding both the black hole and the disc. The whole system is embedded in a spherical bulge. We follow the hydrodynamics of the gas using the three dimensional SPH/N-body code developed by \cite{Springel05gadget}, \textsc{Gadget-2}. 
We use $N_{\rm disc} = N_{\rm shell} = 0.5 \times 10^6$ SPH particles of mass 
$m_{\rm part} = 100 \msun$
for modelling the disc and shell respectively
(i.e. the total number of SPH particles is $N_{\rm tot} =10^6$). The smoothing lengths are fully 
adaptive, giving a smallest resolvable scale of $h_{
\rm min}=2.8 \times 10^{-2} \pc$. The number of SPH neighbours was fixed at $N_{\rm neigh} = 40$.

The black hole, at the relevant scales for our study, can be safely modelled as a Keplerian potential.
We follow Hobbs11 for setting up the scenario in which the gas 
evolves.  The background potential is fixed to be that of an isothermal non-singular sphere with 
core radius $\rcore$. Superimposed is the Keplerian potential of a central black hole of mass $\mbh$. 
For this model, the mass enclosed within radius $r$ is
given by:

\begin{equation}\label{potential}
M(r) = \mbh + 
 \begin{cases}
   \mcore (r / \rcore)^3, &r < \rcore \\
    M_{\rm bulge} (r/r_{\rm bulge}),  &r \geq \rcore,
  \end{cases}
\end{equation}
with  $r_{\rm bulge} = 1 \,\kpc$ and $M_{\rm bulge} = 10^{10} \msun$ being the characteristic length and mass scale for the spherical bulge. We use these as the units for our code. Accordingly,
$\mcore = 2 \times 10^{-2}$,
$\mbh = 10^{-2}$; 
$\rcore = 2 \times 10^{-2}$.
The unit of time is defined as $T_{u}= (r_{\rm bulge}^3/GM_{\rm bulge})^{1/2} \simeq 5 \, \myr$, and the velocity is in units of $V_{u} \simeq 208 \kms$. In these units the minimum smoothing length of the SPH particles $h_{\rm min}=2.8\times 10^{-5}$ and the gravitational constant is automatically set $G = 1$. The mass of the black hole, together with the mass distribution given by equation (\ref{potential}) are not active components of the simulation in the sense that they just set the background potential in which the gas is moving and do not change in time.

The self-gravity of the gas is neglected, so that the interaction among gas particles is purely hydrodynamical. Cooling has been 
modelled with an isothermal equation of state as we expect the cooling time scale to be much shorter 
than the dynamical time scale of the simulations (e.g. \citealt{Hopkins10sim, Thompson05}). We also set up an
accretion radius, $\racc$, such that every particle that 
comes inside that radius with low enough orbital energy and angular momentum is regarded as \textit{eventually 
accreted} and is immediately removed from the simulation. The accretion radius is, in code units, $\racc = 
10^{-3}$, corresponding to one parsec in physical units. This accretion radius approach is often implemented for 
sink particles in star formation simulations (see \citealt{Bate95}). Gas particles forming 
the shell are distinguished from particles forming the disc by adding a different tag to each set of particles, and we
keep a record of the number of accreted particles to compute the accretion rate and make a comparison among
the different simulations.

\section{Setting up the Initial Conditions}\label{sec: ic}

\begin{figure}
  \includegraphics[width=0.49\textwidth]{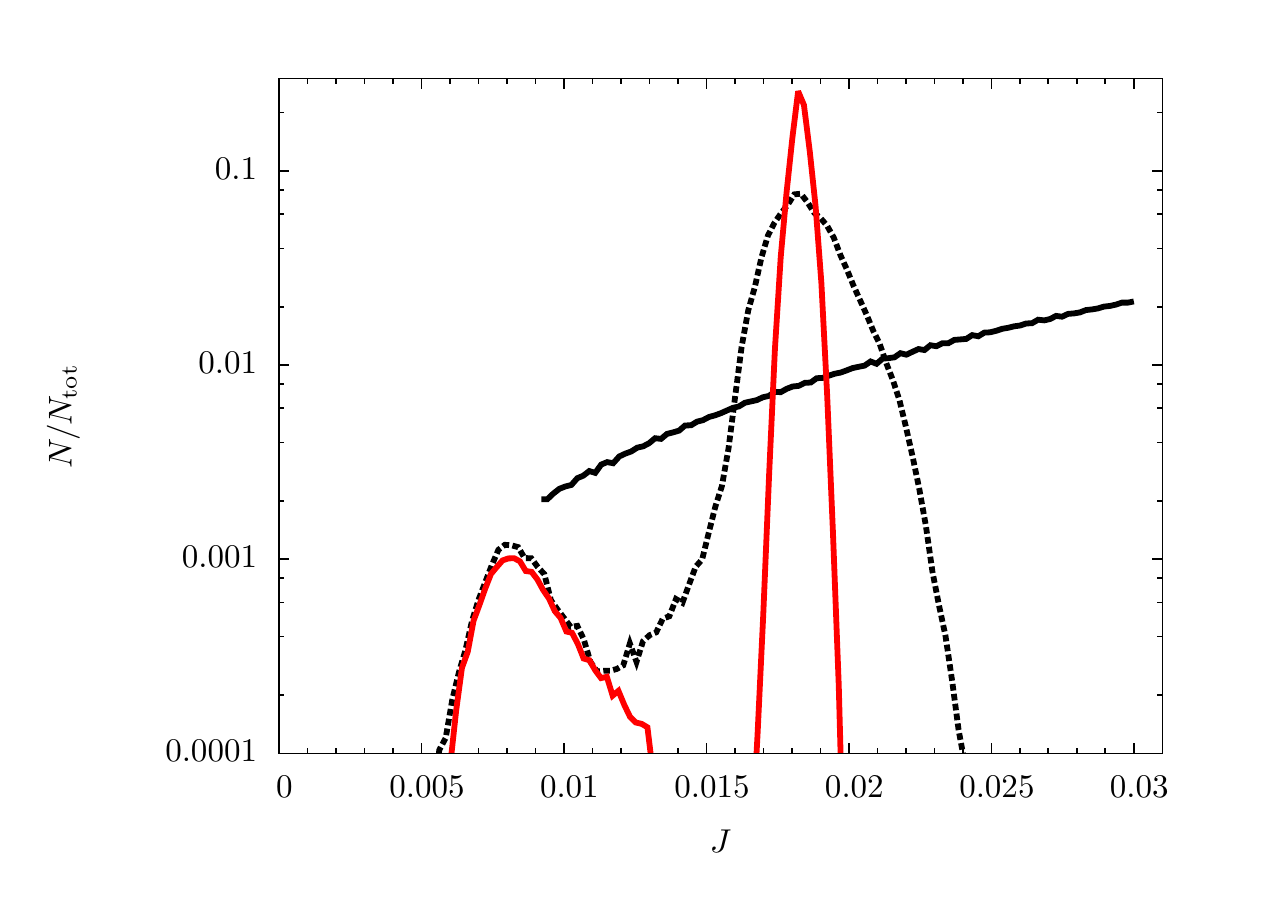}
  \caption{Evolution of the distribution of the specific angular momentum moduli 
 of the shell particles which will evolve into the primitive disc. The initial condition (solid black line) being that of a uniform shell rotating around the $z$ axis.  The black dotted line and the red line correspond to $t = 0.15$ and $t = 0.5$ respectively.}
  \label{fig: Fig5Hobbs}
\end{figure}

\begin{figure}
  \includegraphics[width=0.49\textwidth]{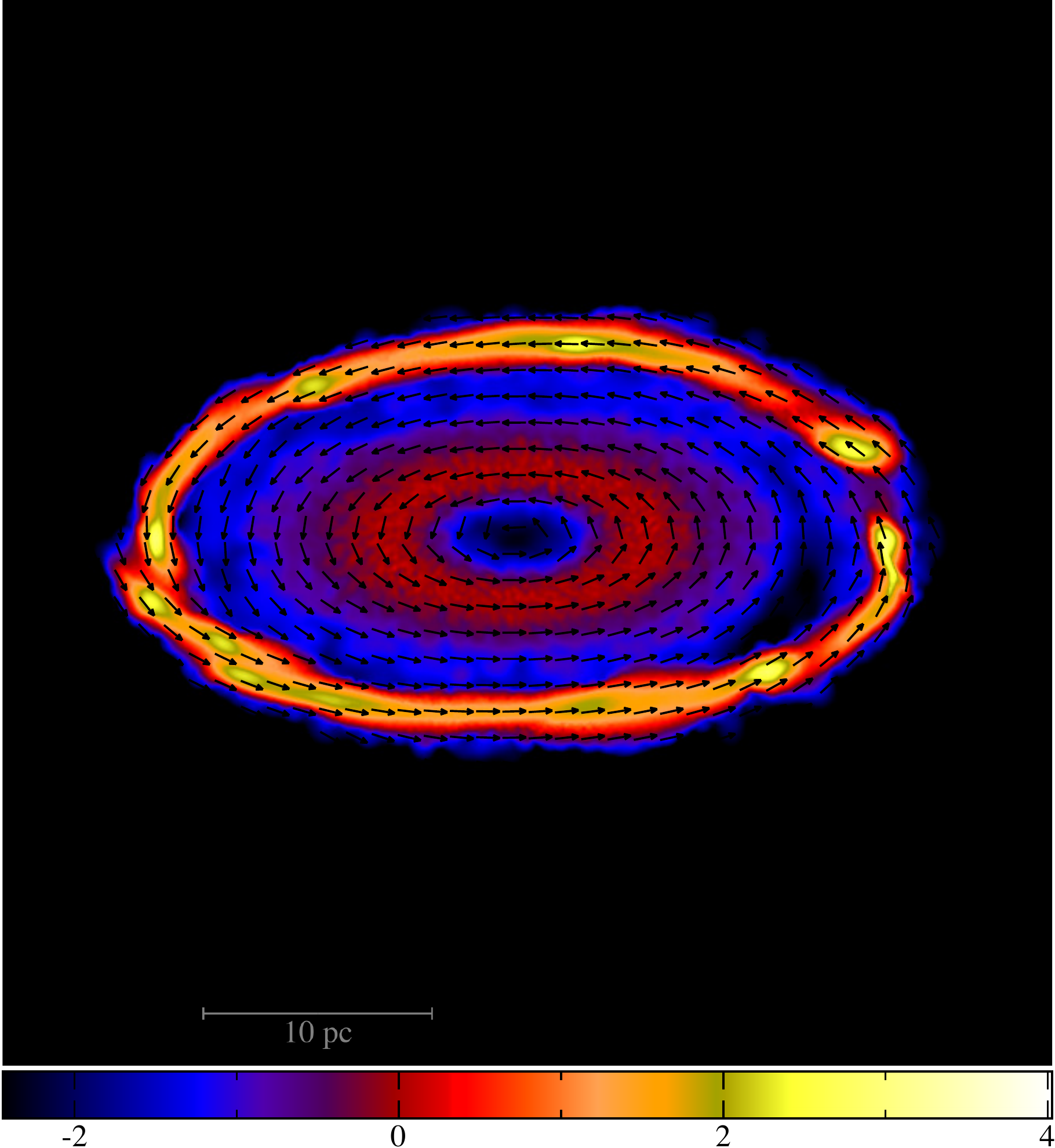}
  \caption{Disc that forms after  $t = 0.5$ of the evolution of a rotating homogeneous gaseous shell. This disc is the inner primitive disc used for the initial conditions in all our simulations. The colour map shows $\log \Sigma$ in code units, where $\Sigma$ is the projected column density. The arrows are not representative of the velocity profile, they just illustrate the flow direction.}
  \label{fig: prim disc}
\end{figure}

The system under study consists of a primitive gaseous disc of size $\sim 16 \pc$ around the central black hole, 
plus a spherical shell gas cloud of size $\sim 100 \pc$ centred also at the black hole, 
the two immersed in the galactic bulge + BH potential of 
equation (\ref{potential}). The inner disc is rotating in a stable configuration, while the
shell is initially set into rotation with a cylindrical velocity profile, not in virial equilibrium. The shell will start 
collapsing toward the central black hole due to the action of gravity and will interact with the disc.
 
The primitive disc is generated following Hobbs11, simulating the collapse of an initially rotating spherical gaseous 
shell to mimic the evolution and outcome of an earlier accretion event (hereafter we relax the meaning of 
``accretion'' to a generic inflow of gas towards $r_{\rm acc}$ ). We carve a spherical shell of $0.5 \times 10^6$ SPH 
particles from a cube of glass-like distributed particles, as this reduces the noise in the pressure force in SPH 
particles (see e.g. \citealt{Diehl12, Price2012}). The mass of the shell, $M_{\rm shell} = 5.1\times 10^{-3}$, is 
distributed homogeneously from $r_{\rm in} = 0.03$ to 
$r_{\rm out} = 0.1$ and is set into rotation around the $z$ axis with a cylindrical velocity profile 
$\vrot = 0.3$.

The  unstable shell is allowed to evolve under the gravitational influence
of the potential described in equation (\ref{potential}). We stress that the velocity profile, together with the density profile are idealized in order to capture the physics of the problem and to avoid unnecessary computational expense.

The rotating gas shell, falling towards the centre,  evolves into a disc-like structure.
In the first stages, the angular momentum has a quite flat distribution (Figure \ref{fig: Fig5Hobbs}). As the gas starts mixing, angular momentum is transported among the particles and, due to the symmetry of the problem, the $x$ and $y$ angular momentum components cancel out. At $t = 0.5$, the gas has already settled into a stable disc-like structure rotating around the centre of the potential. As angular momentum is conserved, $\Jdisc$ points in the $z$ direction, with a modulus $J_{\rm disc} = 0.018$ (in code units). Figure \ref{fig: prim disc} shows the state of the gas at this time, in circular orbits at a radius $r_{\rm circ} \simeq 0.016$ (16 pc in physical units). The gas has undergone an {\it angular momentum redistribution shock} (in agreement with figures 1 and 5 from Hobbs11). The final configuration, as noted by Hobbs11, is closer to a ring rather than a disc. For simplicity we will refer to this structure as primitive disc.

Once the disc is formed in the first place, we superimpose a new gaseous shell, replica of the pristine shell, this time adding a tilt to it and varying its initial $\vrot$. This shell mimics a second accretion event, uncorrelated with the first one.
Hereafter, $\vrot$ will denote the initial rotational velocity of the second shell infalling toward the central black hole, and $\tilt$ the initial angle between the disc and shell total angular momentum vectors, denoted as $\Ldisc$ and $\Lshell$ respectively.

We explore six different cases, resulting from scenarios in which the infalling shell has a smaller, equal and greater angular momentum than that of the disc combined with its two possible senses of rotation (co- or counter-rotation) with respect to the disc. We build each of these scenarios by setting the initial shell $\vrot = 0.2, 0.3, 0.7$ and a tilt angle $\tilt = 60^\circ, 150^\circ$. This will enable us to asses the impact of the sense of rotation and angular momentum on the dynamics of the composite system. The simulations are run for more than 10 dynamical times measured at $r = \rcore$, and throughout them, the accreted mass is $M_{\rm acc} < 0.1 M_{\rm sys}$ even for the most active event, so we will regard $M_{\rm disc}$,  $M_{\rm shell}$ and the total mass of the system $M_{\rm sys}$ as constants. 

In the following we introduce the specific angular momentum of each SPH-particle, ${\mathbf j}_i = {\mathbf r}_i \wedge {\mathbf v}_i $, and the total specific angular momentum of the disc,

\begin{equation}
\Jdisc = \frac{1}{N_{\rm disc}} \sum_{i = 0}^{N_{\rm disc}}{\mathbf j}_{i,{\rm disc}} ~,
\end{equation}

\noindent with modulus $J_{\rm disc}$. 
 We remark that initially the primitive disc has  $J_{\rm disc}= 0.018$ (in code units) in all simulations.
Similar notation is used to distinguish and describe the shell particles. 

In addition we introduce the mean angular momentum,
\begin{equation}
\left< {j_{\rm disc}} \right> = \frac{1}{N_{\rm disc}}\sum_{i = 0}^{N_{\rm disc}} j_{i,{\rm disc}}~,
\end{equation}

\noindent for the disc particles, with an equivalent definition for the shell particles describing the angular momentum content of each component of the system in terms of the moduli $j_i = |{\mathbf j}_i|$.

\section{The simulations}\label{sec: the simulations}

\begin{figure*}
\includegraphics[width=0.24\textwidth]{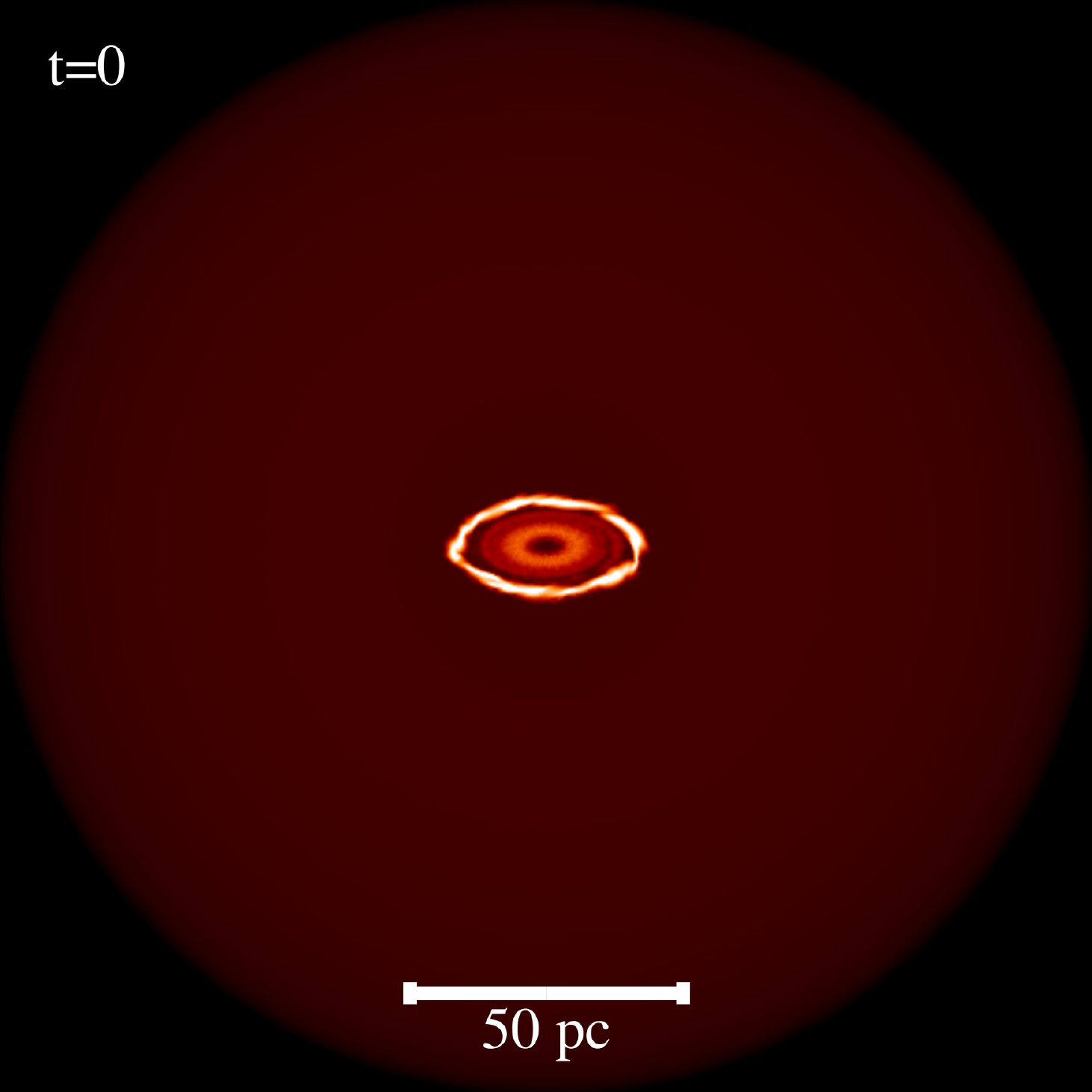}
\includegraphics[width=0.24\textwidth]{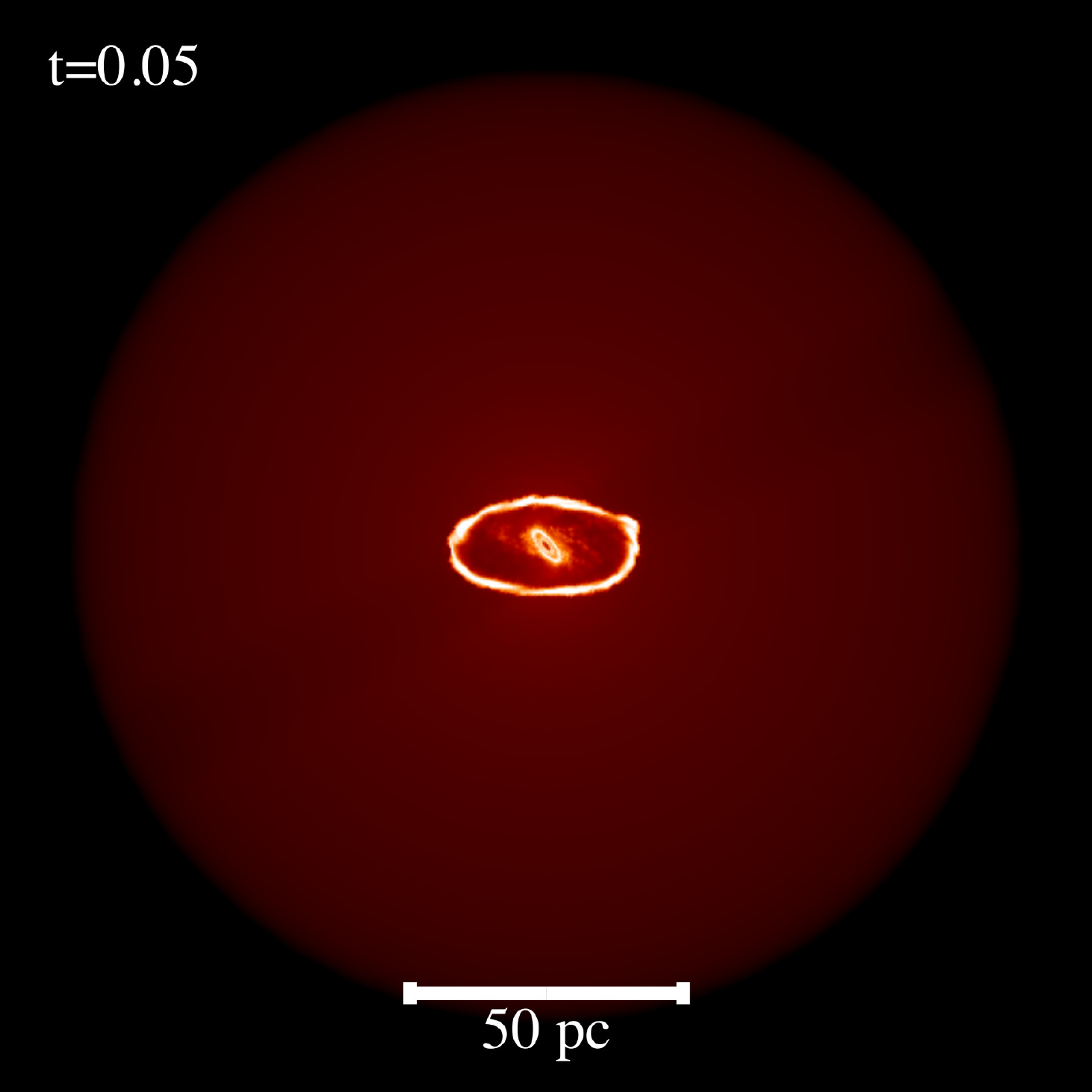}
\includegraphics[width=0.24\textwidth]{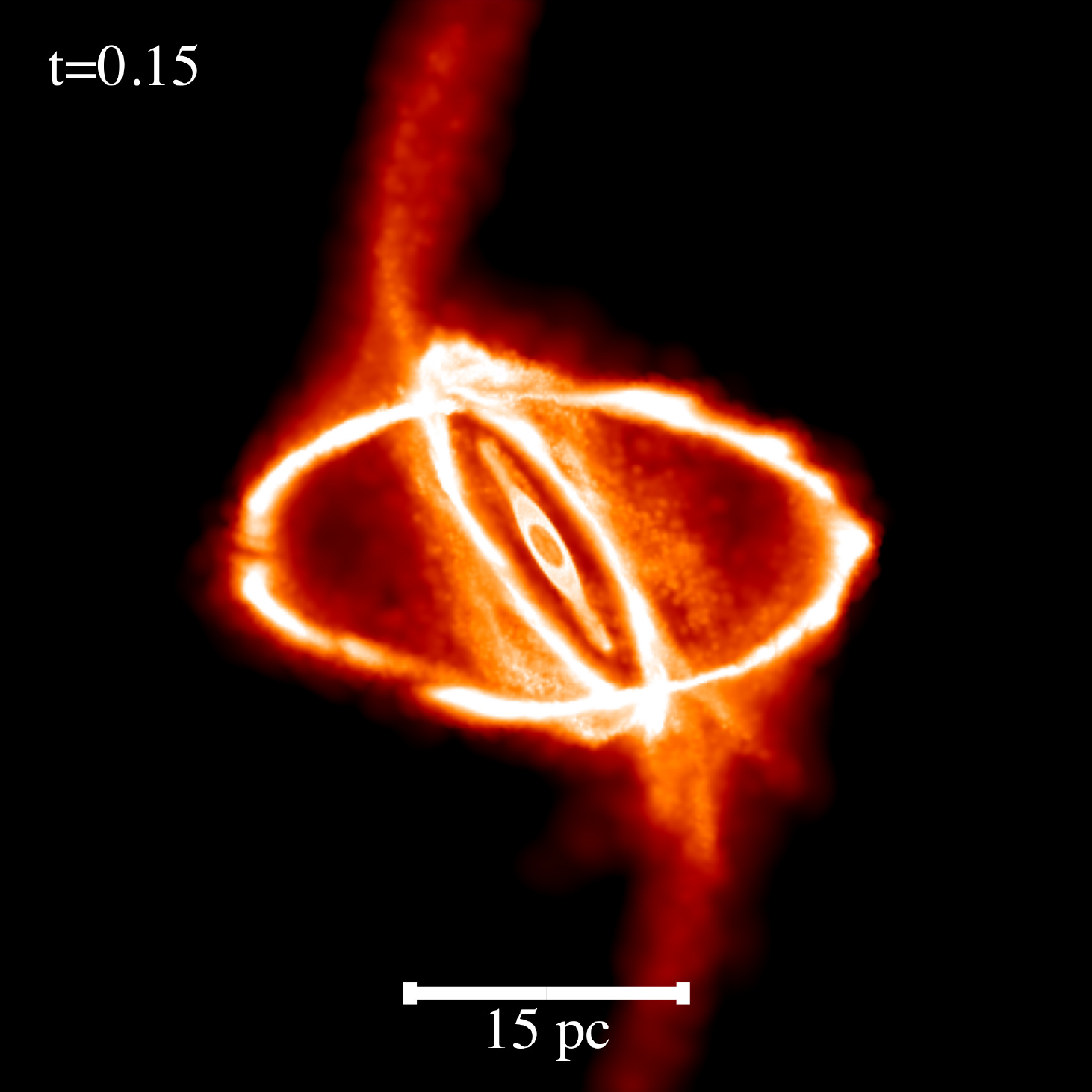}
\includegraphics[width=0.24\textwidth]{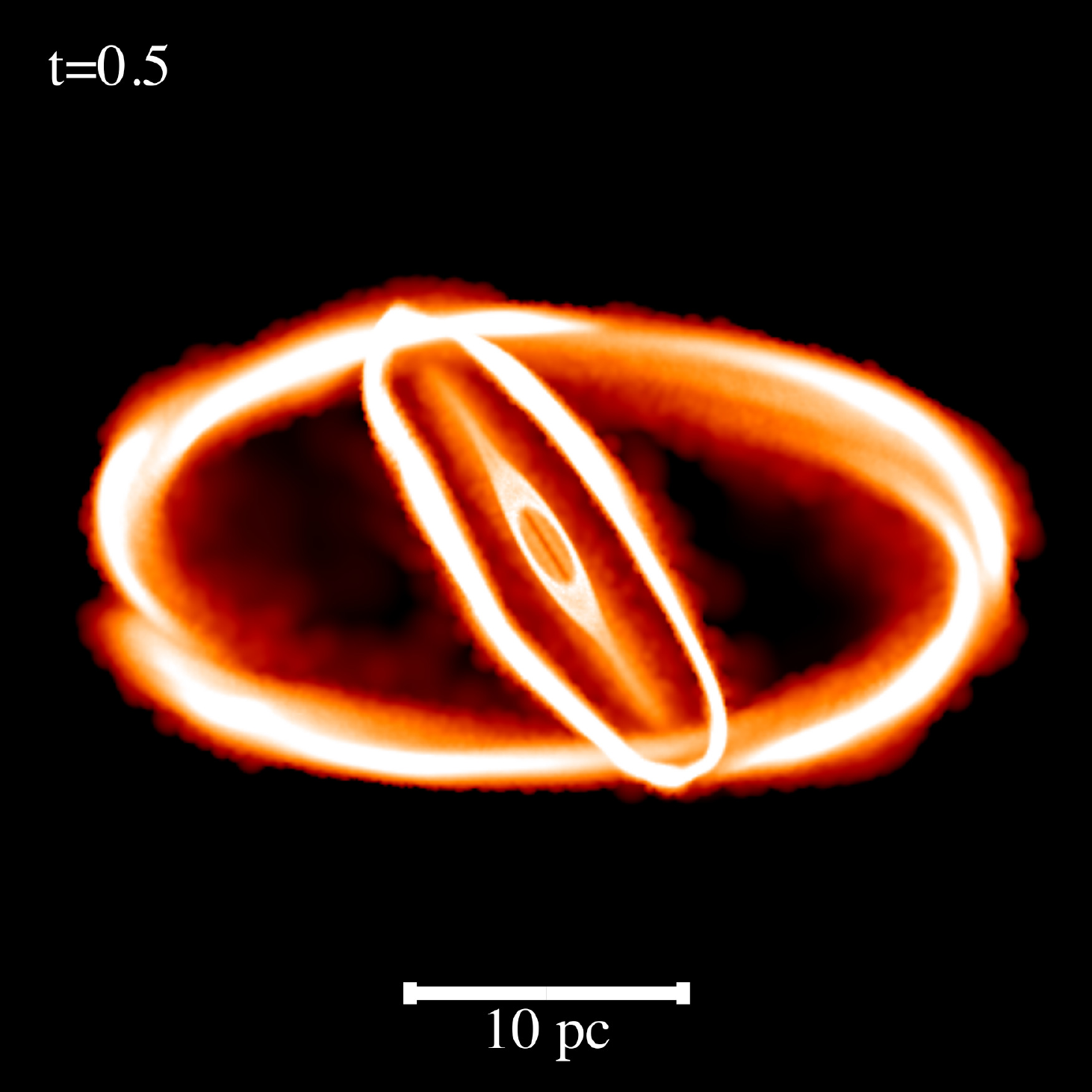}\\

\includegraphics[width=0.24\textwidth]{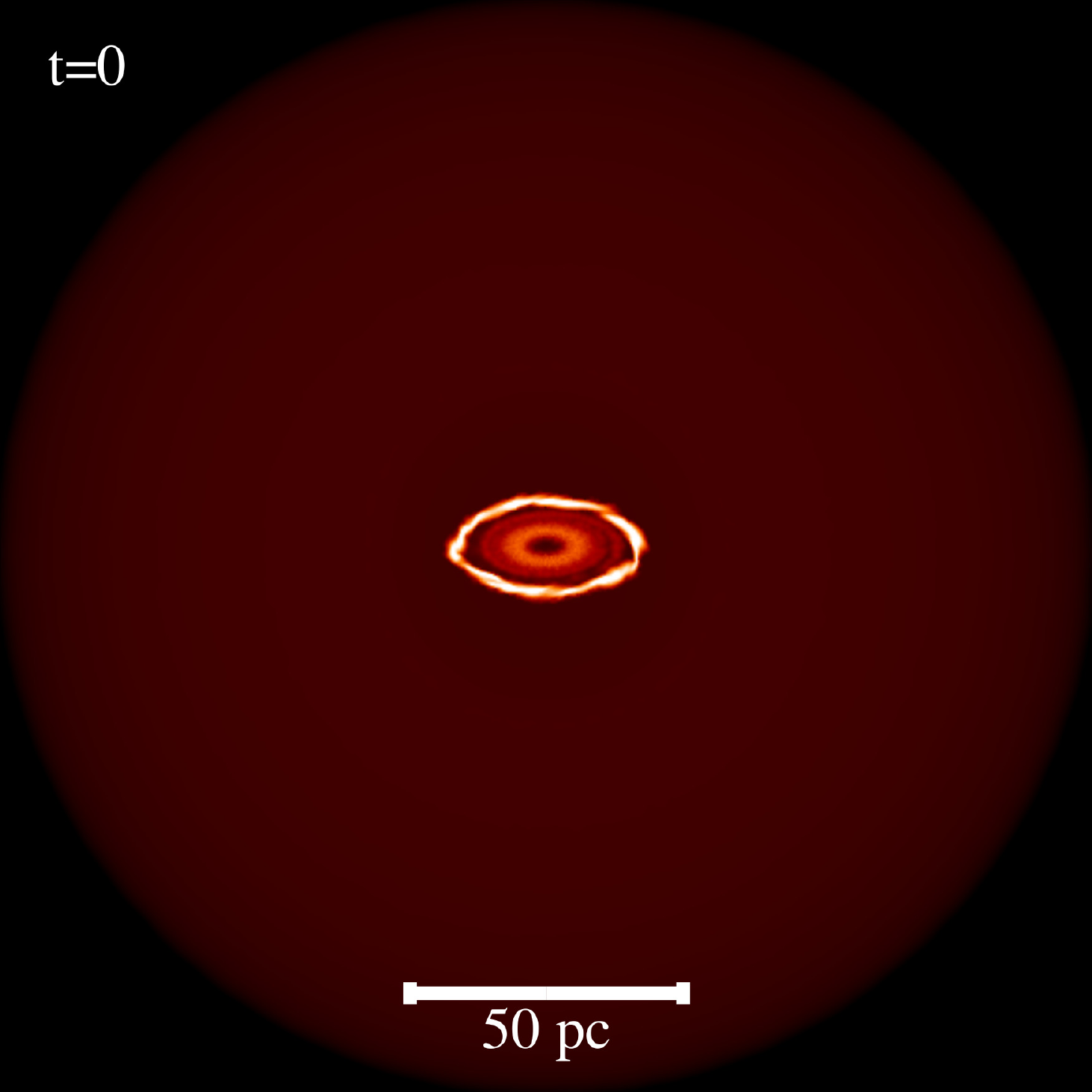}
\includegraphics[width=0.24\textwidth]{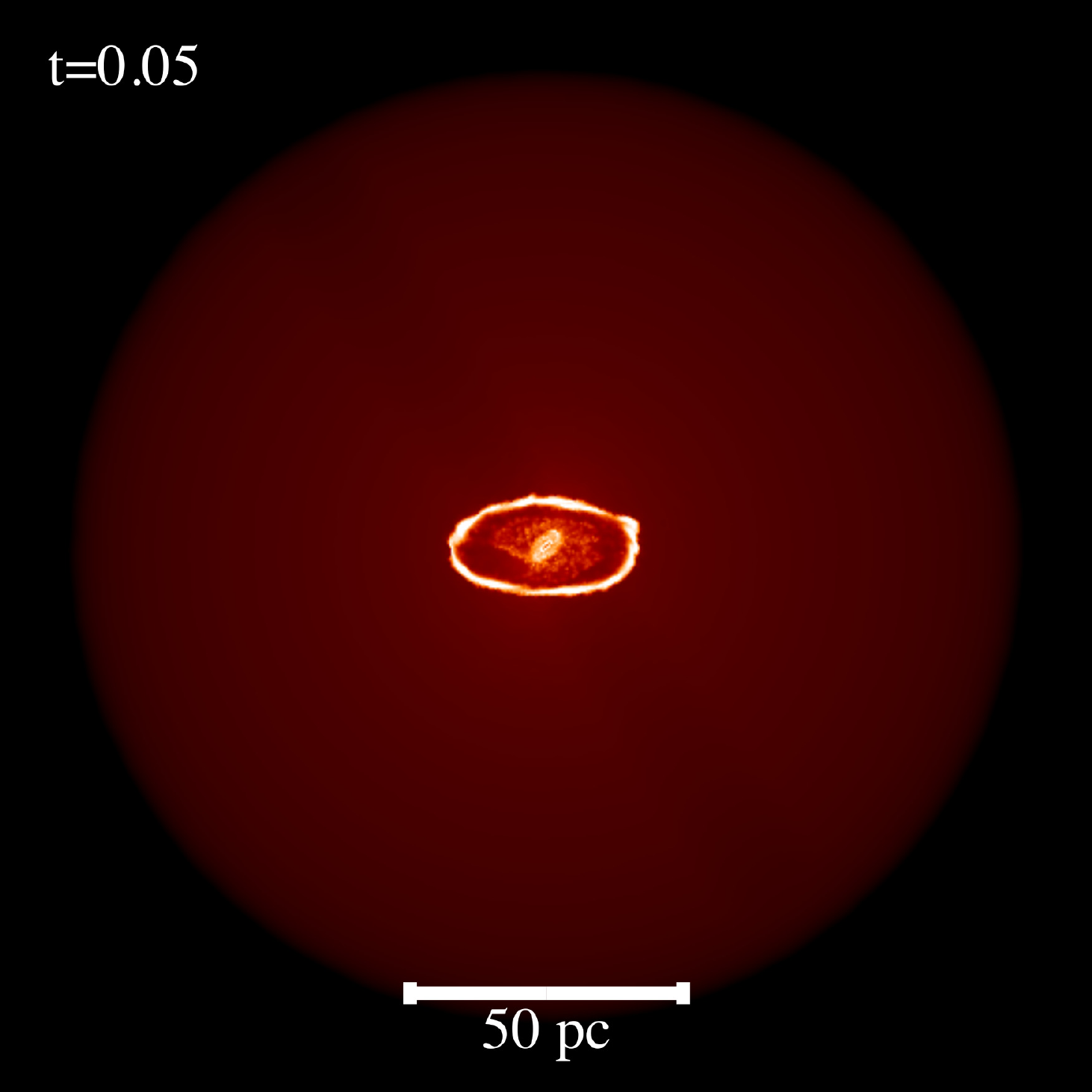}
\includegraphics[width=0.24\textwidth]{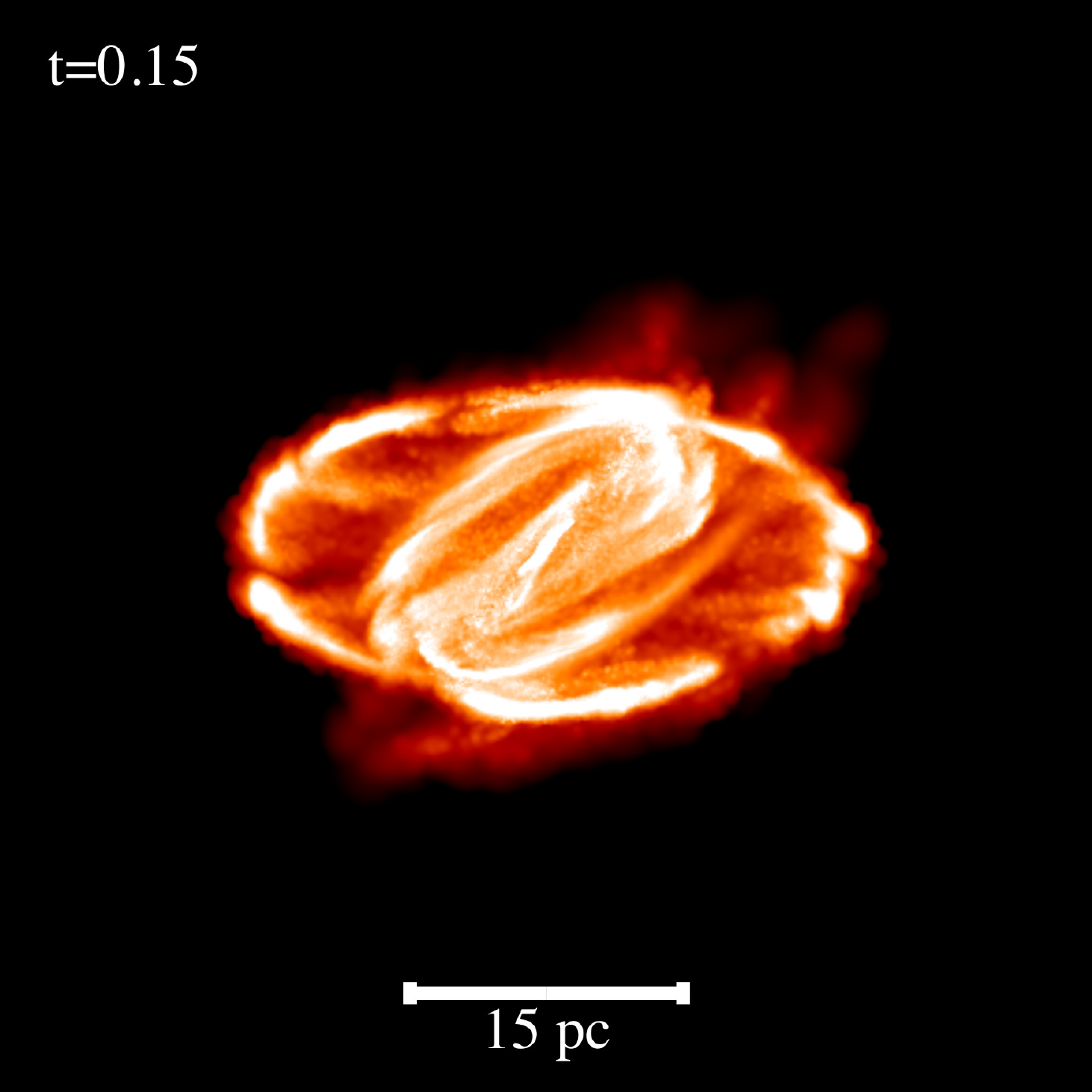}
\includegraphics[width=0.24\textwidth]{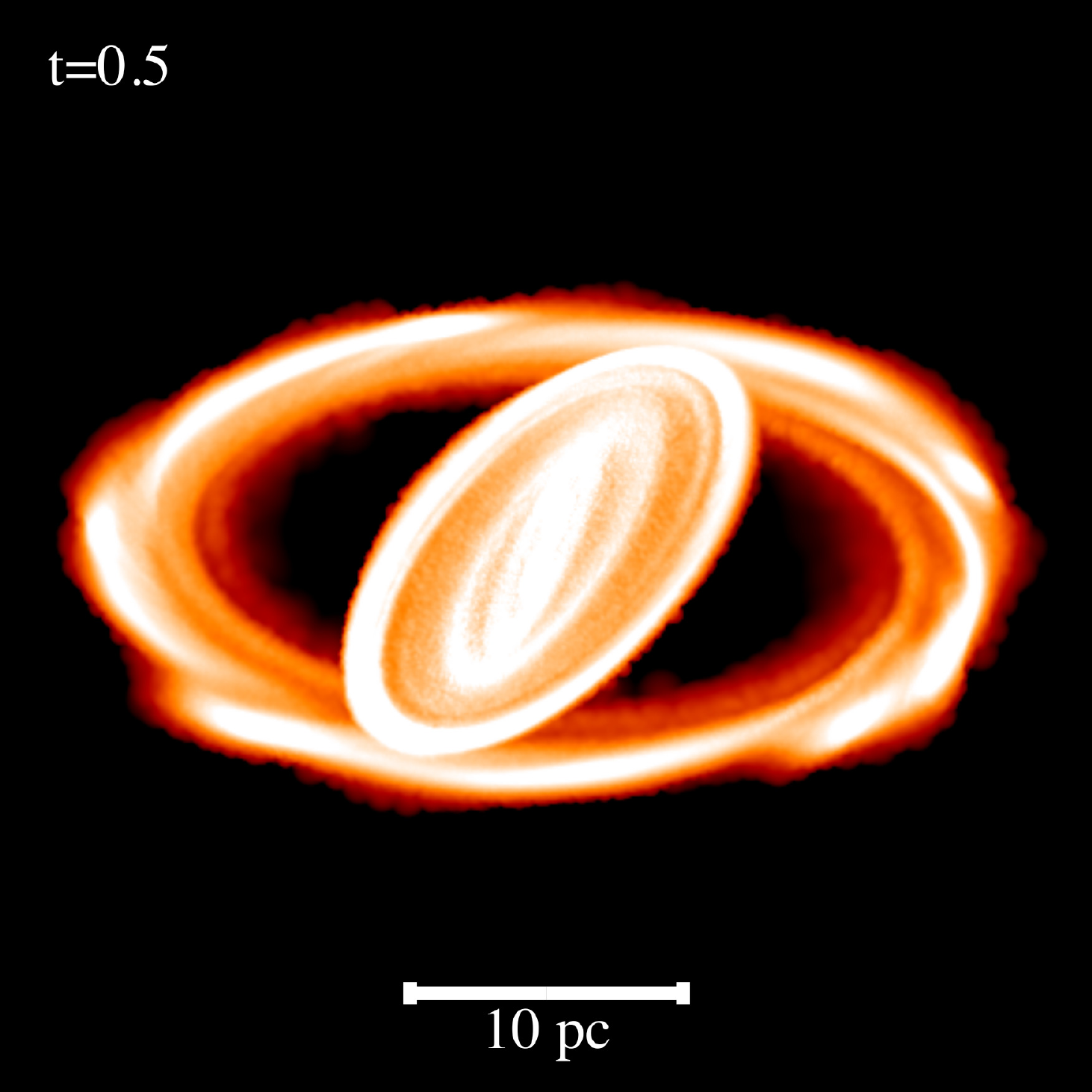}\\
\caption{Evolution of nested accretion events with $J_{\rm shell} < J_{\rm disc}$, corresponding to $\vrot = 0.2$. The upper and lower rows correspond to shell initial tilt angles of $\tilt = 60\degree$ and $150 \degree,$ respectively. From left to right, the time at which snapshots are taken are $t = 0, 0.05, 0.15, 0.5$.  See text for details. Note the change of scale between the different snapshots: 50 pc in the first two, 15 pc and 10 pc in the last two.}
\label{fig: evolutions v02}
\end{figure*}

Figures \ref{fig: evolutions v02}, \ref{fig: evolutions v03} and \ref{fig: evolutions v07} show four different stages in the evolution of the disc+shell simulated 
systems, from the beginning to the new final state of equilibrium after interaction 
has subsided. In all the six cases, the shell is initially collapsing to form a disc from the 
inside out, as particles with lower angular momentum (i.e. smaller circularization 
radius) fall first. These particles will be the first to interact with the primitive disc. Mixing of shell and 
disc particles occurs with different strength depending on $\vrot$ and $\tilt$. We 
describe the final fate of each system separately in the following subsections. For 
sake of clarity, we take $t = 0$ as the time the second shell starts falling towards 
the previously formed disc. In all the cases, the shell particles are exterior to the primitive disc,
at the time of initial infall \footnote{Videos of the simulations can be found at \url{http://people.sissa.it/~jcarmona/my_videos.php}. The snapshots shown in the paper are made using \textsc{Splash} (\citealt{PriceSplash07}).}.

\subsection{Smaller angular momentum event: $J_{\rm shell} < J_{\rm disc}$}

We first consider a shell with $\vrot = 0.2$, corresponding to $J_{\rm shell} = 0.012 < J_{\rm disc}$. Under these initial conditions the shell has less rotational support against gravity than the shell that gave rise to the primitive disc. If the primitive disc were not present, the outcome of these two simulations would be a tilted copy of the primitive disc, this time with tighter orbits. We next describe the evolution for the shell having $\tilt = 60 \degree$ and $\tilt = 150\degree.$

\subsubsection{Co-rotating shell, $\tilt = 60 \degree$}
Figure \ref{fig: evolutions v02} shows the evolution of the system into a configuration of nested discs. 
First the shell stops falling and two arms develop due to the shell particles impinging on the primitive disc at an 
angle. We notice that these arms do not develop in the other simulations, meaning that they form due to the high 
density flows from the shell that impact the lower density inner regions of the primitive disc, kicking the gas in 
wider orbits that are observed as arms. However, this is just a transient signature, as the arms completely 
disappear after less than two orbits, leaving two tilted discs with a ring like structure each, the innermost one 
rotating in the sense of the shell.

\subsubsection{Counter-rotating shell, $\tilt = 150 \degree$}
In the counter-rotating case depicted in Figure \ref{fig: evolutions v02}, the shell starts falling toward the centre and very rapidly forms an inner tiny dense disc. The shell drags in its way some gas from the primitive disc. After that, the tiny disc starts twisting and soon a very tight spiral pattern develops, forming a wider disc that extends from the centre to some region inside the primitive disc. This disc develops a wider structure (as opposed to the two ring-structure formed in the previous case) due to the interaction of opposite flows. As gas from the shell was falling, it started breaking via shocks with gas from the disc, giving to the particles a wide spectrum of angular momenta and thus, making them able to form an inner counter-rotating disc and a ring-like outer disc comprising most of the particles of the primitive disc.

\subsection{Equal angular momentum event: $J_{\rm shell} = J_{\rm disc}$}

\begin{figure*}
\includegraphics[width=0.24\textwidth]{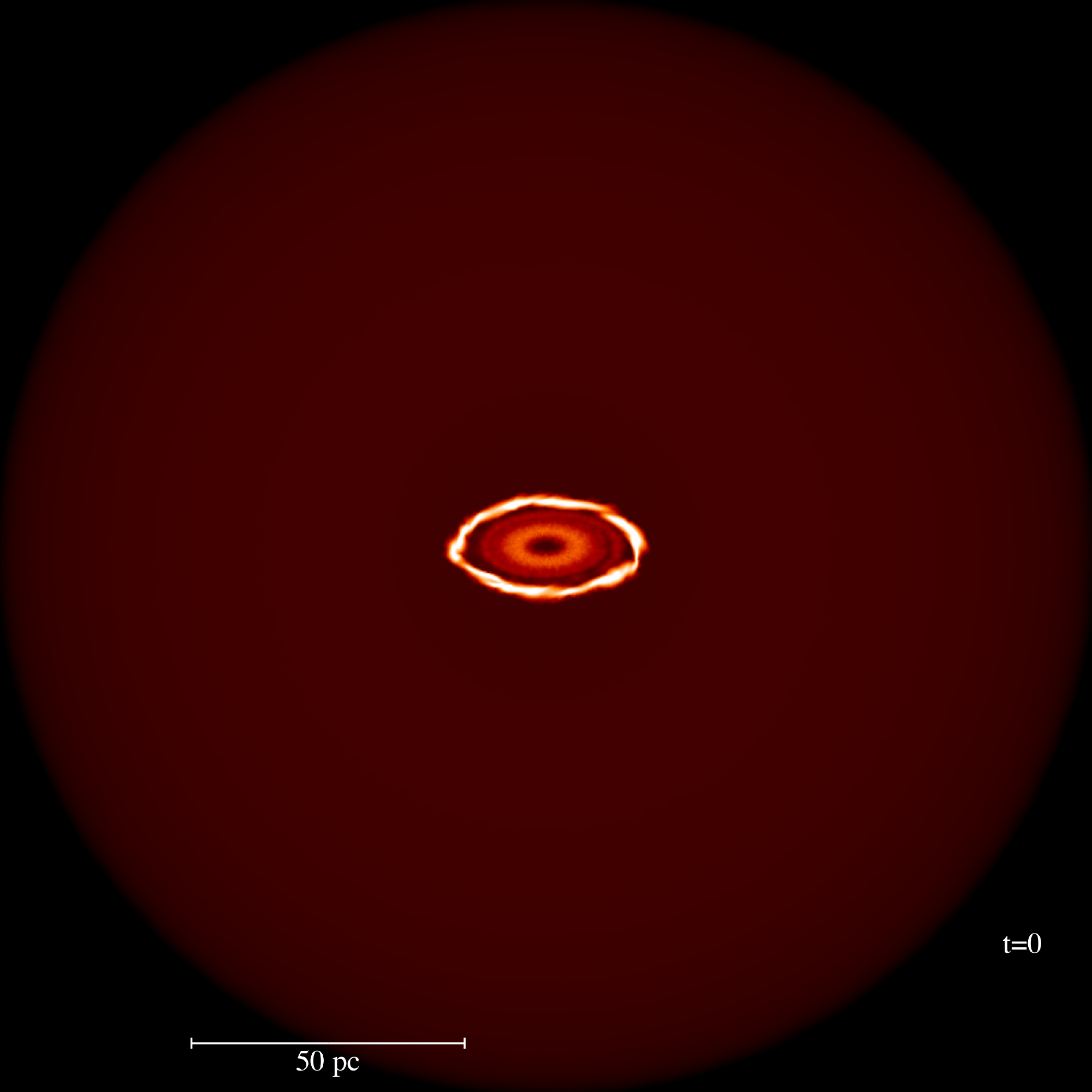}
\includegraphics[width=0.24\textwidth]{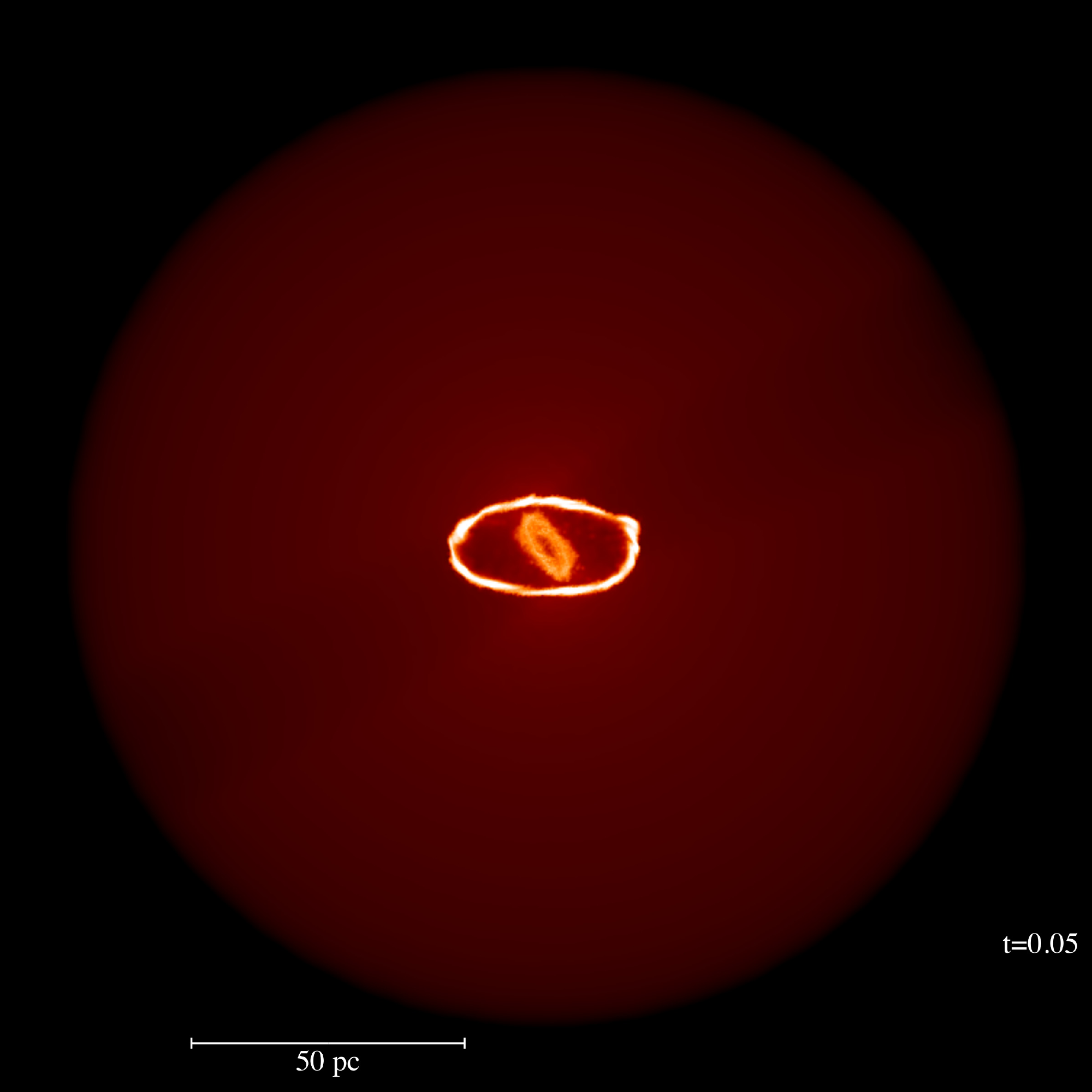}
\includegraphics[width=0.24\textwidth]{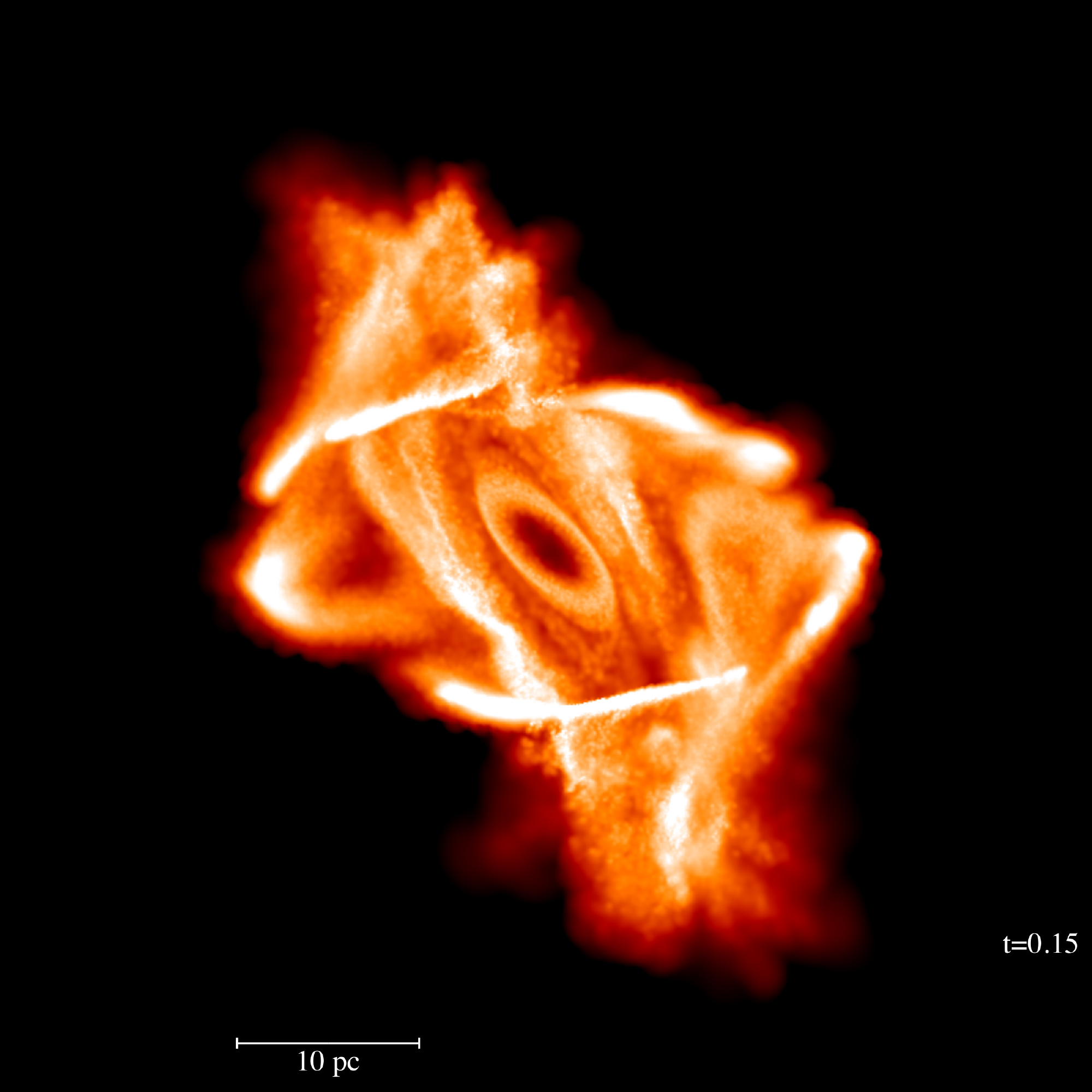}
\includegraphics[width=0.24\textwidth]{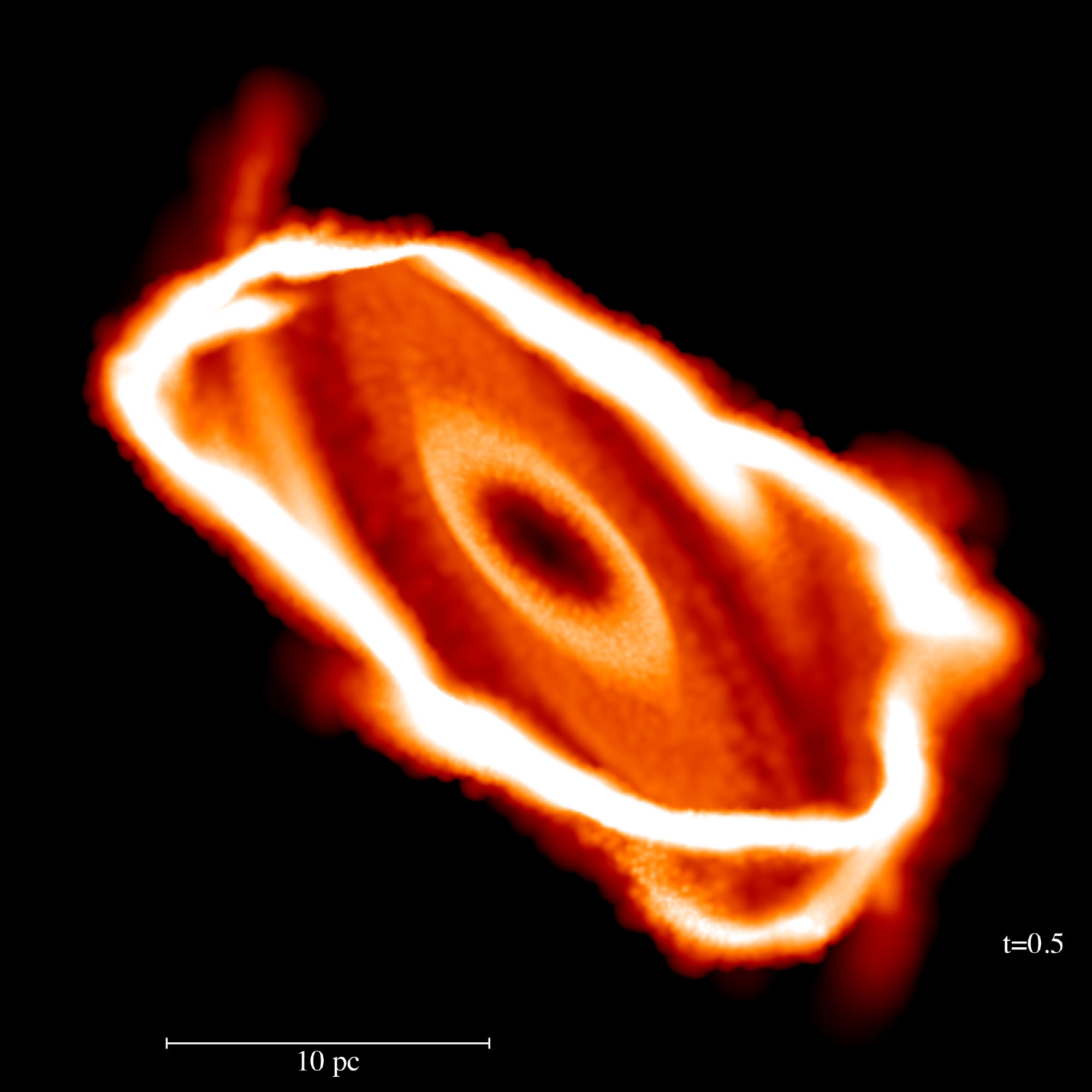}\\

\includegraphics[width=0.24\textwidth]{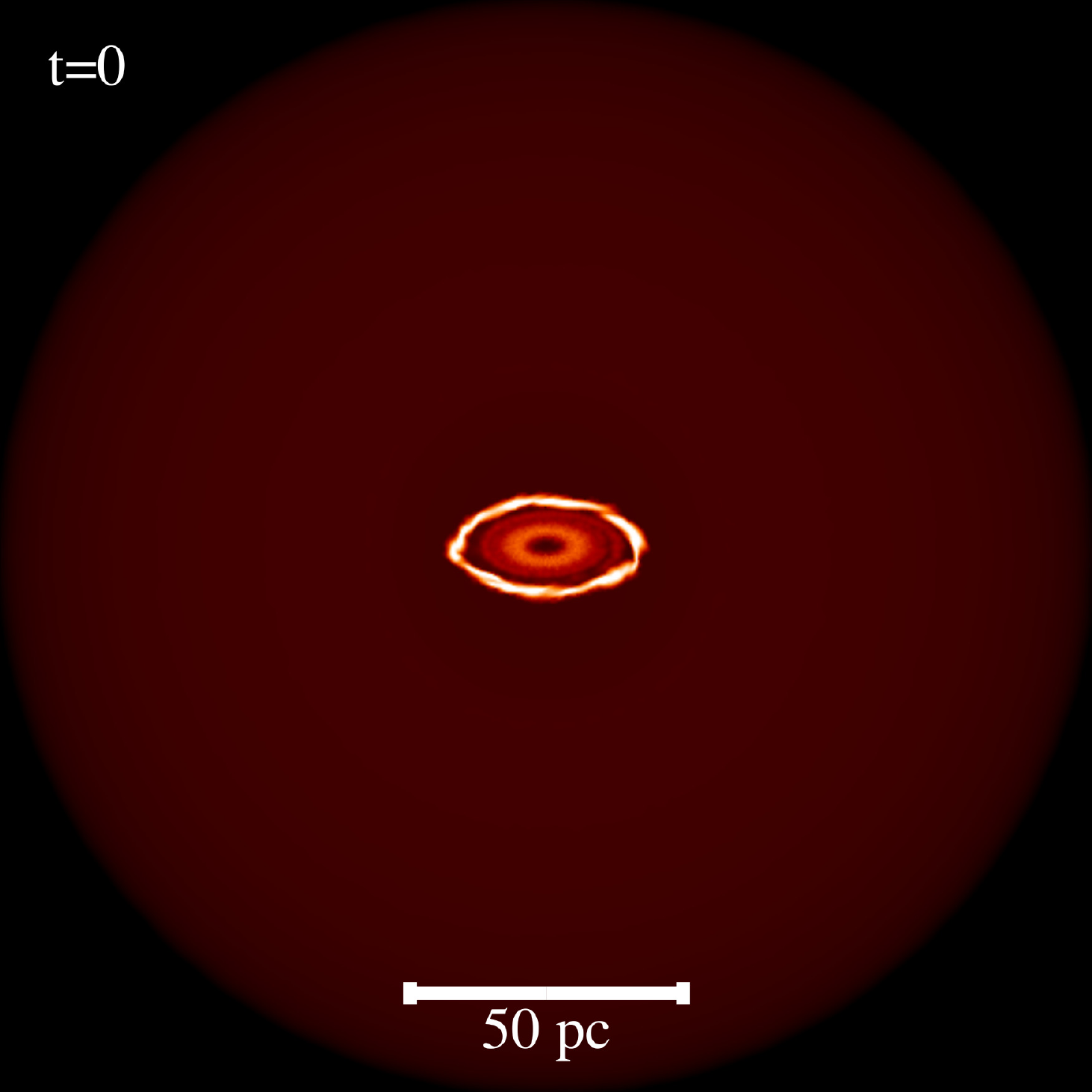}
\includegraphics[width=0.24\textwidth]{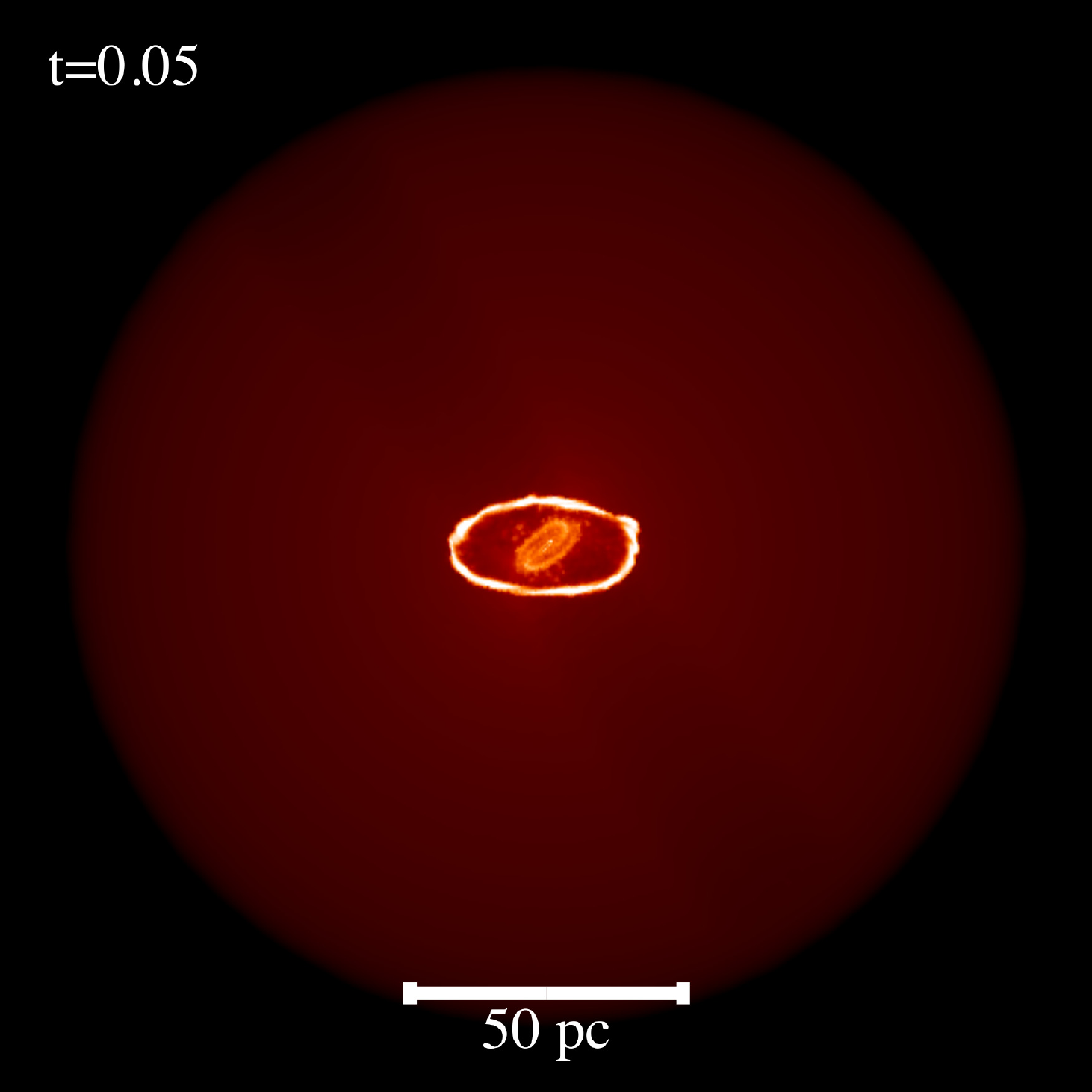}
\includegraphics[width=0.24\textwidth]{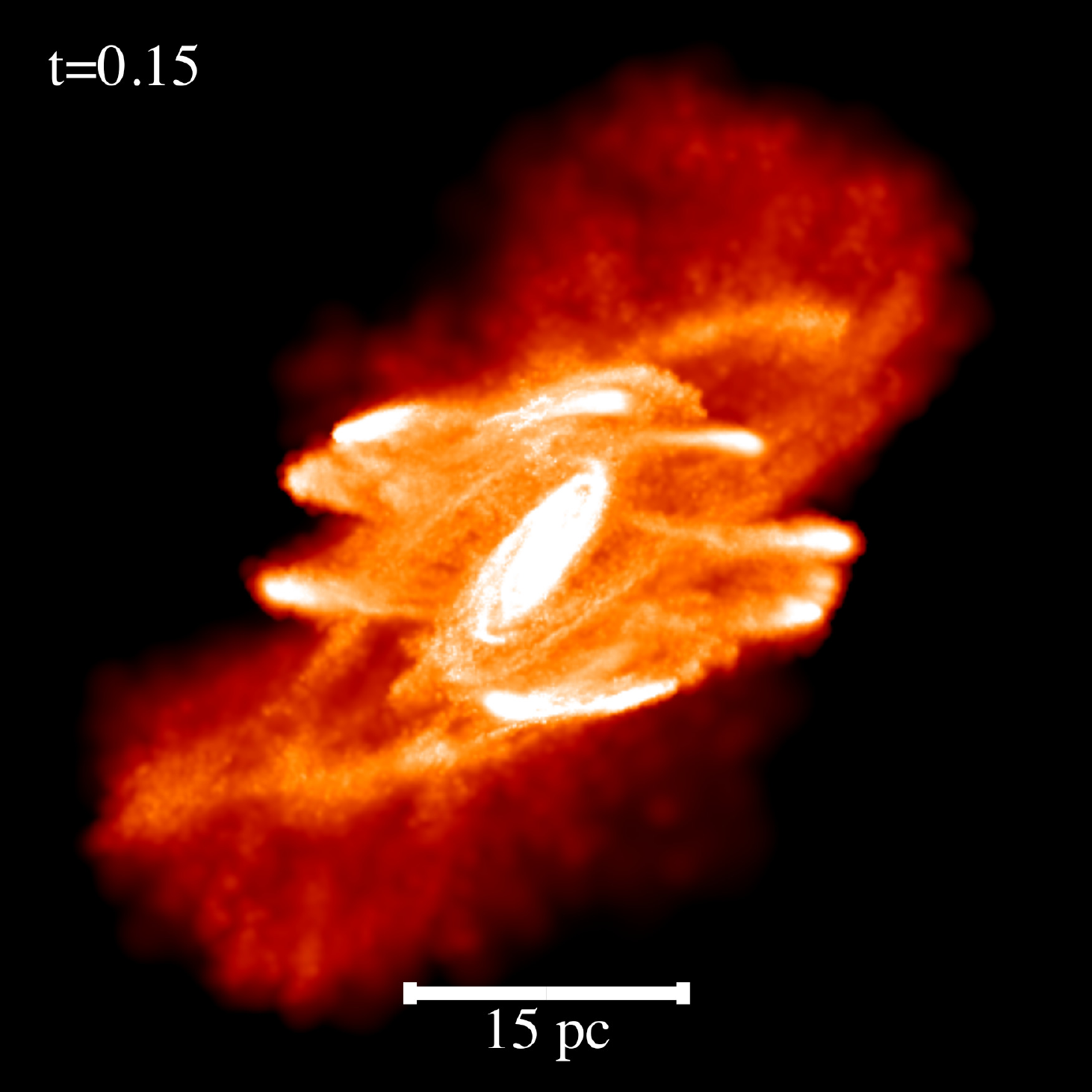}
\includegraphics[width=0.24\textwidth]{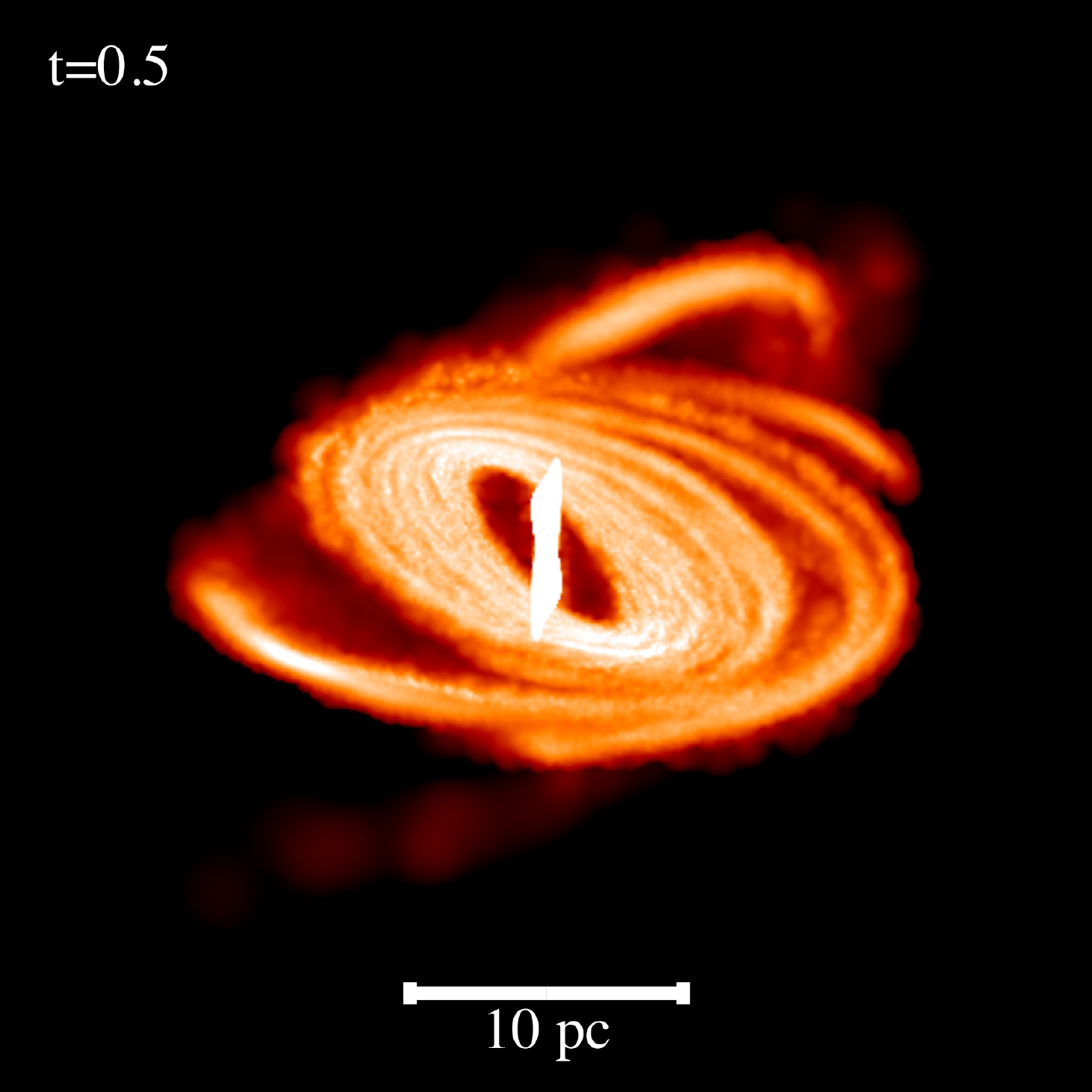}\\
\caption{Evolution of nested accretion events with $J_{\rm shell} = J_{\rm disc}$, corresponding to $\vrot = 0.3$. The upper and lower rows correspond to shell initial tilt angles of $\tilt = 60\degree$ and $150 \degree,$ respectively. From left to right, the time at which snapshots are taken are $t = 0, 0.05, 0.15, 0.5$. See text for details. Note the change of scale between the different snapshots: 50 pc in the first two, 15 pc and 10 pc in the last two.}
\label{fig: evolutions v03}
\end{figure*}

We now consider the case of an outer shell with an initial velocity $\vrot = 0.3$ (Figure \ref{fig: evolutions v03}), corresponding to a total angular momentum per unit mass $J_{\rm shell} = J_{\rm disc} = 0.018$. Under these conditions, the shell interacts with the pristine disc that, before forming, had the same initial angular momentum distribution and mass content.

\subsubsection{Co-rotating shell, $\tilt = 60 \degree$}

The first row of Figure \ref{fig: evolutions v03} shows the evolution of the system in which the
shell co-rotates relative to the disc with $\tilt  = 60\degree$.
Shell particles crossing the plane orthogonal to $\Lshell$ from above interact via shocks with shell particles 
crossing from below, simultaneously interacting with the  disc formed in the first accretion event. 
Those particles that lost angular momentum via shocks would flow toward the centre of the potential. However, 
the interaction with the primitive disc favours the formation of an inner  tilted disc that would not have formed 
otherwise. This occurs at about a time $t=0.05$. As the shell continues to infall, the innermost recently formed 
disc remains unaffected. The remaining shell particles start forming a new disc that would evolve into a copy of the 
primitive disc if the latter were not present. At $t = 0.15$, the shell stops infalling and forms a disc-like structure 
tilted at $\sim 60^\circ$ with respect to the primitive disc. The interaction at this point becomes very violent, the 
gas from both accretion events is shock-mixed and starts oscillating around a plane tilted at $~30^\circ$ with 
respect to the $x-y$ plane of the primitive disc, until a state of equilibrium is reached at around $t = 0.5$. The 
gas now settles quietly into a disc-like configuration of radius $\sim 10 \pc$ with an angular momentum 
distribution peaked at around $30^\circ$, as described in section \ref{sec: v03tilt060}. Memory is lost of the 
previous orientations of the disc and shell.

\subsubsection{Counter-rotating shell, $\tilt = 150 \degree$}
Consider now the case of a shell with rotation axis tilted by $\tilt = 150^\circ$ relative to $\Jdisc$. 
The interaction between the two flows soon develops an inner dense disc rotating in the sense of rotation of the shell. Contrary to the co-rotating equal-angular-momentum case, at $t = 0.05$, this inner disc continues growing until $t \sim 0.15$. This is due to the cancellation of angular momentum between the events: shell particles that would otherwise remain on larger orbits (in absence of the primitive disc) are braked into tighter orbits, adding  mass to the inner disc. This example is considerably more violent than the previous one, and the outcome is a complex structure: a very tight and dense disc orbiting the black hole in an orbit almost coplanar with the $y-z$ plane (i.e. tilted $\sim 90^\circ$ relative to $\Jdisc$), a larger disc mainly formed by particles from the primitive disc and a tail of shell particles counter-rotating and tilted (relative to $\Ldisc$) by $\sim140\degree$.
In addition we note the presence of a spiral pattern in the bigger disc rotating around the $z$ axis.

\subsection{Larger angular momentum event: $J_{\rm shell} > J_{\rm disc}$}

\begin{figure*}
\includegraphics[width=0.24\textwidth]{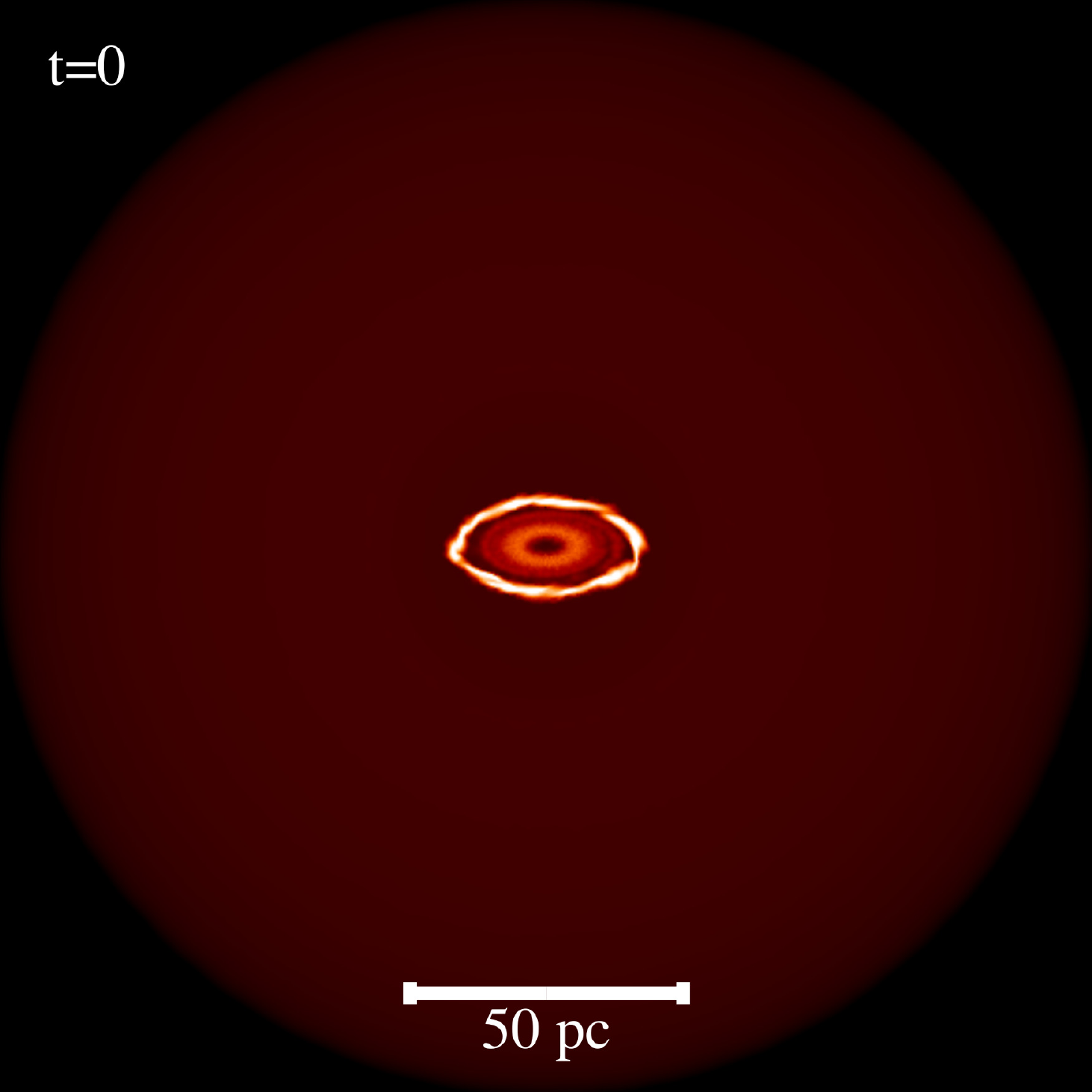}
\includegraphics[width=0.24\textwidth]{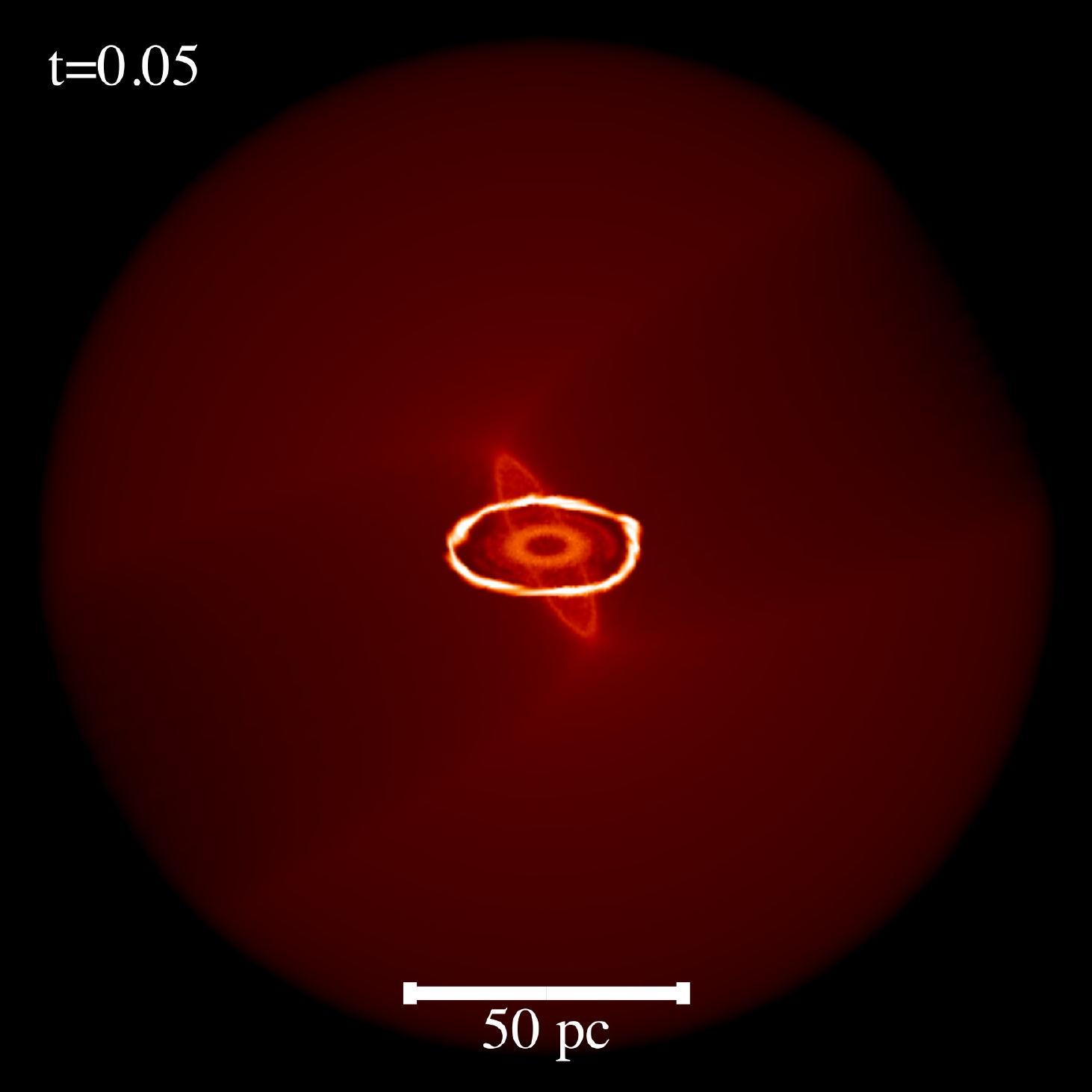}
\includegraphics[width=0.24\textwidth]{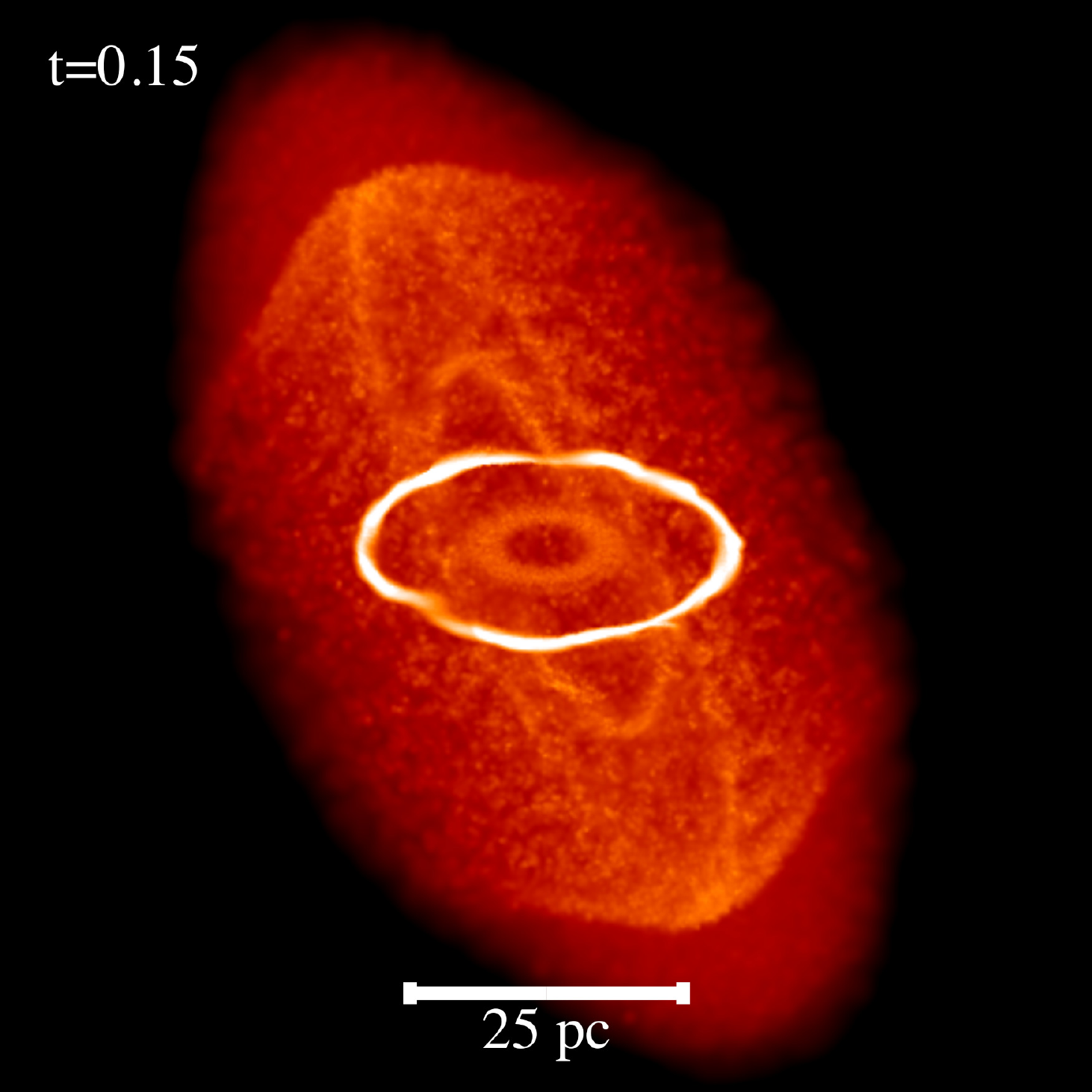}
\includegraphics[width=0.24\textwidth]{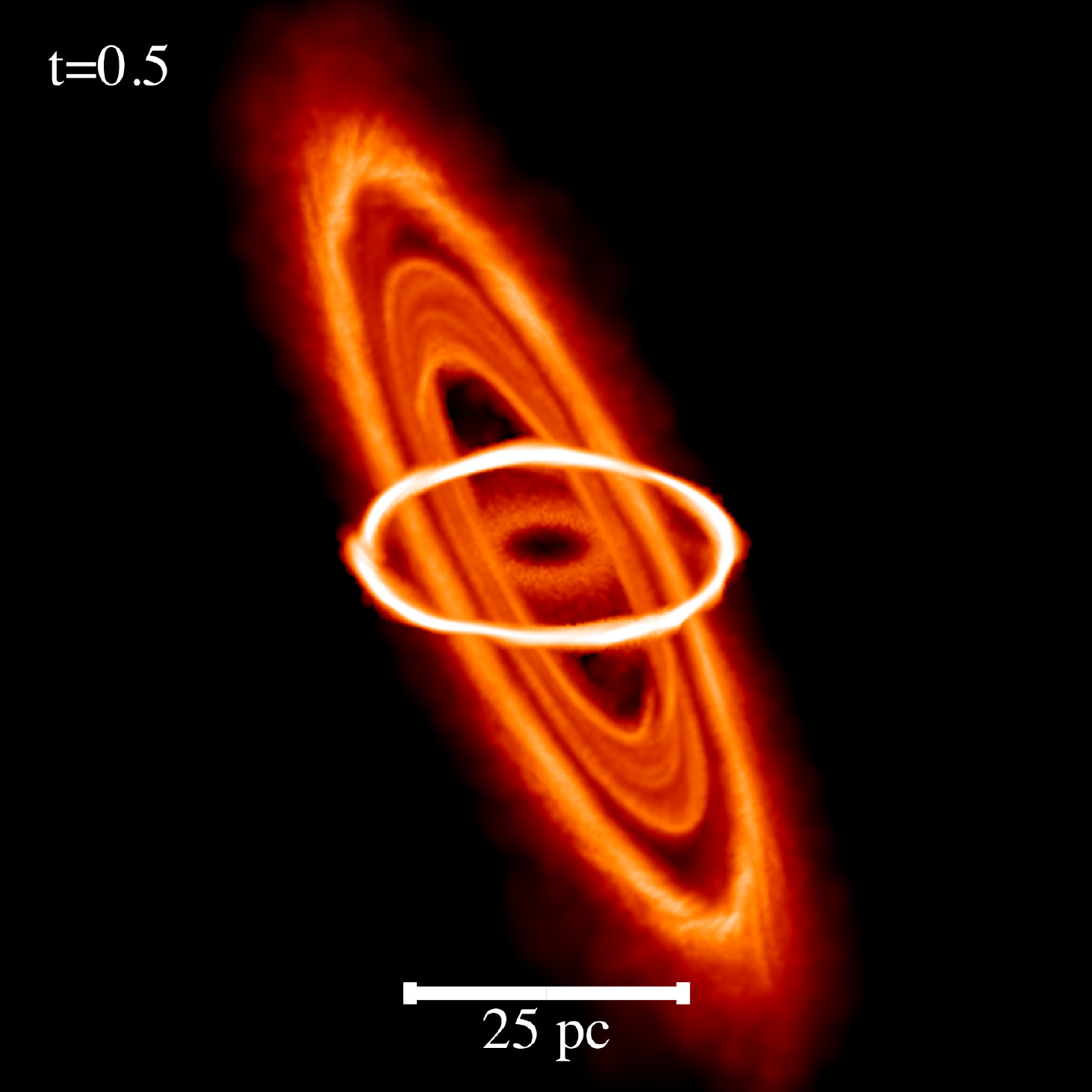}\\

\includegraphics[width=0.24\textwidth]{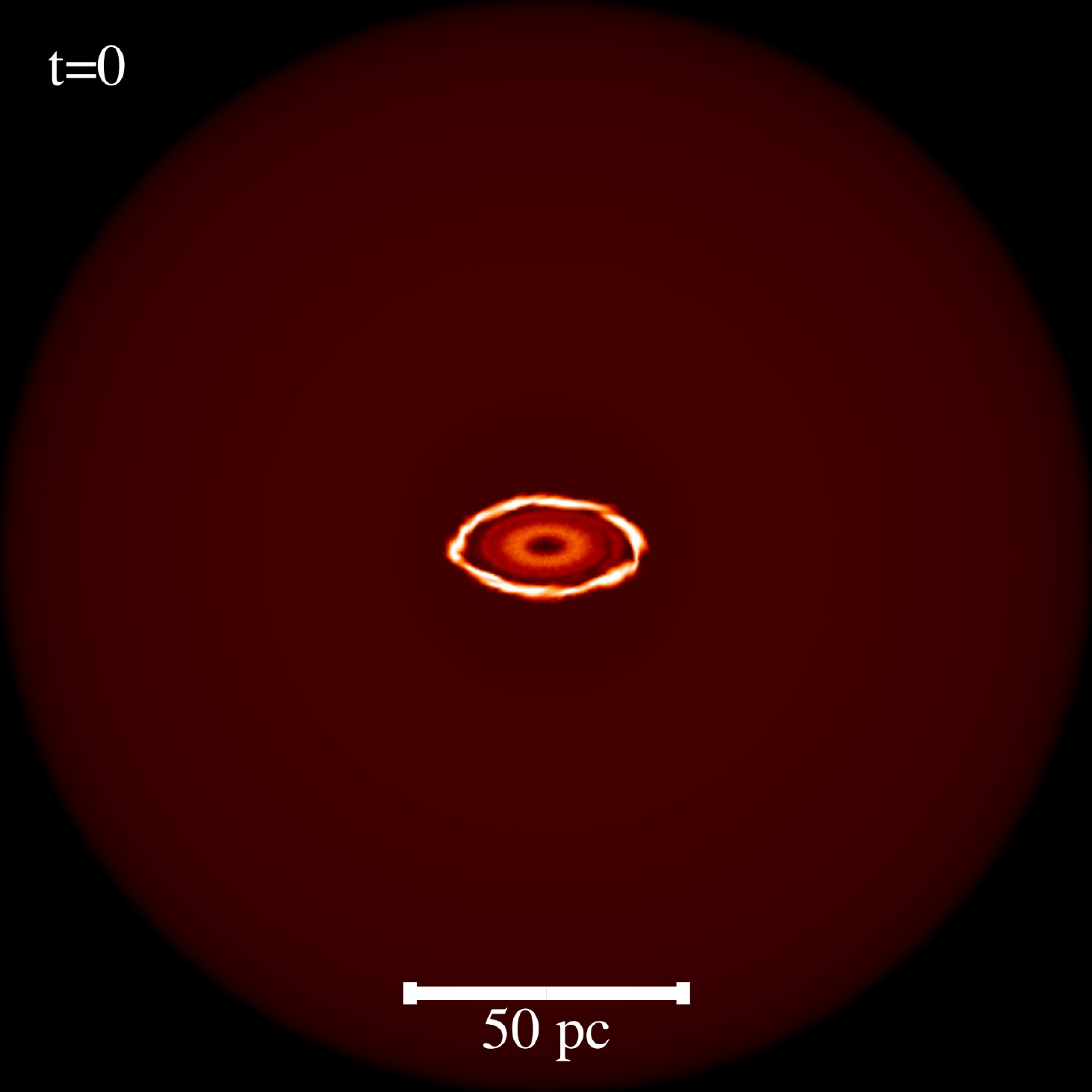}
\includegraphics[width=0.24\textwidth]{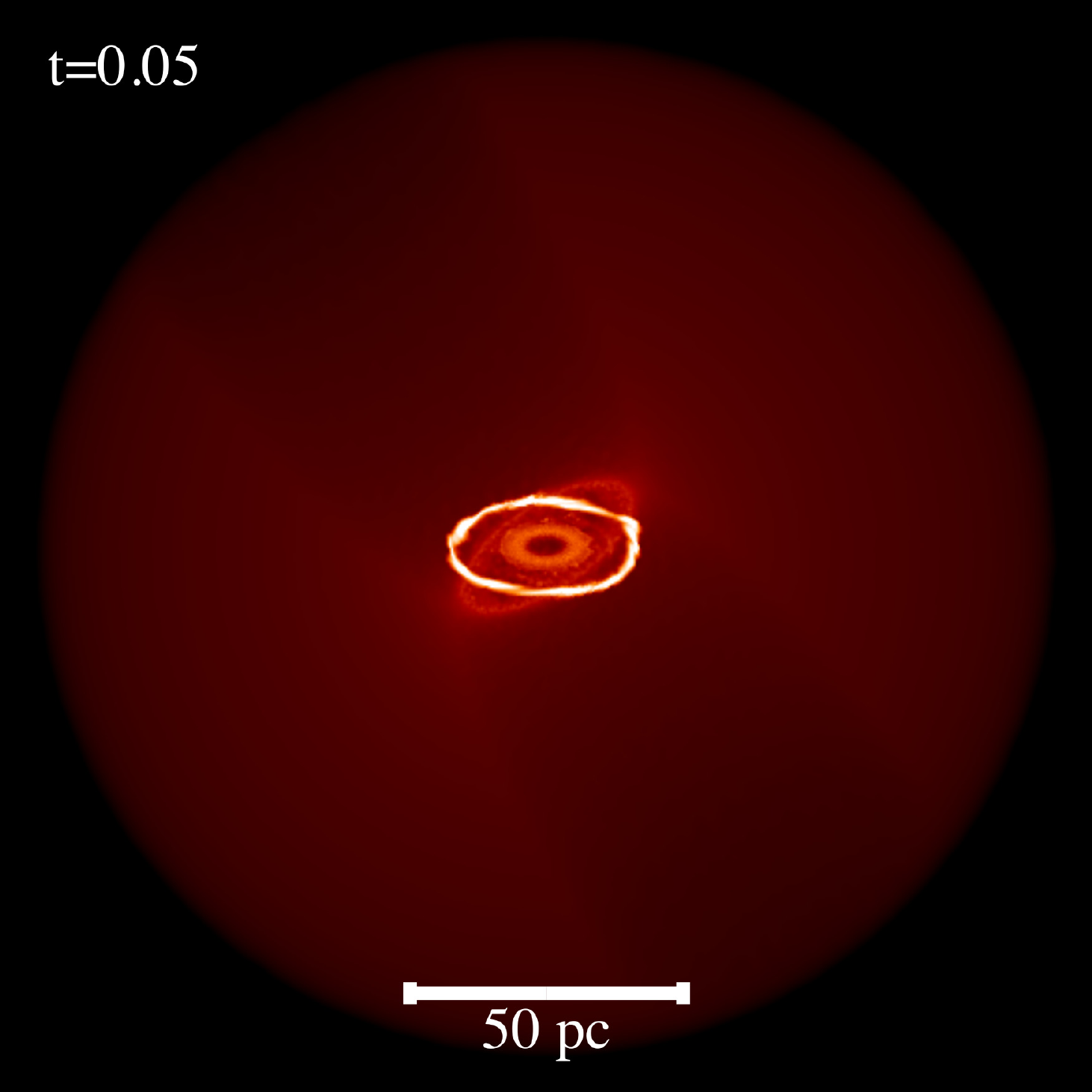}
\includegraphics[width=0.24\textwidth]{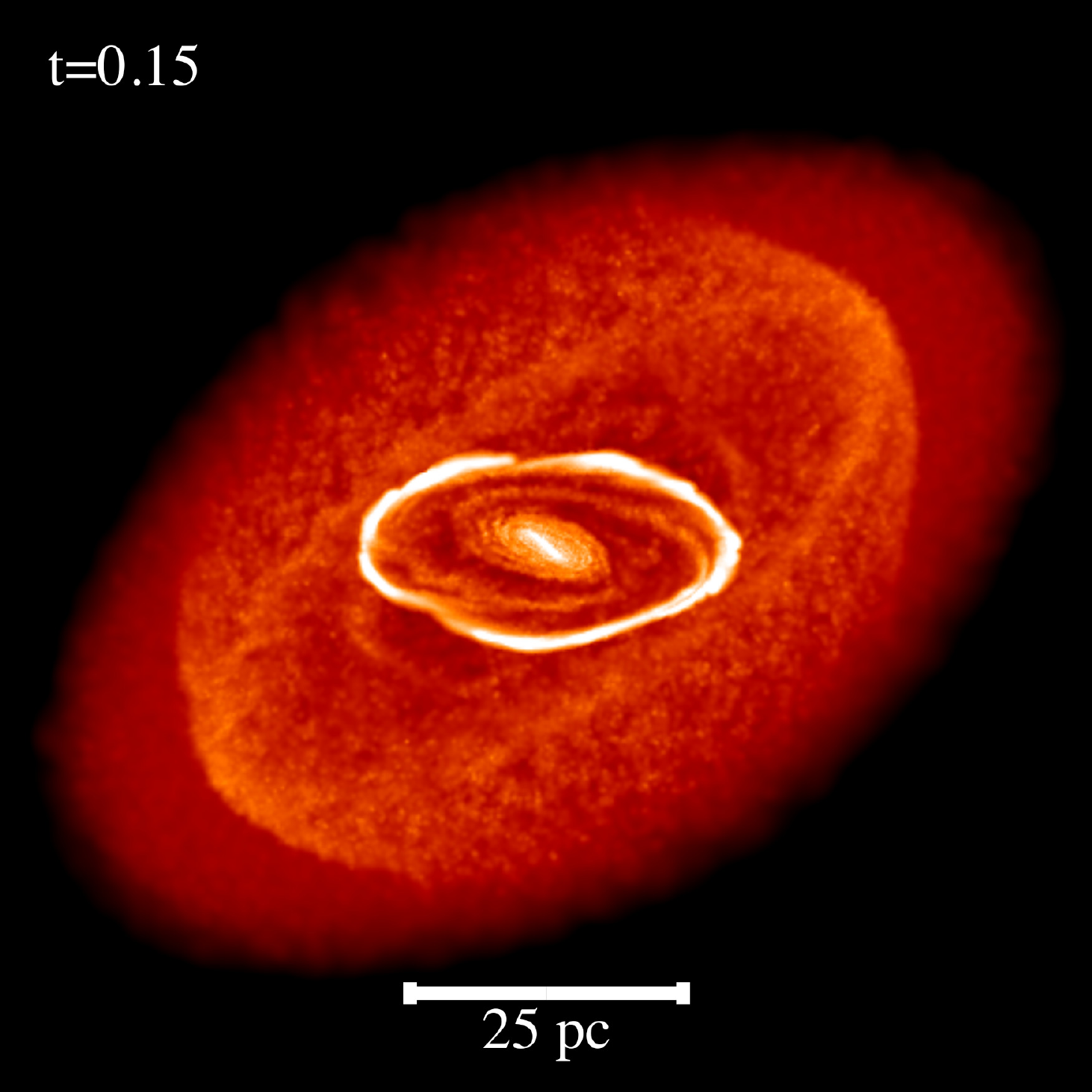}
\includegraphics[width=0.24\textwidth]{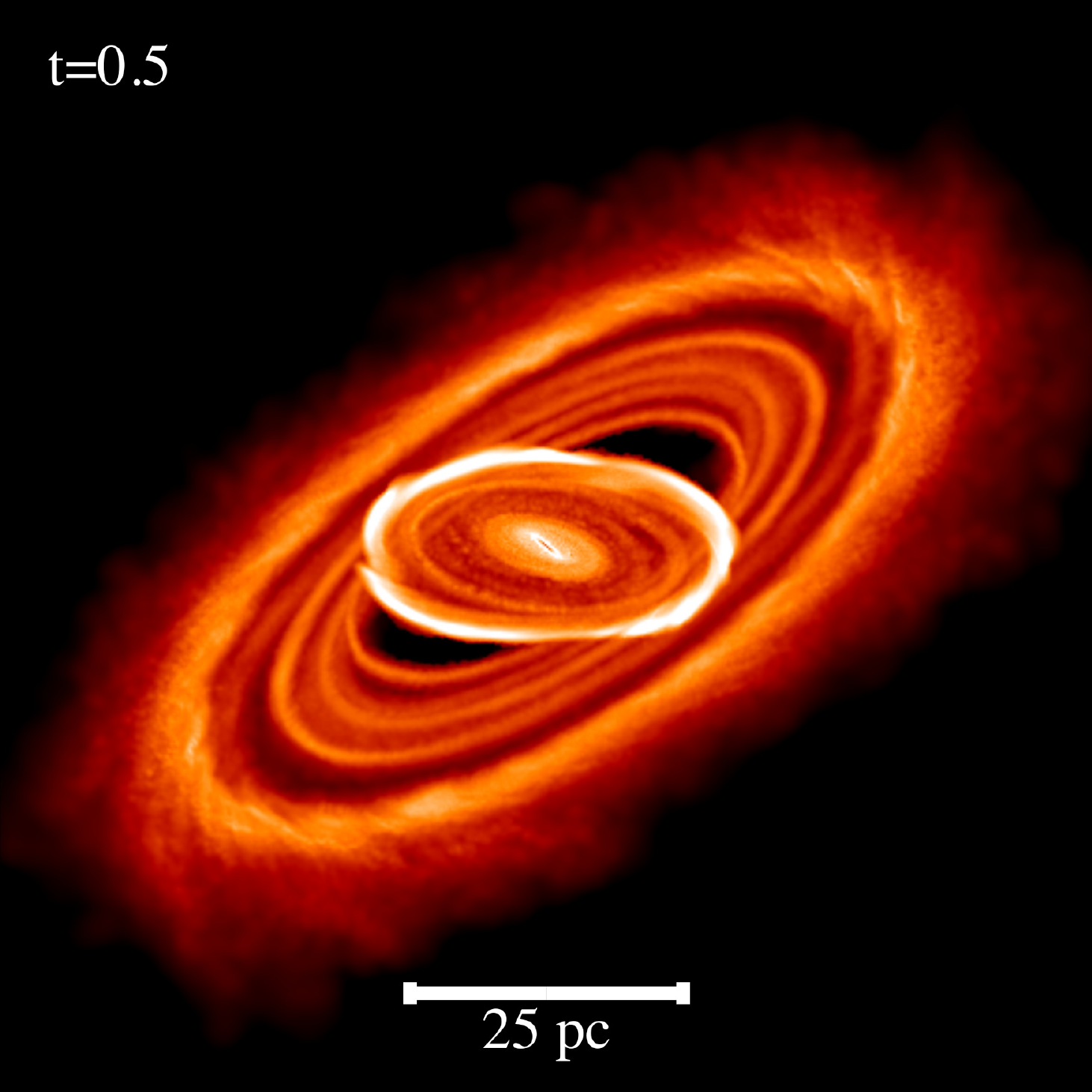}
\caption{Evolution of nested accretion events with $J_{\rm shell} > J_{\rm disc}$, corresponding to $\vrot = 0.7$. The upper and lower rows correspond to shell initial tilt angles of $\tilt = 60\degree$ and $150 \degree,$ respectively. From left to right, the time at which snapshots are taken are $t = 0, 0.05, 0.15, 0.5$. See text for details.Note the change of scale between the different snapshots: 50 pc in the first two and 25 pc in the last two.}
\label{fig: evolutions v07}
\end{figure*}

To build this case, we set an initial $\vrot = 0.7$ for the shell corresponding to an angular momentum per unit mass $J_{\rm shell} = 0.042 > J_{\rm disc}$. If the primitive disc were not present, the outcome of these two simulations would be a tilted copy of the primitive disc with wider orbits (with an
averaged circularization radius $r_{\rm circ}\sim 0.04$).The evolution of these cases is shown in Figure \ref{fig: evolutions v07}.

\subsubsection{Co-rotating shell, $\tilt = 60 \degree$}
The evolution of this larger angular momentum shell with $\vrot = 0.7$ and tilted at $\tilt = 60^\circ$ is rather simple. The angular momentum of almost all of the gas forming the shell is large enough for
particles  to remain in wide orbits and leave the primitive disc unaffected. The final configuration is the superposition of the primitive disc and a larger co-rotating disc formed from shell particles and tilted at the same tilt angle that the shell initially had. The evolution is illustrated in Figure \ref{fig: evolutions v07}.

\subsubsection{Counter-rotating shell, $\tilt = 150 \degree$}
Under these conditions, the shell carries a larger angular momentum than the 
primitive disc and is counter-rotating. One would expect to observe in this case an 
evolution similar to the previous case, i.e. the formation of an outer, counter-
rotating disc misaligned relative to the inner, unperturbed disc. However, Figure 
\ref{fig: evolutions v07} shows a different evolution.  As the disc 
forms from the inside out, the infalling shell particles with 
lower angular momentum try to form a counter-rotating disc of size comparable 
to that of the primitive disc. Cancellation of angular momentum
causes these particles to move further inwards changing also their sense of 
rotation. 
An inner denser  disc in co-rotation forms that can be identified as a density 
enhancement at the 
centre of the last two snapshots shown in the bottom right panels of Figure \ref{fig: 
evolutions v07}. 
The shell particles with larger angular momentum settle into a larger, counter-rotating disc which surrounds the inner co-rotating disc.

In the next section we explore the evolution of the angular momentum of the gas particles in a quantitative way, to explain the observed outcomes more thoroughly.

\section{Angular momentum evolution} \label{sec: ang mom evol}

In the previous section we showed  how the infalling shell perturbs the primitive disc, and how the system settles into a new stationary state after a violent interaction. To assess the reciprocal influence of the two flows it is necessary to have a quantitative understanding of the evolution of the angular momentum of the composite system and for each component separately. To this purpose, we keep track of the specific angular momentum distribution of the gas during the simulations, as this allows us to compare the rotational support of each component of the system against the potential in which it is immersed.

In the analysis of the simulated data, we observe angular momentum cancellation in some
fraction of the gas for the cases in which the second accretion event is in counter-rotation, a change in the flow direction for the equal angular momentum co-rotating case and an almost null impact for the larger angular momentum co-rotating case. 
As shown in Figures \ref{fig: evolutions v02}, \ref{fig: evolutions v03} and \ref{fig: evolutions v07}, there are some cases in which, after mixing, the flows merge into a single disc, some others in which the equilibrium state is formed by nested disc-like structures and some others in which less regular structures develop. To better understand the structures formed by the gas we compute a quantity which will give us information about the sense of rotation of each part of the flow, 

\begin{equation}\label{eq: stheta}
 S(\theta) = \frac{1}{N_\text{part}}\sum_{i=1}^{N_\text{part}} \Theta( \hat{l}_i \cdot \hat{u}_\theta ),
\end{equation}

\noindent where $\hat{l}_i$ points in the direction of the angular momentum of the particle $i$,
$\hat{u}_\theta =\sin(\theta) \, \mathbf { \hat{i}} + \cos(\theta) \, \mathbf { \hat{k}}$ 
is a unit vector pointing in the direction given by the $\theta$ angle in the $zx$ plane and  the function 
$\Theta(\hat{l}_i \cdot \hat{u}_\theta) = 1$ 
when $\hat{l}_i$ is within one degree of $\hat{u}_\theta$ and $\Theta(\hat{l}_i \cdot \hat{u}_\theta) = 0$ otherwise.
Thus $S(\theta)$ is the fraction of gas whose angular momentum points within one degree of the direction of $\theta$. 

A narrow peak in the function $S(\theta)$ corresponds to  a well defined plane of rotation for the 
gas. Thus, if a sizeable fraction of gas particles clusters around a specific value of $S(\theta)$ 
and the distribution of the angular momentum moduli is relatively broad, the configuration 
corresponds to a {\it planar disc}. A broad $S(\theta)$ refers instead  to a broad distribution of the 
angular momentum orientation and the configuration can be referred to as a {\it warped disc}. In 
this case, the distribution of angular momenta is expected to be wide. A {\it ring} would 
correspond to a narrow distribution for $j$ (i.e. the specific angular momentum modulus) with 
the function $S(\theta)$ peaking around a particular angle. 
The analysis of the equilibrium states is illustrated in Figures \ref{fig: final state v02}, \ref{fig: final state v03} and \ref{fig: final state v07}, where
we give the colour coded density map of the final structure (viewed at angles chosen to appreciate the details),
the distribution of the specific angular momentum moduli of disc (red) and shell (green) particles, and
the function $S(\theta)$ again for disc and shell particles.

\subsection{Lower angular momentum event: $J_{\rm shell} < J_{\rm disc}$}
\begin{figure*}
\includegraphics[width=0.25\textwidth]{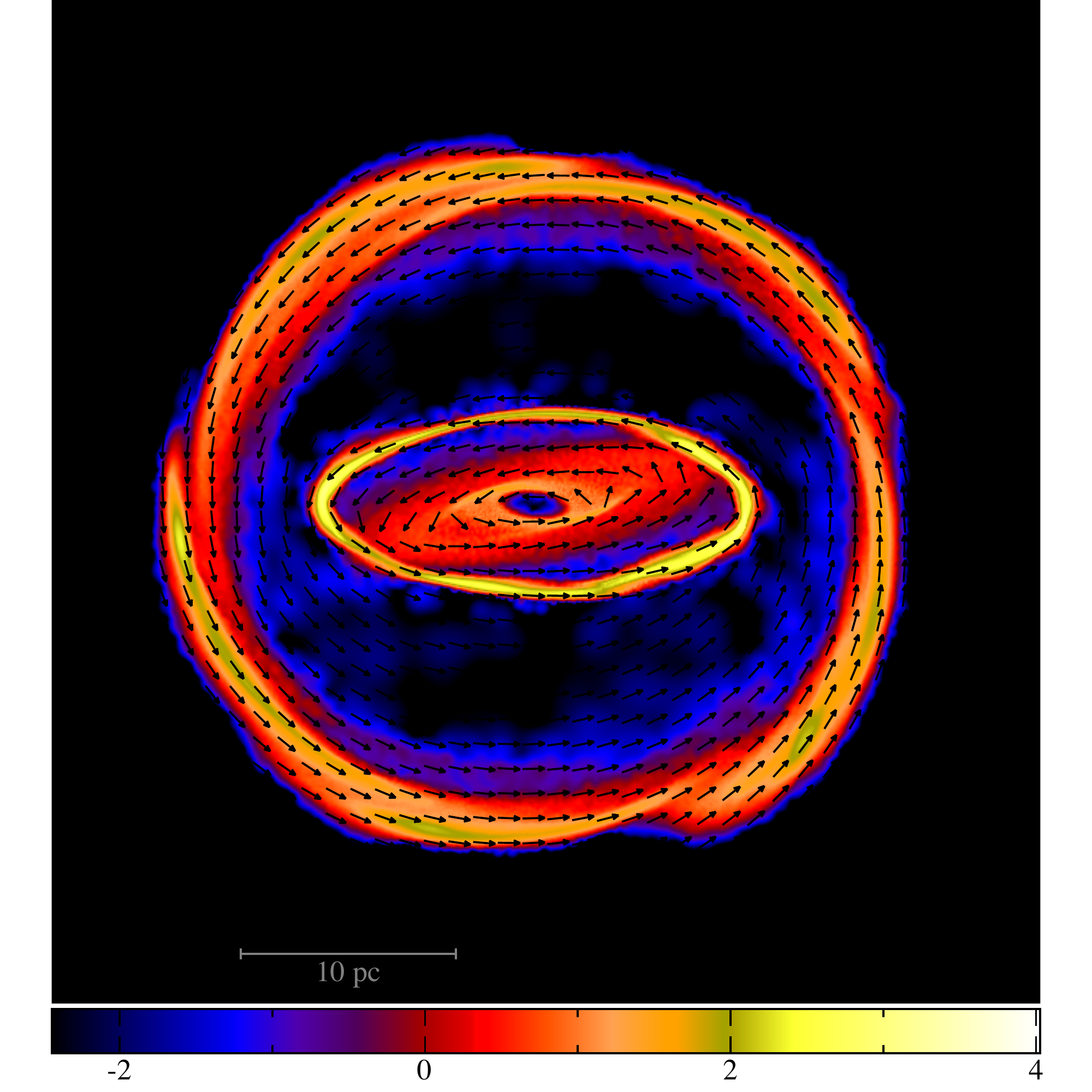}
\includegraphics[width=0.37\textwidth]{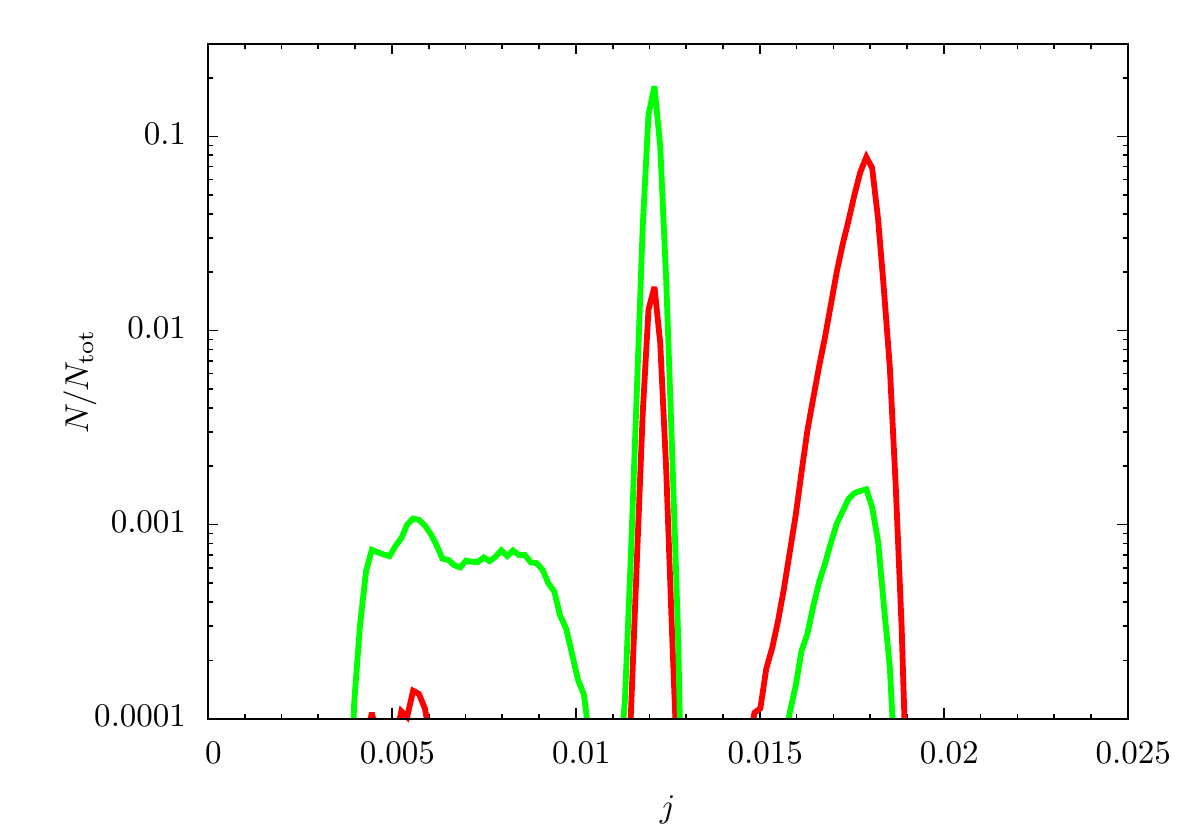}
\includegraphics[width=0.37\textwidth]{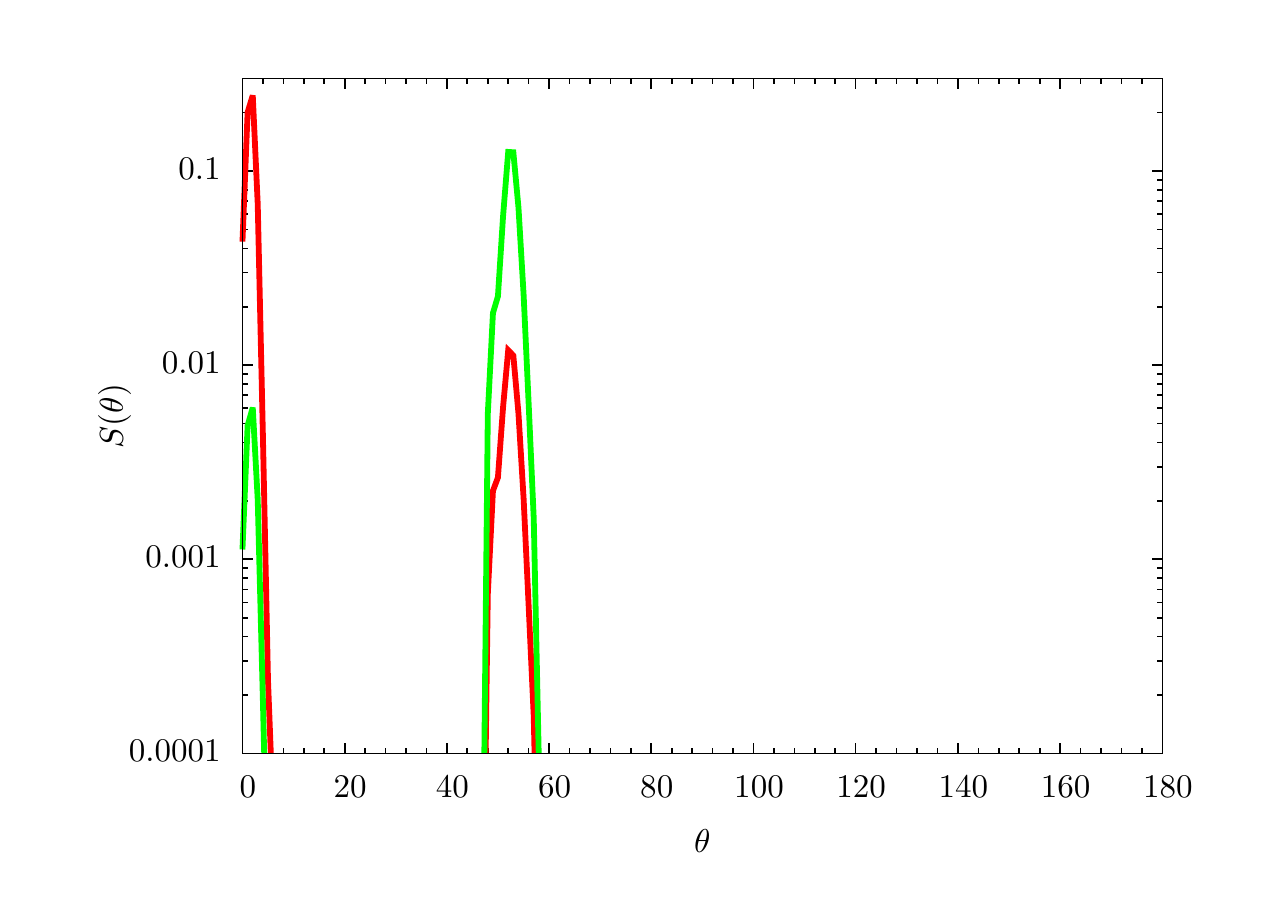}\\
\includegraphics[width=0.25\textwidth]{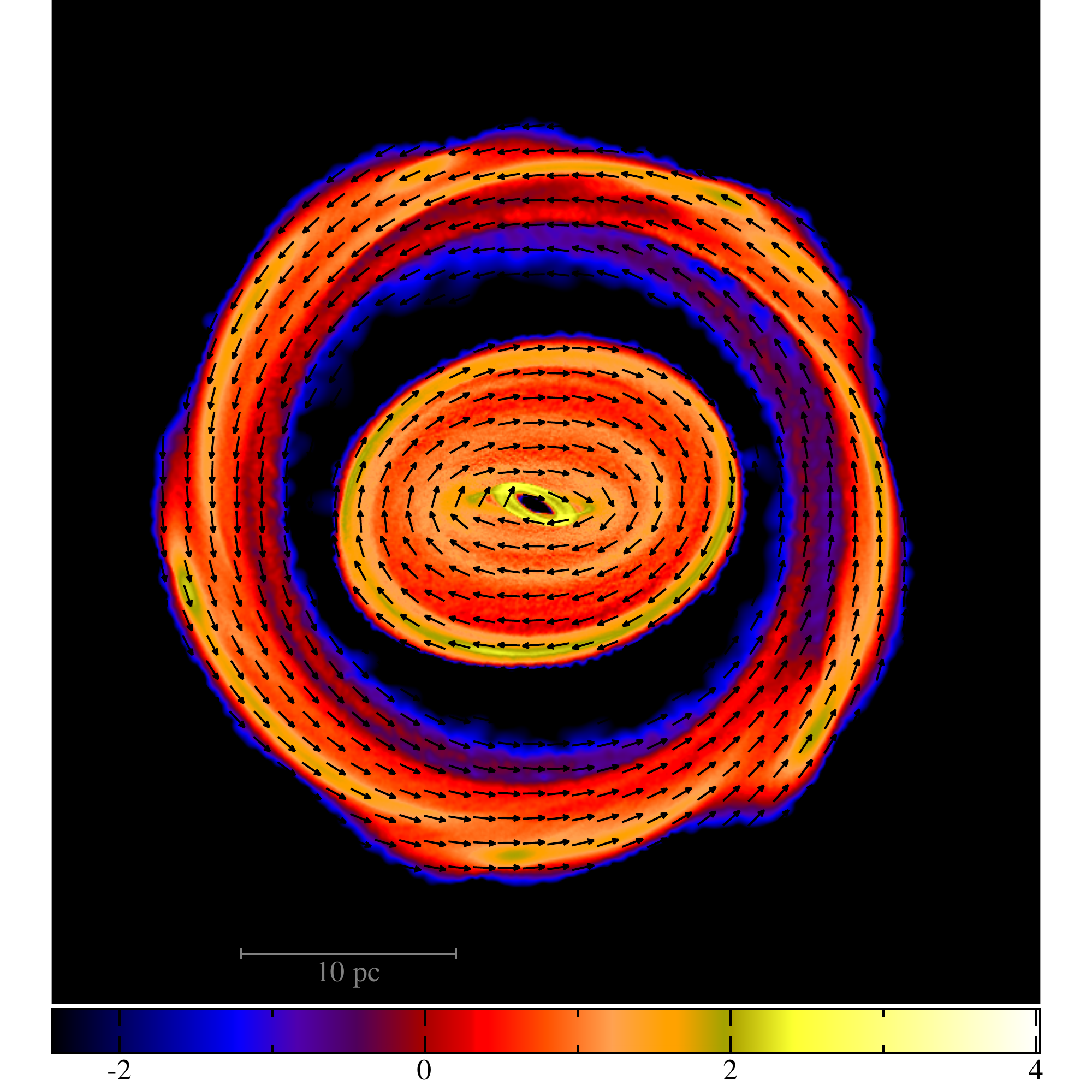}
\includegraphics[width=0.37\textwidth]{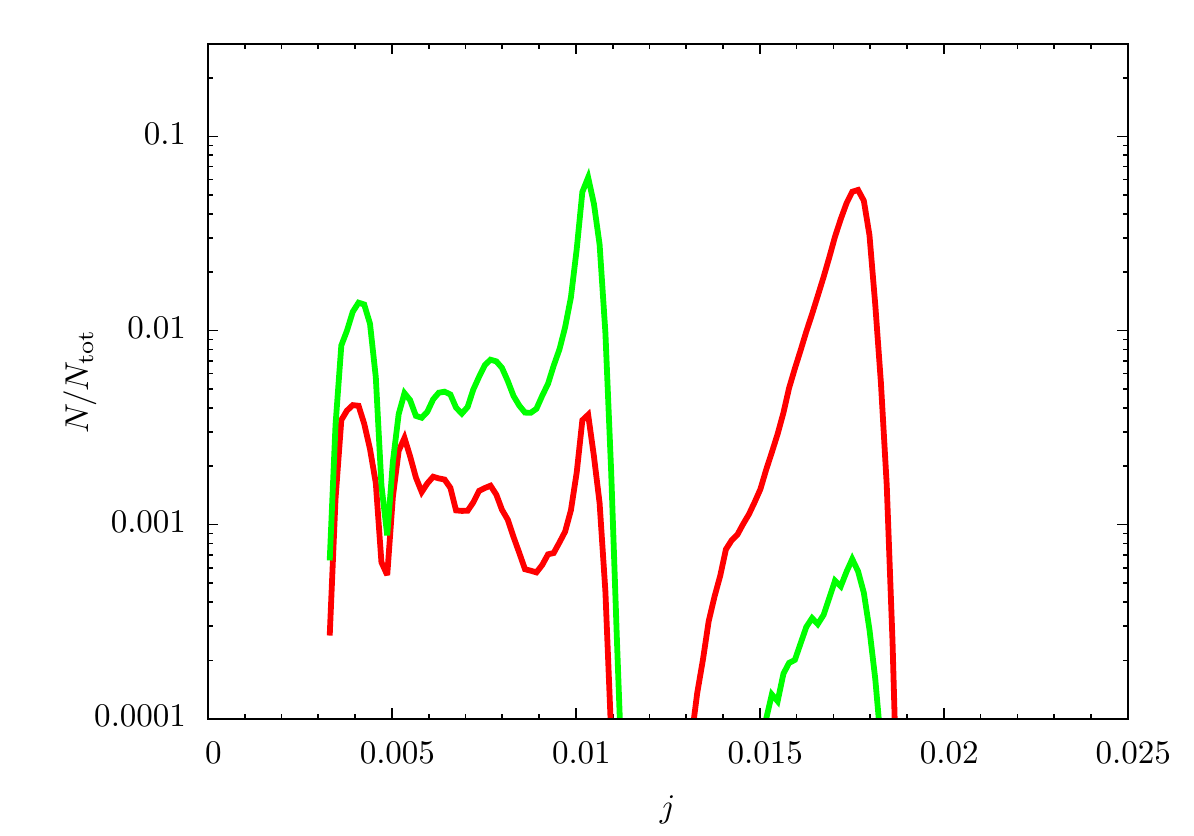}
\includegraphics[width=0.37\textwidth]{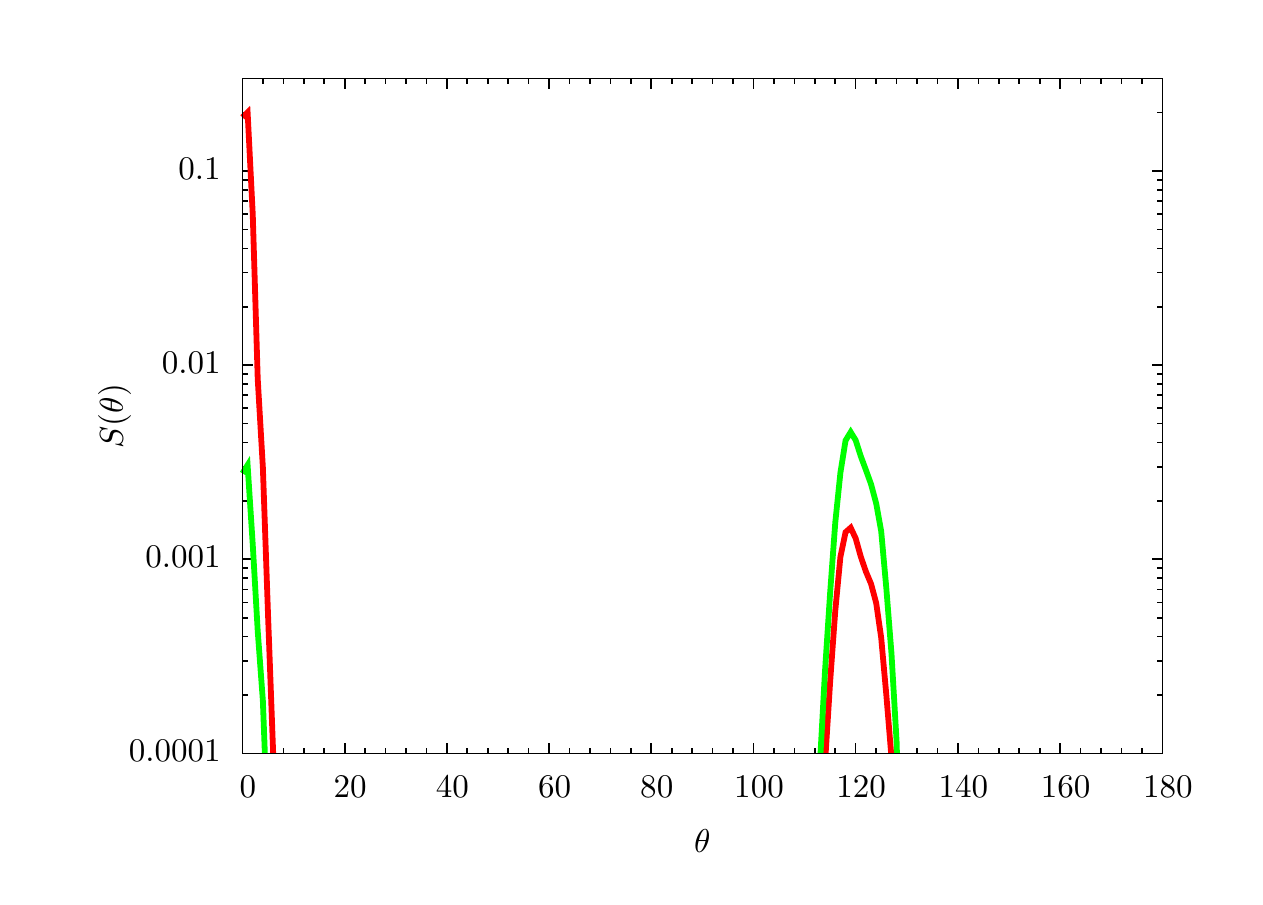}
\caption{
Final state of the nested accretion events with $J_{\rm shell} < J_{\rm disc}$, corresponding to $\vrot = 0.2$. The
upper and lower rows correspond to shell initial tilt angles of $\tilt = 60\degree$ and $150 \degree,$ respectively. 
The left column shows colour coded maps of $\log \Sigma$ in code units, where $\Sigma$ is the projected 
column density, with arrows indicating the sense of rotation of the flow. The middle column shows the
distributions of the angular momentum moduli for the disc (red 
lines) and shell (green lines) particles, while the right column shows plots of $S(\theta)$ for the disc (red) and shell 
(green) particles. The scale length bar in the left panels is 10 pc size.
}
\label{fig: final state v02}
\end{figure*}

In this case, gas particles belonging to the shell  pass through the ring before reaching their circularization radius, interacting with the disc and and then leaving it.  This leads to the redistribution of the angular momentum depicted in Figure \ref{fig: final state v02}.

\subsubsection{Co-rotating shell, $\tilt = 60 \degree$}
\label{sec: v02tilt060}

The upper central panel of Figure \ref{fig: final state v02} shows the specific angular momentum distribution among disc and shell particles after the interaction, at $t = 0.5$. We observe two peaks and one broader region in the distribution. The first peak lies at $j = J_{\rm disc} = 0.018$, and is formed by disc particles and by some smaller amount of shell particles which have been captured during the interaction; the second peak lies at $j = J_{\rm shell} = 0.012$ and is predominantly formed by shell particles, although there is also a non-negligible amount of disc particles. These disc particles were not there before the interaction, which means that they have been dragged there by the infalling shell particles. The third interesting region is  located between $j \simeq 0.005$ and $j \simeq 0.009$ and is broader and less dense.
It is formed by shell particles that have exchanged angular momentum with disc particles that
experience an outward drag, rather than an inward one. From the angular momentum distribution, we describe the structure as a lower density disc surrounded by two ring-like structures.

The top right panel of Figure \ref{fig: final state v02} for $S(\theta)$ highlights the presence of two well defined peaks, meaning that the structures that result from the interaction are planar. One is rotating in the plane of the primitive disc, and the other one is rotating at $\theta \simeq 50\degree$,
and comprises the inner ring and innermost disc. The two peaks have almost equal height and inverted proportions of the mix between disc and shell particles.

\subsubsection{Counter-rotating shell, $\tilt = 150 \degree$}
\label{sec: v02tilt150}

This case is shown in the lower row of Figure \ref{fig: final state v02}. The distribution shown is the result of a rather violent interaction in which a significant amount of gas from both the shell and the disc is dragged all the way down to the accretion radius. We observe two populated regions: The first one, a broad inner region extending from $j \simeq 0.004$ to $j \simeq 0.01$, is a mix between disc particles and a greater amount of shell particles. The outer peak, even if still at $j = J_{\rm disc} = 0.018$, has widened, and is composed by a mix of particles belonging to the two components of the system with a greater proportion of disc particles.  The lower right panel  shows $S(\theta)$ displaying two peaks, the first one at $\theta \simeq 0\degree$ and the second, shorter and broader, at $\theta \simeq 120\degree$. The first peak has a major proportion of disc particles, while the opposite is true for the second peak.
The tilted shell shifts to a lower tilt angle and a significant amount of disc particles end in the
inner disc which is counter-rotating relative to the primitive disc, being dragged by the interaction with shell particles.

\subsection{Equal angular momentum event: $J_{\rm shell} = J_{\rm disc}$}

\begin{figure*}
\includegraphics[width=0.25\textwidth]{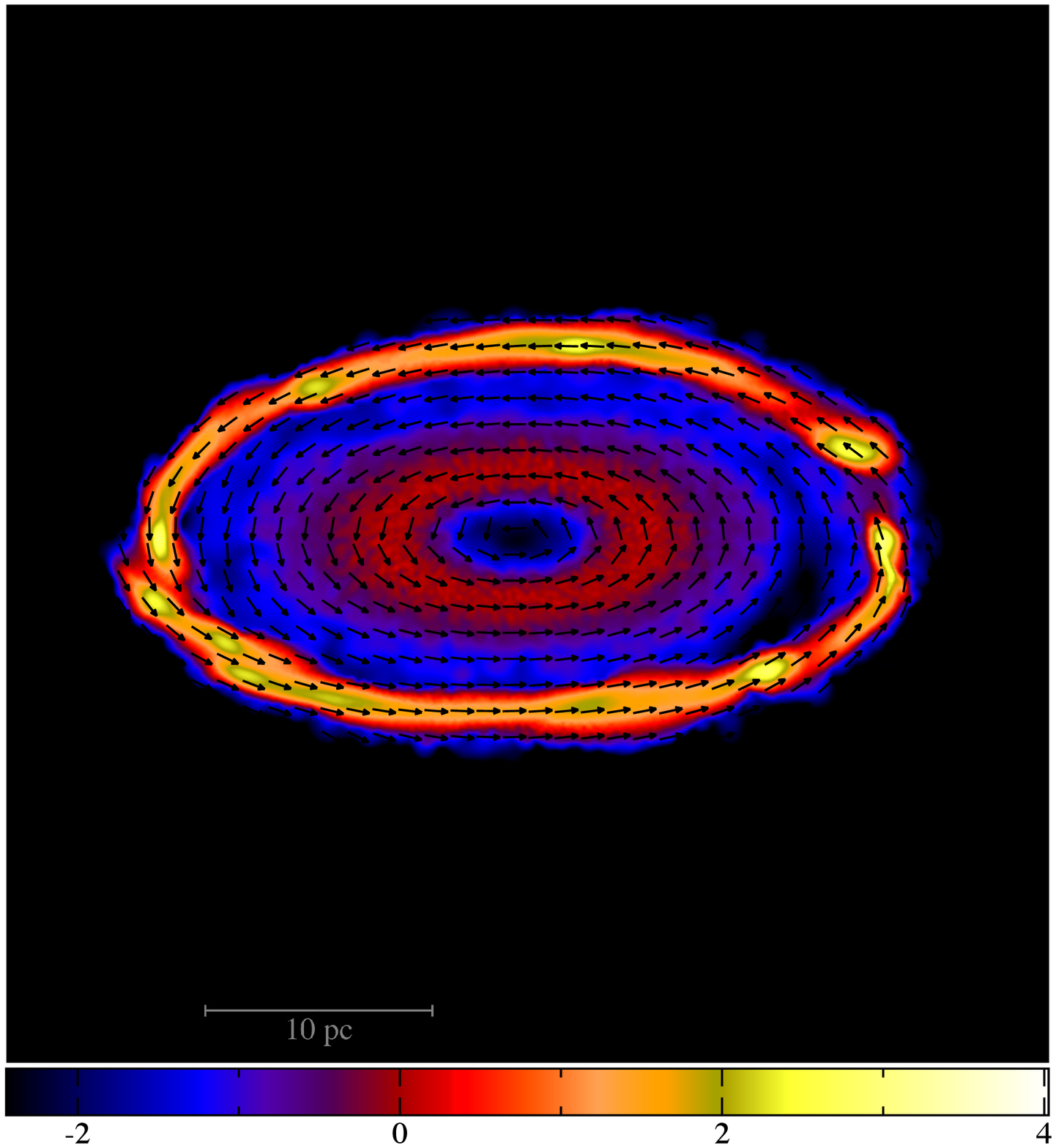}
\includegraphics[width=0.37\textwidth]{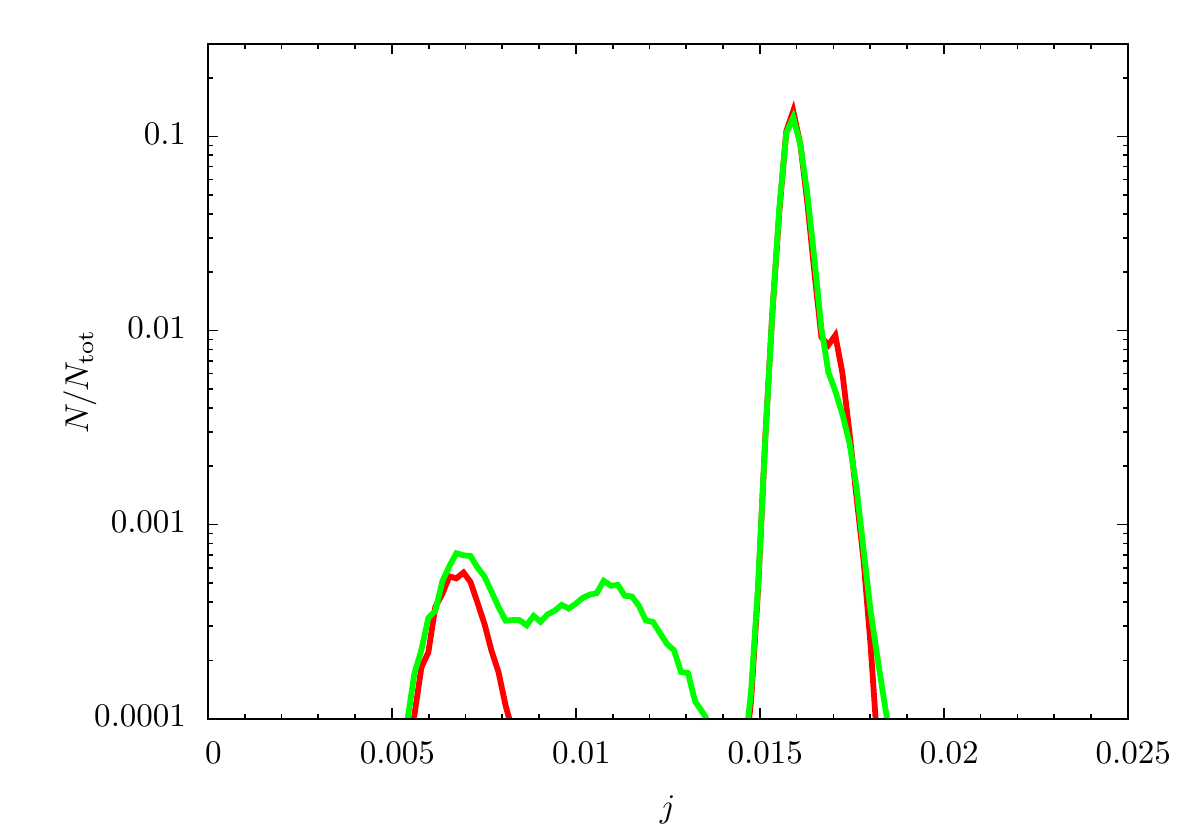}
\includegraphics[width=0.37\textwidth]{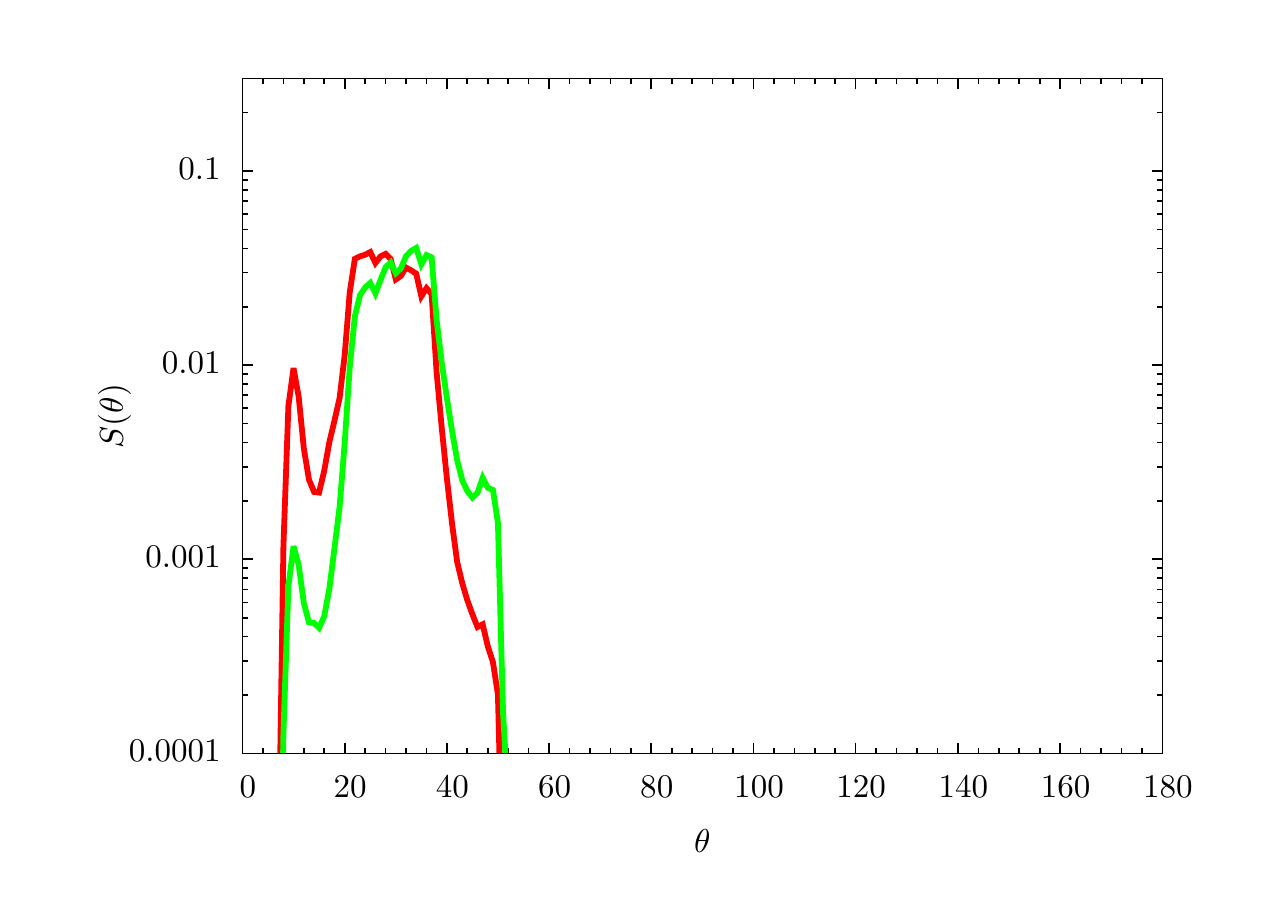}\\
\includegraphics[width=0.25\textwidth]{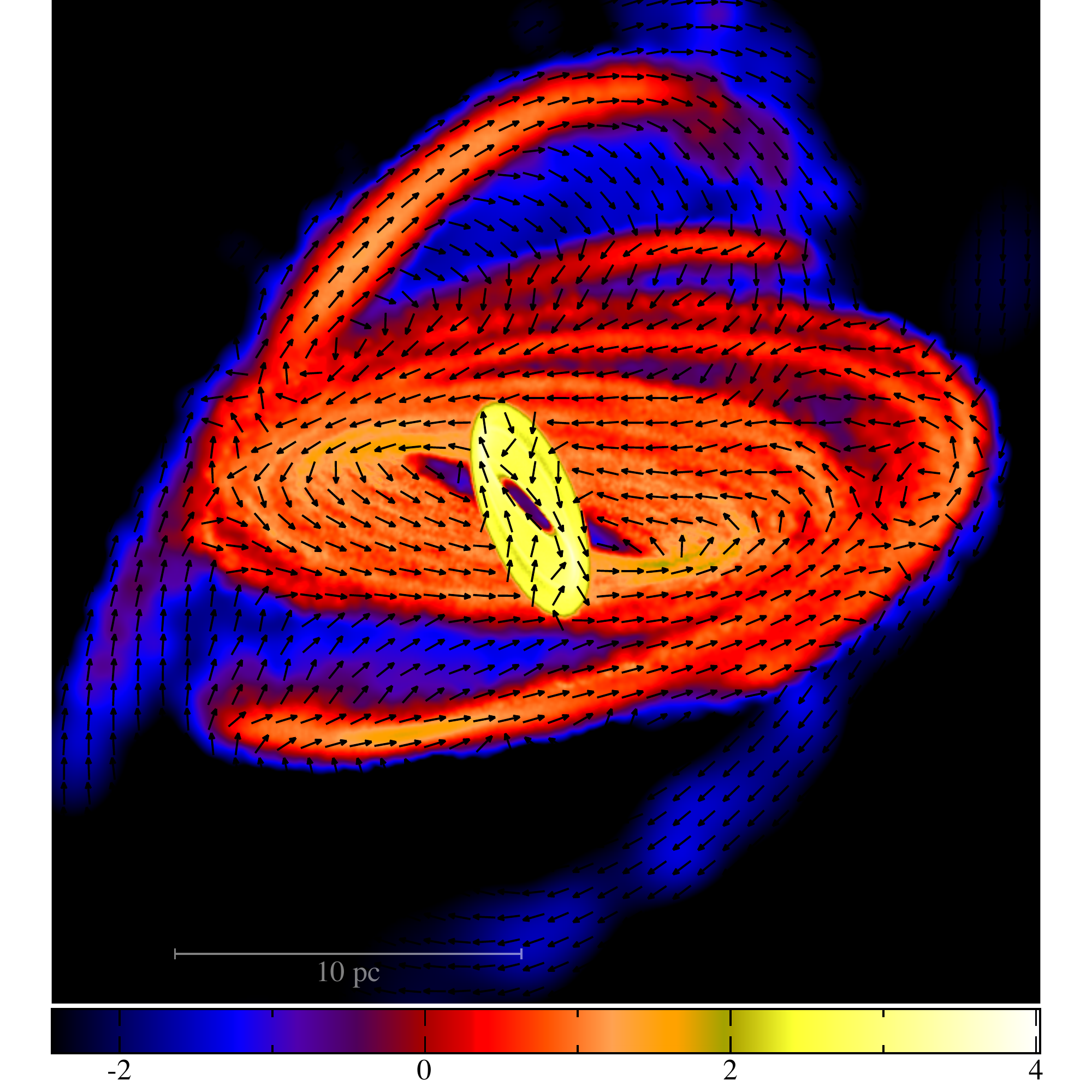}
\includegraphics[width=0.37\textwidth]{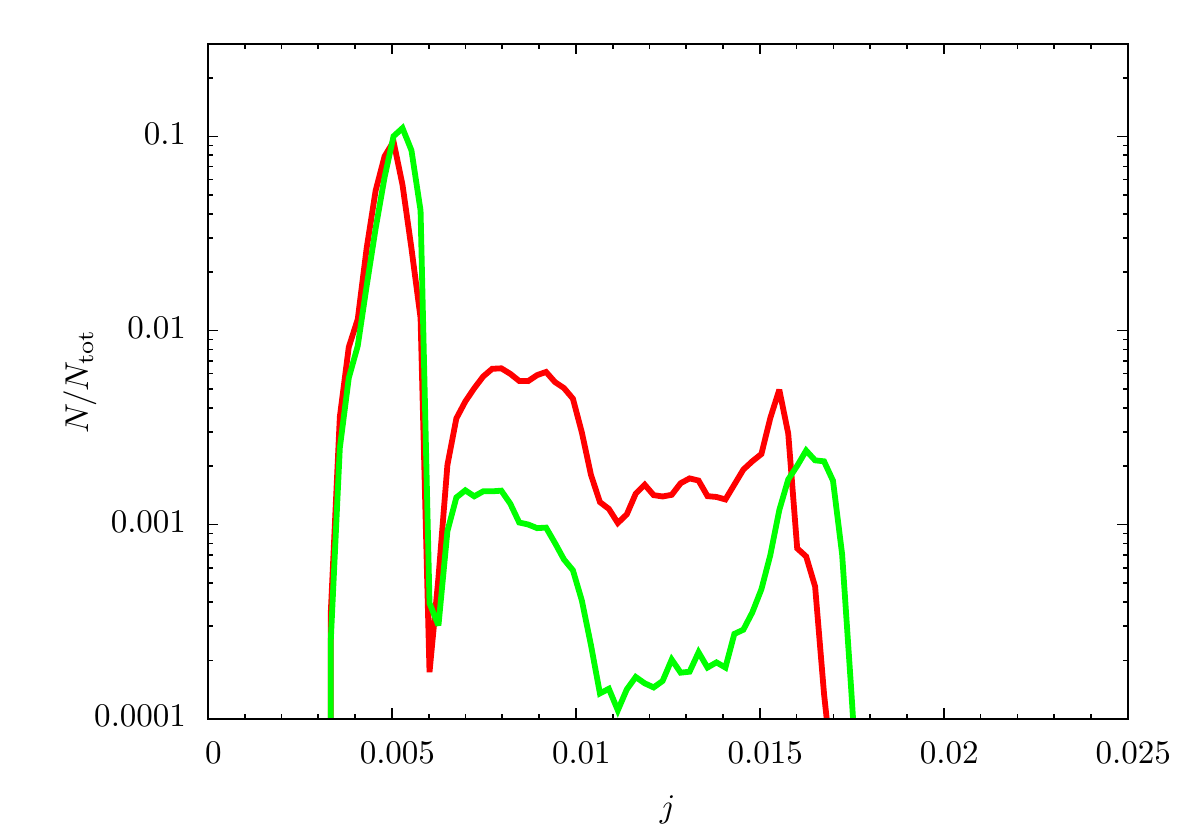}
\includegraphics[width=0.37\textwidth]{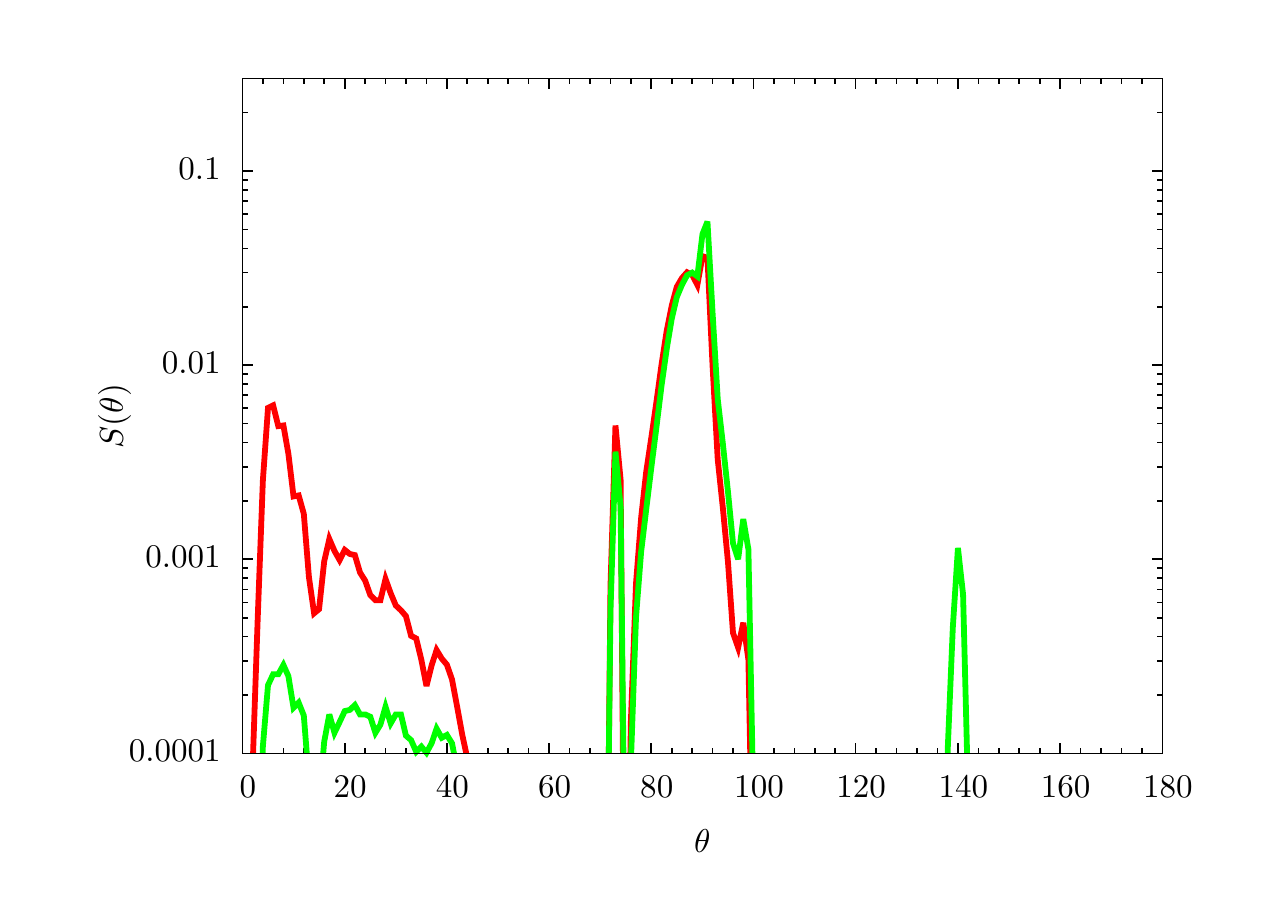}
\caption{
Final state of the nested accretion events with $J_{\rm shell} = J_{\rm disc}$, corresponding to $\vrot = 0.3$. The
upper and lower rows correspond to shell initial tilt angles of $\tilt = 60\degree$ and $150 \degree,$ respectively. 
The left column shows colour coded maps of $\log \Sigma$ in code units, where $\Sigma$ is the projected 
column density, with arrows indicating the sense of rotation of the flow. The middle column shows the
distributions of the angular momentum moduli for the disc (red 
lines) and shell (green lines) particles, while the right column shows plots of $S(\theta)$ for the disc (red) and shell 
(green) particles. The scale length bar in the left panels is 10 pc size.
}
\label{fig: final state v03}
\end{figure*}

In this accretion event the orbits of shell particles will circularize at a radius equal 
to the radius of the primitive disc, thus, it is expected that gas belonging to the 
shell will strongly interact with the primitive disc.

\subsubsection{Co-rotating shell, $\tilt = 60 \degree$}
\label{sec: v03tilt060}

The upper row of Figure \ref{fig: final state v03}  shows the angular momentum distribution and  $S(\theta)$ at $t = 0.5$, when the system has reached a state of stationary equilibrium. 
The disc particle distribution now peaks at $j \sim 0.016$, implying that $\sim 10 \%$ of the specific angular momentum content has been cancelled during the interaction. 
Although the angular momentum distribution has not been severely affected, the gas mixing have changed the original plane of rotation: After the shell started infalling, shell particles tried to settle into circular orbits at the radius where the primitive disc was located. These particles kicked the disc particles at an angle $\tilt = 60 \degree$ with a speed roughly equal to the speed the disc particles had, killing in the interaction the normal component of their relative velocity while preserving the parallel component, tilted in this case at $\tilt / 2 = 30\degree$ with respect to the $z$ axis. Thus, what we observe in the end is a disc tilted at $\tilt = 30 \degree.$  The surface density distribution (upper left panel) is non homogeneous through this disc-like structure that shows some degree of warping. 

\subsubsection{Counter-rotating shell, $\tilt =150 \degree$}
\label{sec: v03tilt150}

The angular momentum loss in this case is greater than in the previous one. Particles from the shell start falling toward the disc, trying to settle into circular orbits at the same radius of the disc, but contrary to the previous case, the second accretion event is now in counter-rotation, leading to an almost head-on collision between the flows and cancelling out a large amount of their angular momentum. The final  distribution of angular momentum, shown in the lower mid panel of Figure \ref{fig: final state v03}, is broad and extends down to the accretion radius, implying a complete mixing of the two flows. The peak of the angular momentum of disc particles has decreased from $ j = 0.018$ to $ j = 0.005$, which means that there has been a huge net cancellation of angular momentum. 

The lower right panel of Figure \ref{fig: final state v03} shows three main planes of rotation: one at $\sim 90 \degree$, corresponding to gas that has interacted the most, forming in the end a dense disc in tight orbits around the centre (see figure \ref{fig: evolutions v03}). This disc contains gas from both components of the system in almost equal proportions. The second plane of rotation, at $\sim 140 \degree$, corresponds to gas initially belonging to the shell that presumably fell last, when the major part of the interaction had already happened and most of the gas had already gone into smaller orbits. This gas is observed in the left lower panel as an irregular tail orbiting the black hole. The last structure rotates between 0 and $\sim 40 \degree$ and is mainly composed of particles initially belonging to the primitive disc.  

\subsection{Larger angular momentum event: $J_{\rm shell} > J_{\rm disc}$}

\begin{figure*}
\includegraphics[width=0.25\textwidth]{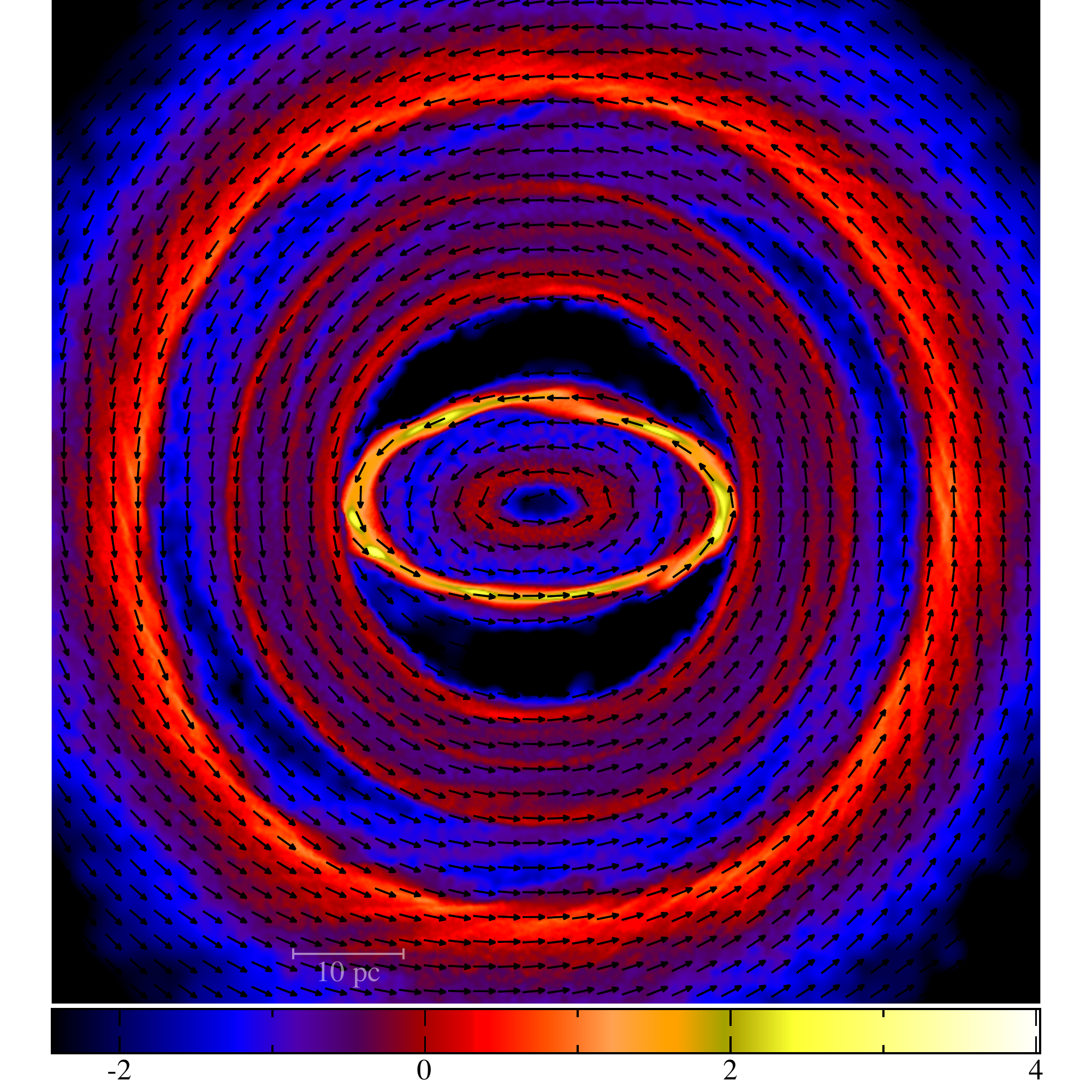}
\includegraphics[width=0.37\textwidth]{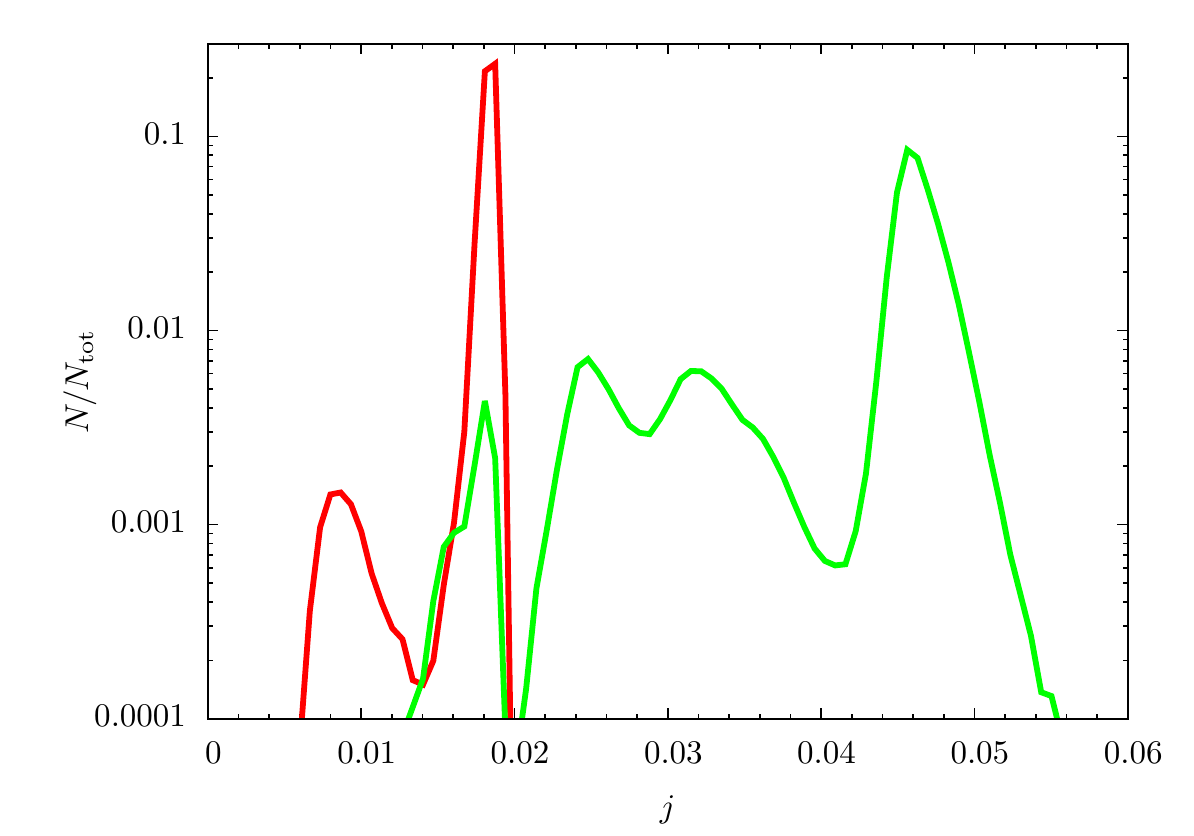}
\includegraphics[width=0.37\textwidth]{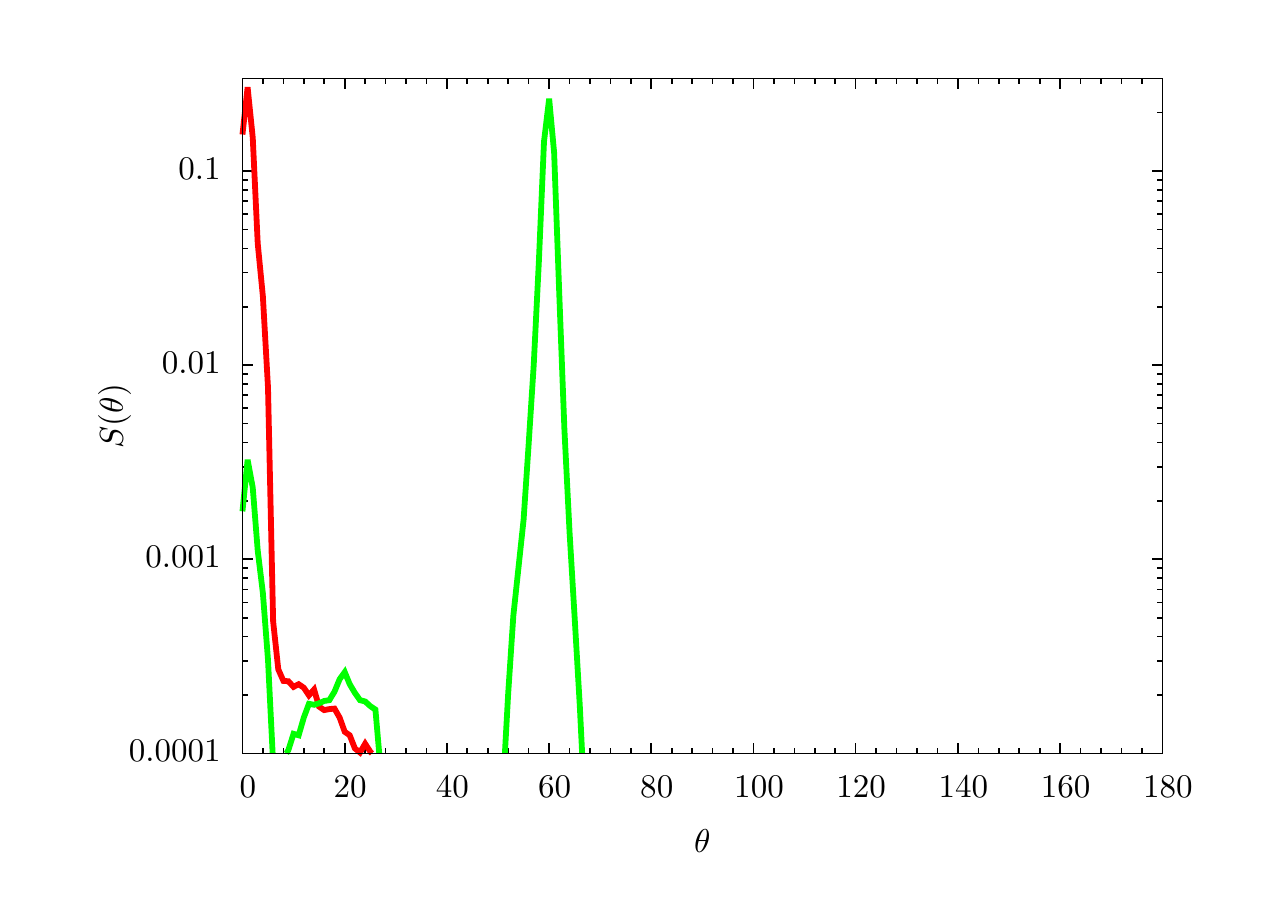}\\
\includegraphics[width=0.25\textwidth]{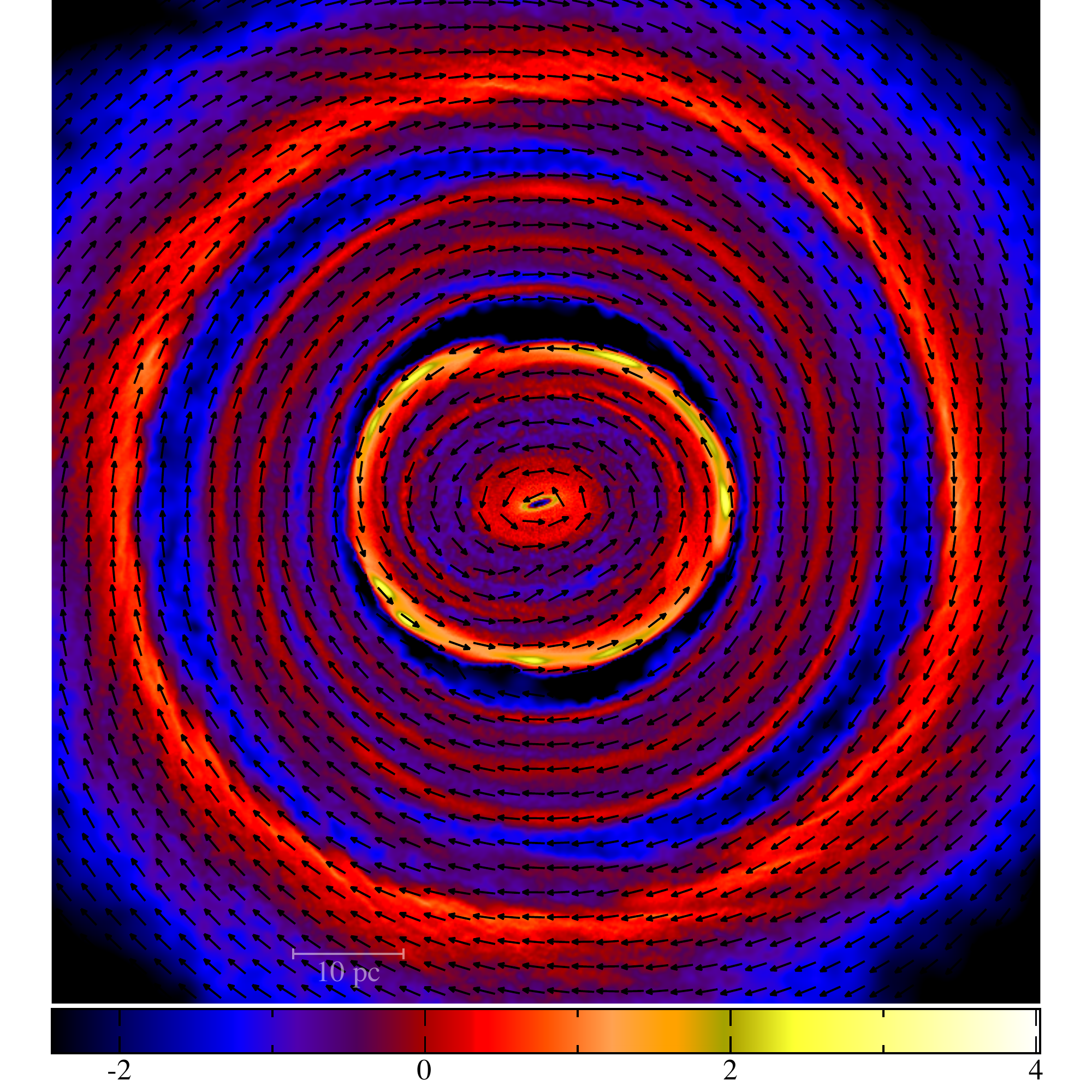}
\includegraphics[width=0.37\textwidth]{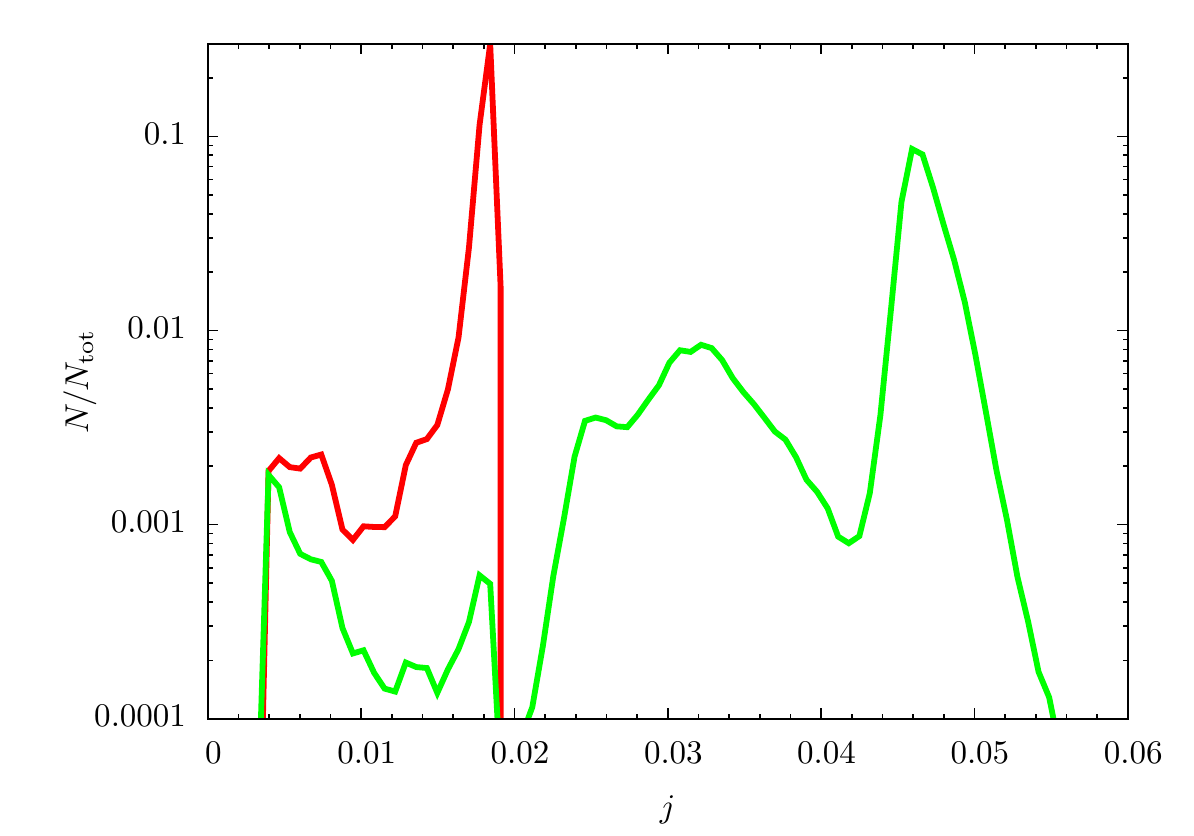}
\includegraphics[width=0.37\textwidth]{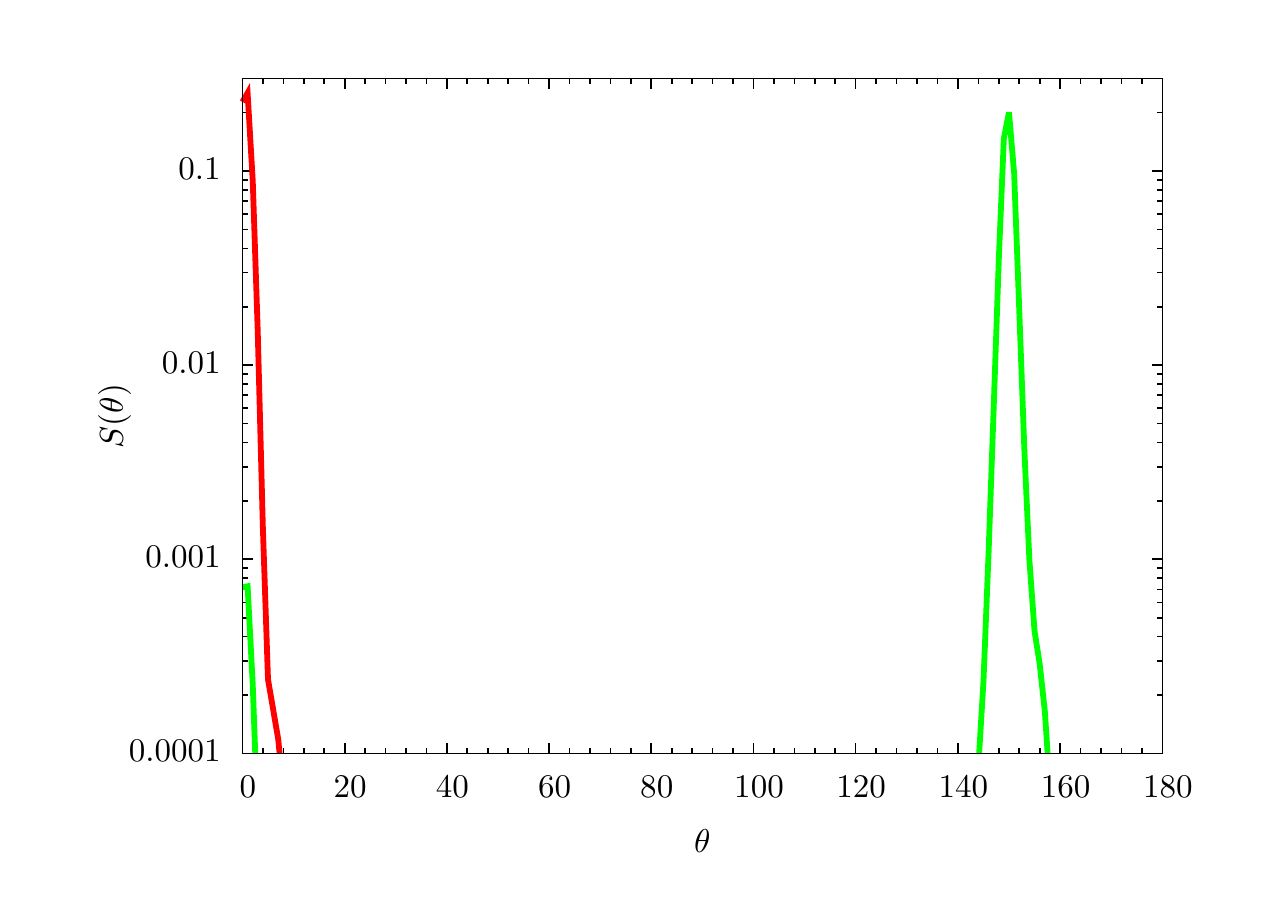}
\caption{
Final state of the nested accretion events with $J_{\rm shell} > J_{\rm disc}$, corresponding to $\vrot = 0.3$. The
upper and lower rows correspond to shell initial tilt angles of $\tilt = 60\degree$ and $150 \degree,$ respectively. 
The left column shows colour coded maps of $\log \Sigma$ in code units, where $\Sigma$ is the projected 
column density, with arrows indicating the sense of rotation of the flow. The middle column shows the
distributions of the angular momentum moduli for the disc (red 
lines) and shell (green lines) particles, while the right column shows plots of $S(\theta)$ for the disc (red) and shell 
(green) particles.The scale length bar in the left panels is 10 pc size.
}
\label{fig: final state v07}
\end{figure*}

In this case, shell particles have a bigger circularization radius than the radius of the primitive disc, thus
we expect a less violent mixing, as shown in figure \ref{fig: final state v07}.
 
\subsubsection{Co-rotating shell, $\tilt = 60 \degree$}
\label{sec: v07tilt060}

The top row of Figure \ref{fig: final state v07} shows the result for this case in which very low amount of mixing occurs.
The primitive disc is barely affected. It acts as an obstacle for low angular momentum shell particles, 
trapping and assimilating some fraction of them as part of the original disc. As illustrated in the top right panel, 
the main planes of rotation of both flows remain practically unaffected.

\subsubsection{Counter-rotating shell, $\tilt = 150 \degree$}
\label{sec: v07tilt150}

In this case, the interaction of shell particles with the primitive disc during the 
inside out formation of the secondary disc causes a fraction of shell particles to co-rotate
with the disc, as illustrated in the bottom central and right panels of Figure 
\ref{fig: final state v07}. There exist a region of particles populated at low $j$
indicating that an inflow is present across the inner boundary.   The angular momentum 
orientation highlights the presence of two clear orbital planes:
one tilted by $150 \degree$ and one at $0^\circ$,  
 corresponding to the primitive disc with a small amount of shell particles.
\newline
\newline
In summary, we have seen that all second accretion events with equal or smaller angular momentum content than the first one are capable of altering the angular momentum distribution of the system significantly. In the other hand, the interaction between the disc and higher angular momentum shells leaves the angular momentum distribution of each constituent of the system basically unaltered.

\section{Angular momentum conservation}
\label{sec: ang mom cons}

\begin{table}
\caption{Angular momentum lost, $L_{\rm lost}$, after the interaction of the shell with the primitive disc. 
$\vrot$ denotes the rotational velocity of the shell, and $\tilt$ the tilt angle between 
$\Ldisc$ and $\Lshell$. $L_{\rm lost}$ for the system should be zero, however, in our simulations a net mass loss is present due to inflow.}
\centering
\begin{tabular}{|c|c|c|c|}\hline
$\vrot$ & $\tilt$ &flow & $ \% ~ L_{\rm lost}$ \\ 
\hline
\hline
0.2 & $60 \degree$   & disc  &  7.7  \\ 
0.2 & $150 \degree$ & disc  &  31       \\
0.3 & $60 \degree$   & disc  &  13      \\ 
0.3 & $150 \degree$ & disc  &  75      \\
0.7 & $60 \degree$   & disc  &    1.1      \\
0.7 & $150 \degree$ & disc  &    5.5    \\
      &                         &          &         \\
0.2 & $60 \degree$   & shell  & 2.9    \\ 
0.2 & $150 \degree$ & shell  &  47 \\
0.3 & $60 \degree$   & shell  &  14 \\ 
0.3 & $150 \degree$ & shell  &  74  \\
0.7 & $60 \degree$   & shell  &  0.77\\
0.7 & $150 \degree$ & shell  &  2.4\\
      &                         &          &                \\
0.2 & $60 \degree$   & system  &    0.28      \\ 
0.2 & $150 \degree$ & system  &    1.1 \\
0.3 & $60 \degree$   & system  &    $< 0.1$\\
0.3 & $150 \degree$ & system  &    4.6  \\
0.7 & $60 \degree$   & system  &   $< 0.1$\\
0.7 & $150 \degree$ & system  &   $< 0.1$\\
\hline
\end{tabular}
\begin{flushleft}
\end{flushleft}
\label{tab: momloss}
\end{table}

In the last section it was shown that events in counter-rotation modify 
the angular momentum distribution of the system more than events in 
co-rotation, in particular when the angular momentum of the shell and of the primitive disc are comparable, initially. In 
table \ref{tab: momloss} we assess this behaviour quantitatively by computing the angular 
momentum loss in each case for the disc, the shell, and the disc+shell composite system. We observe that counter-rotating infalling shells can cancel out up to $75 \%$ of the angular momentum in each component of the system, thus catalysing significantly the inflow towards the accretion radius. Co-rotating events are not as effective, as they can just induce an angular momentum cancellation of less than $15 \%$. 
If the angular momentum content of the shell is large compared to that of the disc, the amount of gas mixing between the shell and the disc is so small that the angular momentum loss is of less than $6 \%$. 

The angular momentum loss of the system as a whole is due to the loss of mass through the accretion radius, that is higher for the equal angular momentum counter-rotating events.

\begin{figure*}
\includegraphics[width=0.88\textwidth]{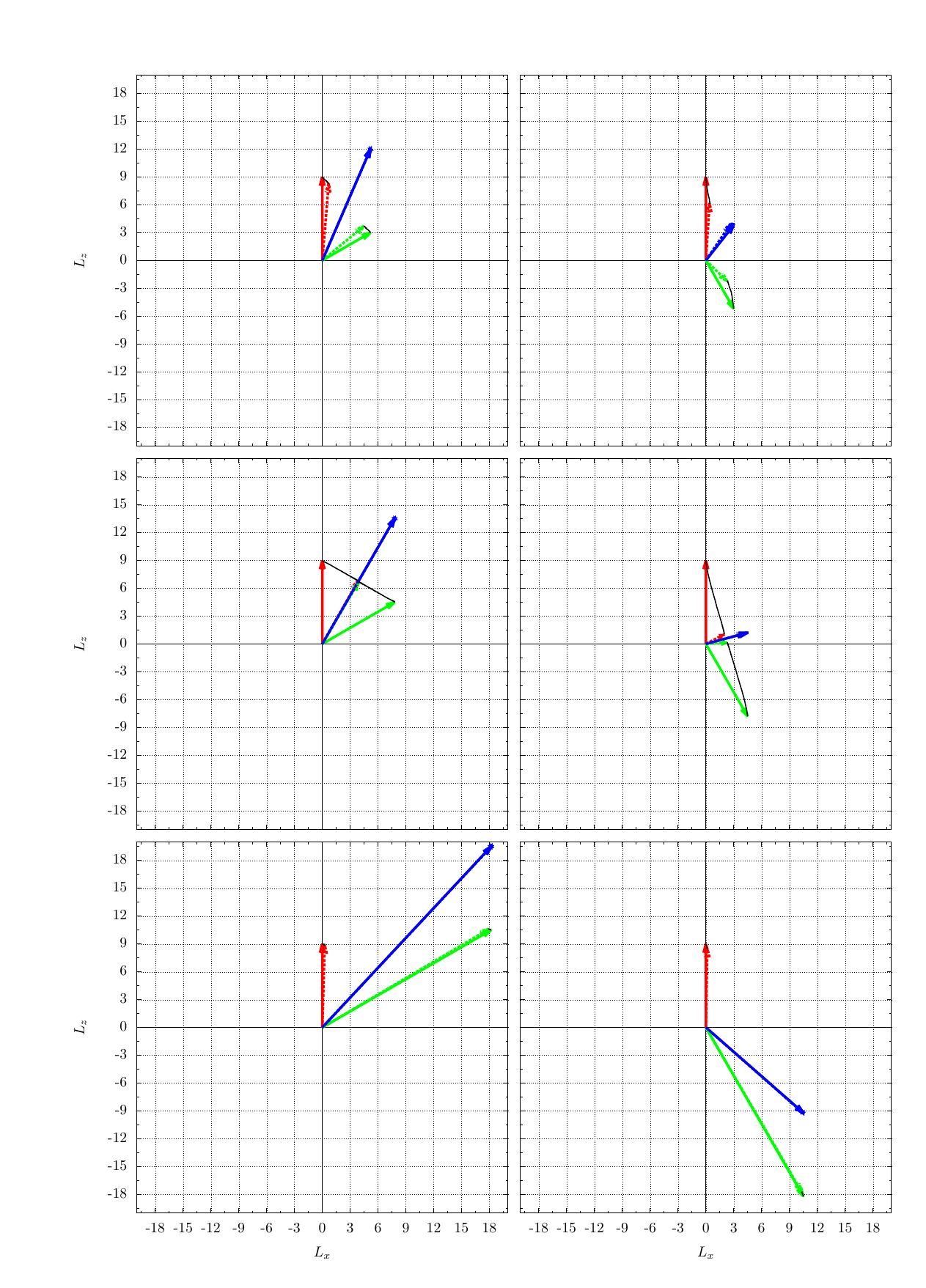}
\caption{Initial (continuous arrows) and final (dashed arrows) states of the angular momentum for each component of the system: Disc (red arrows) and Shell (green arrows). The total angular momentum of the system is also shown (blue arrows).  Each row corresponds to different shell angular momentum content, from top to bottom, $\vrot=$ 0.2, 0.3, 0.7, while the left and right columns correspond to the shell tilt angle $\tilt =60\degree$ and $150\degree$ respectively. In al cases $L_y = 0$. $L_x$ and $L_z$ are in units of $2.08 \times 10^7 \msun \, {\rm km} \, {\rm s}^{-1} \, {\rm kpc} $.}
\label{fig: L-arrows}
\end{figure*}

Figure \ref{fig: L-arrows} shows the angular momentum of each constituent of the system at the beginning and at 
the end of the interaction. It makes clear how dramatically the angular momentum is altered for the cases in which 
the shell and the disc have the same amount of angular momentum ($\vrot = 0.3$). In these cases the interaction 
succeeds in significantly modifying the angular momentum modulus and orientation of both the shell and the 
disc. In the rest of the cases, the angular momentum of each component does not suffer a significant evolution, 
although in the case in which the shell is counter-rotating with smaller angular momentum ($\vrot = 0.2, \tilt = 
150\degree$) the moduli of the angular momentum of the shell and the disc are perceptively reduced. For the 
cases with higher angular momentum ($\vrot =0.7$), the impact of the interaction on the angular momentum of 
each flow is negligible. However, as we will see in the next section, even when the bulk of the gas 
remains unaffected in these cases, this situation does not prevent the inflow rates from being boosted by several orders 
of magnitude, thus effectively catalysing the inflow.


\section{Gas inflows} \label{sec: inflow}

\begin{figure}
\includegraphics[width=0.5\textwidth]{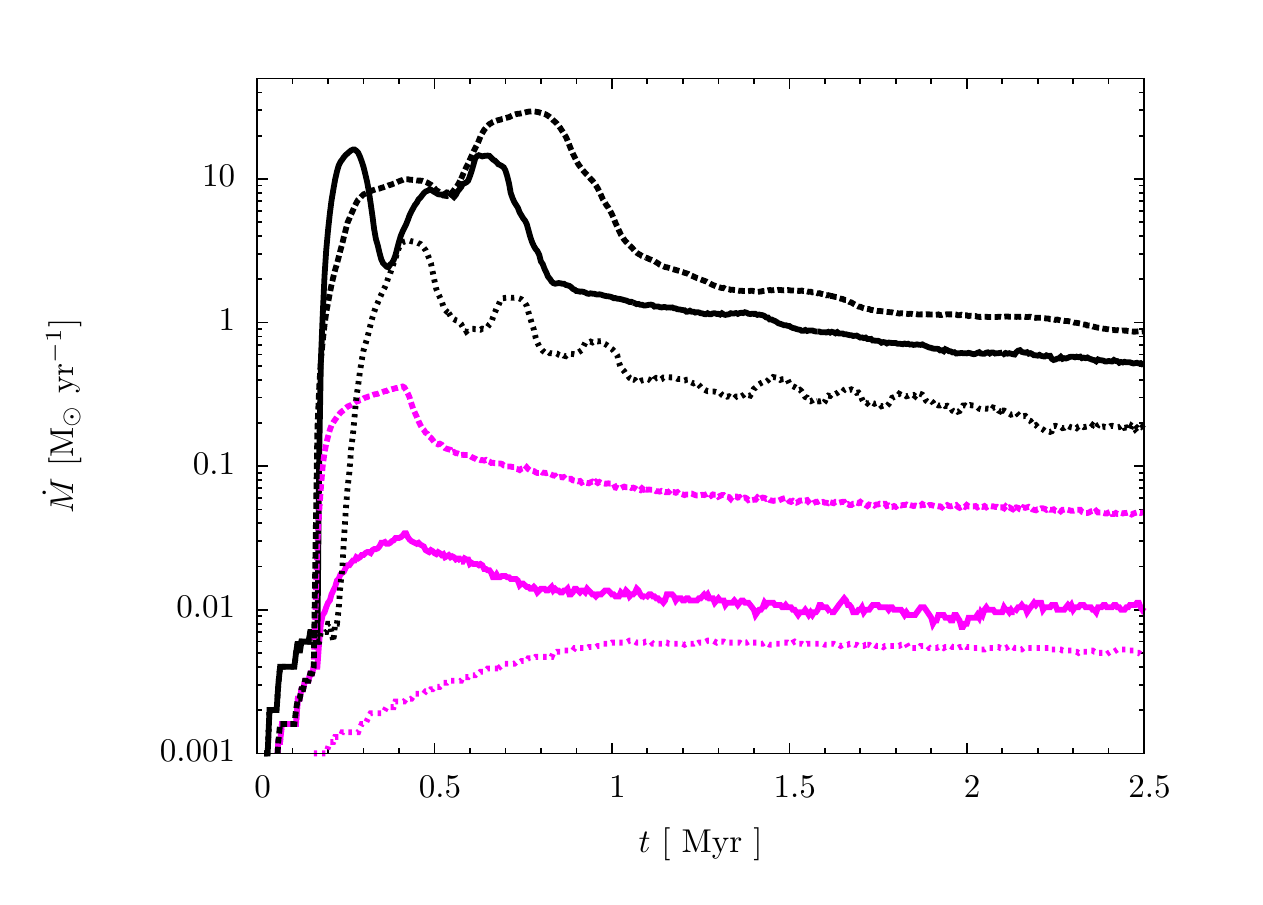}
\caption{Inflow rate (in units of $\msun \,\rm yr^{-1}$) versus time (in Myr) for each case of this study. The lines, 
from bottom to top, represent the following cases of initial $(\vrot, \tilt)$ : $(0.7, 60 \degree)$ short dashed 
magenta line;  $(0.3, 60 \degree)$ magenta solid line; $(0.2,60 \degree)$ long dashed magenta line; $(0.7, 
150\degree)$ short dashed black line; $(0.3, 150 \degree)$  black solid line; $(0.2, 150\degree)$ long dashed 
black line.}
\label{fig: all acc rate}
\end{figure}

\begin{figure}
\includegraphics[width=0.5\textwidth]{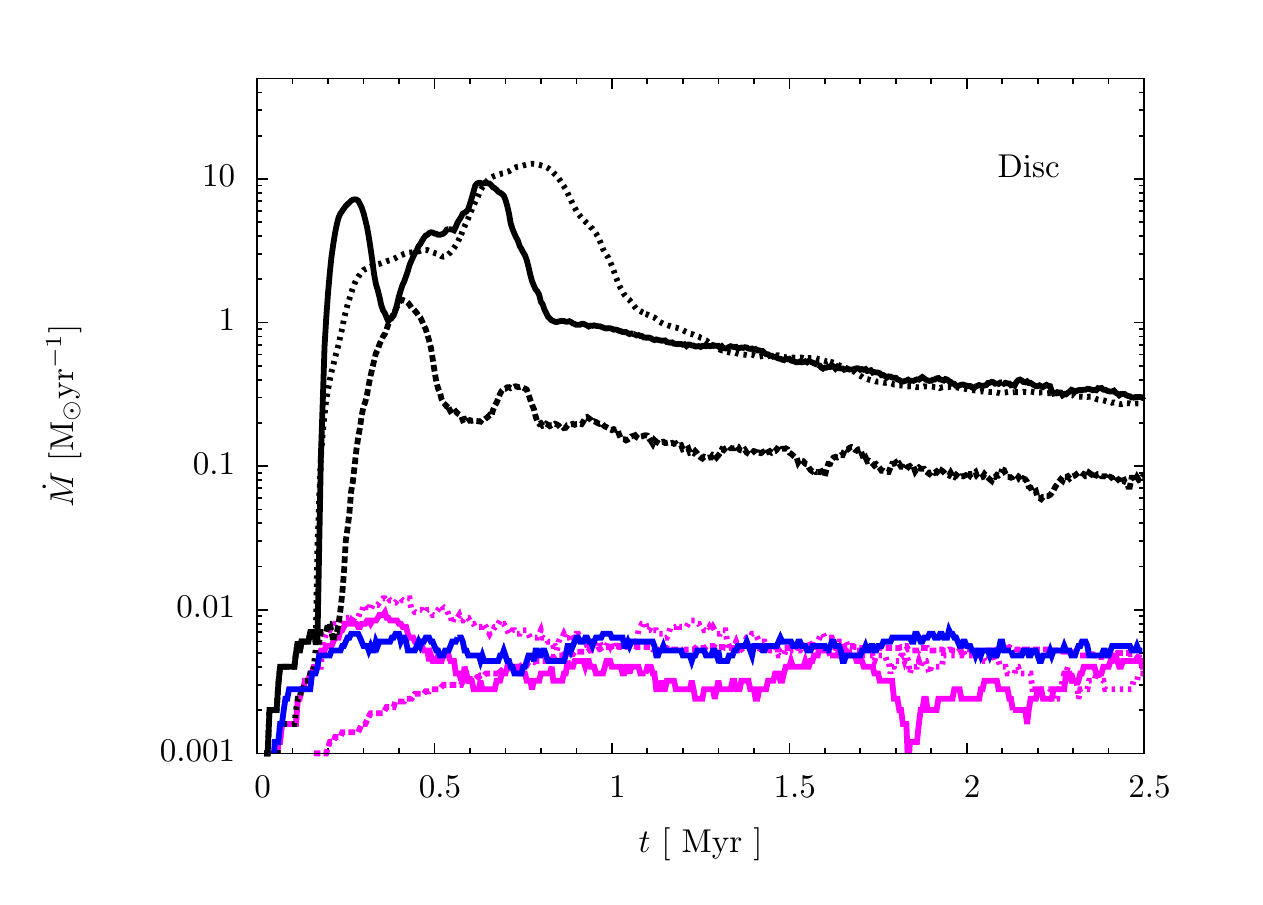}
\caption{Inflow rate (in units of $\msun \,\rm yr^{-1}$) versus time (Myr)  for disc particles only.
Magenta lines and black lines correspond to cases with $\tilt = 60\degree$ and $150\degree$, respectively. Long 
dashed lines correspond to $\vrot = 0.2$, continuous lines correspond to $\vrot = 0.3$ and short dashed lines 
correspond to $\vrot = 0.7$. A control case is also shown (blue line) where the disc is let evolve with no infalling, 
perturbing shell.}
\label{fig: comp prim disc}
\end{figure} 

\begin{figure}
\includegraphics[width=0.5\textwidth]{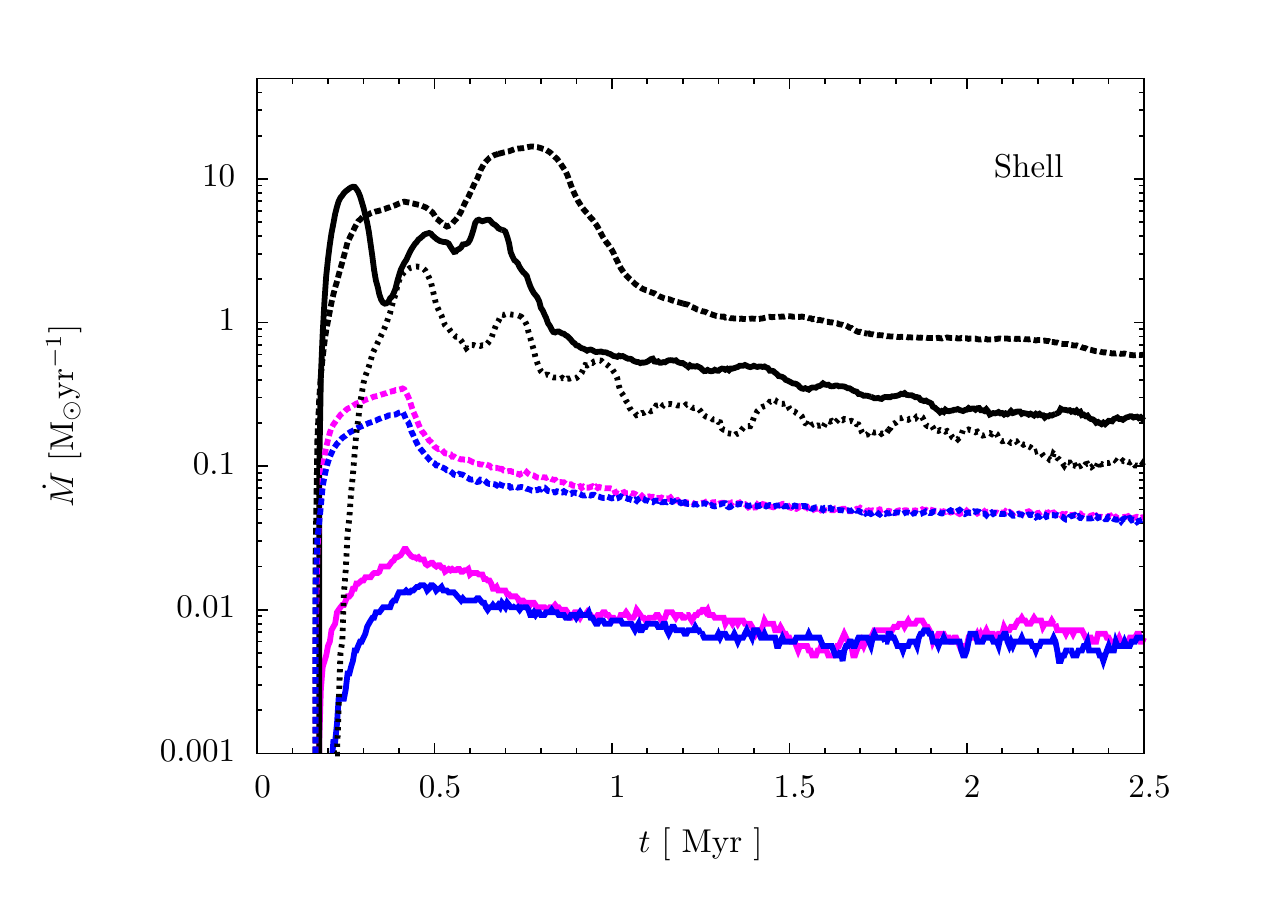}
\caption{Inflow rate (in units of $\msun \,\rm yr^{-1}$) versus time (Myr)  for shell particles only.
Magenta lines and black lines correspond to cases with $\tilt = 60\degree$ and $150\degree$ respectively. Long dashed lines correspond to $\vrot = 0.2$, continuous lines correspond to $\vrot = 0.3$ and short dashed lines correspond to $\vrot = 0.7$. Control cases are also shown (blue lines) where the shells evolve with no primitive disc present ($\mdot < 10^{-5}$ for the $\vrot = 0.7$ case ).}
\label{fig: comp shells}
\end{figure} 

We have studied the hydrodynamics of a rotating gaseous shell falling towards a black hole surrounded by a previously formed $\sim10 \pc$ accretion disc. The particular problem we have been addressing here concerns how the presence of the disc, formed during a previous accretion event, affects the evolution of the shell, and whether the interaction is able to effectively boost or halt the inflow rate. Although the inflow rate which we compute for each system cannot be taken at face value due to the resolution employed, a comparative analysis does give us some insight into the problem. 

Figure \ref{fig: all acc rate} compares the inflow rates corresponding to  the six cases under study.  It shows that counter-rotation between the interacting systems can effectively boost the inflow rate by more than two orders of magnitude at the peak of accretion, and sustain inflow rates  $> 10$ times higher than that obtained in the co-rotating cases at later times.

In order to be on more solid grounds, we have calculated the inflow rate associated to the disc and shell particles, separately,  as a function of time (Figures \ref{fig: comp prim disc} and  \ref{fig: comp shells} respectively). In each figure we have further introduced, for seek of clarity, control runs in which there is no infalling shell to perturb the primitive disc (Figure \ref{fig: comp prim disc}), or there is only an infalling shell with initial $\vrot = 0.2$, $0.3$ and $0.7$ not being perturbed by the primitive disc (Figure \ref{fig: comp shells}). 

Accretion of disc particles is effectively boosted by $\sim 2-3$ orders of magnitude when the shell infalls in counter-rotation, while it is almost unaffected, or even slowed down, when the shell infalls in co-rotation (see figure \ref{fig: comp prim disc}). This is due to the fact that particles falling with high angular momentum kick lower angular momentum disc particles to higher orbits, preventing them from flowing in. Similarly, shell particles are accreted at a higher rate in the counter-rotating events.

Attention is drawn to the fact that even in the case in which the angular momentum distribution and  modulus of each constituent of the system is mildly affected by the interaction, the inflow rate can be 
still highly enhanced.  In the case
$\vrot = 0.7$ and $\tilt = 
150 \degree$, the angular momentum of each component of the system remained practically 
unaffected, as shown by Figures \ref{fig: final state v07} and \ref{fig: L-arrows}, thus we could think that this event would 
produce a negligible enhancement of the accretion rate, however, Figure \ref{fig: comp prim disc}  and Figure 
\ref{fig: comp shells} tell otherwise. In this case, the primitive disc has boosted the 
accretion of the shell by up to 3 orders of magnitude, while the disc itself has lost mass to the black hole at a rate 
$\sim 10-100$ times greater than it would have done if the shell infall had not occurred. Thus, misaligned and 
uncorrelated nested accretion events are, in this way, capable of driving an important amount of gas inflow, acting 
as catalysts.

After the interaction has finished, the inflow rate damps and a stationary state of sustained residual accretion is
established.

\subsection{Equilibrium State}
\begin{figure}
\includegraphics[width=0.5\textwidth]{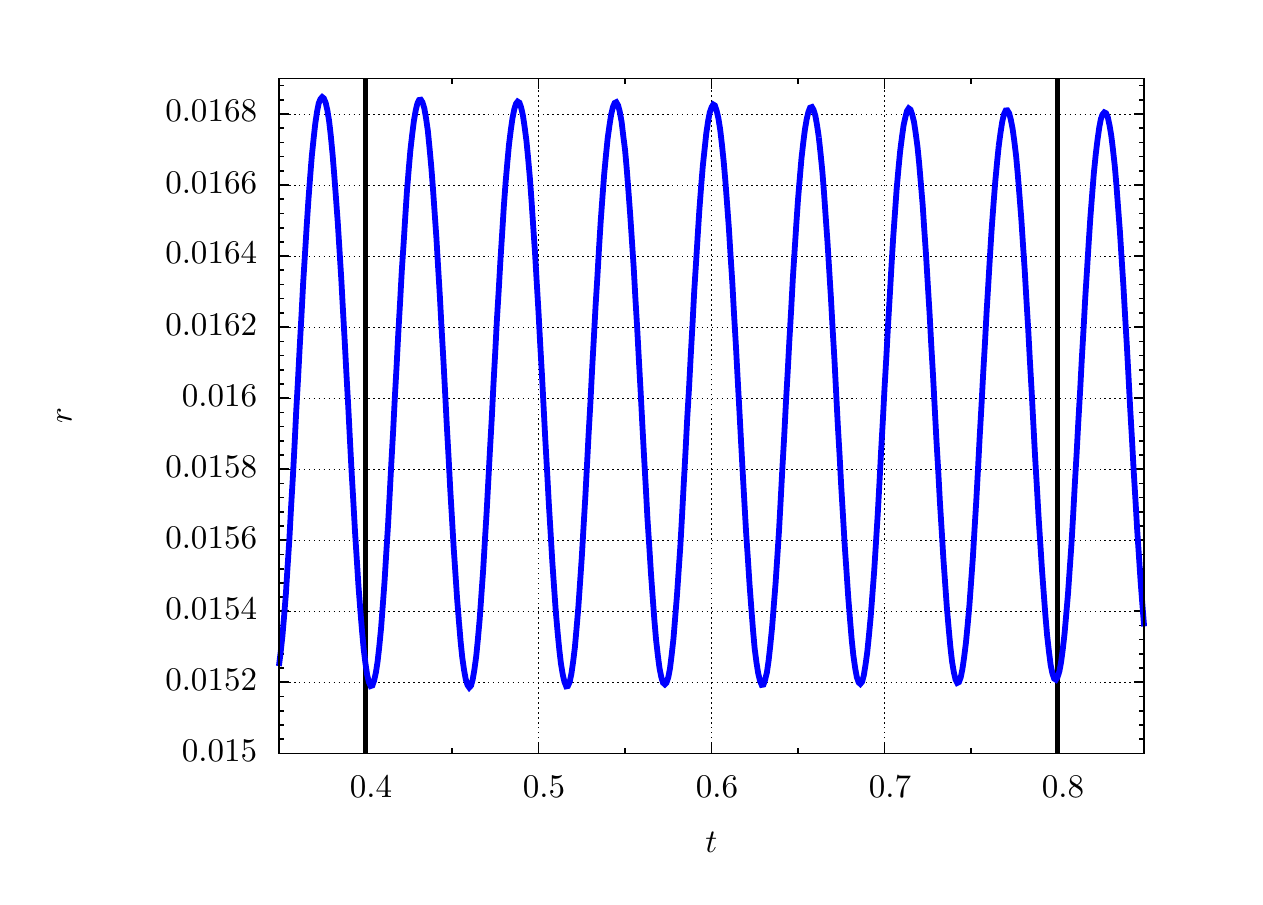}
\caption{
Mean radial distance of the gas particles (in code units) forming the primitive disc to the centre of the potential (i.e. to the black hole) as a function of time (in code units).
}
\label{fig: Wobbling}
\end{figure} 

The analysis of the evolution of the
primitive disc reveals that even when equilibrium is reached, the disc wobbles in the radial direction, this being 
a common behaviour throughout all the nested accretion events of this study.

Figure \ref{fig: Wobbling} shows the oscillation of the mean radius of the particles forming 
the disc in the absence of the infalling shell.  The wobbling of the disc is
due to the presence of residual ellipticity in the orbits of the gas particles with a  periodicity which is consistent
with near Keplerian motion. This dynamical state of the gas causes perturbations that lead to a constant inflow
of particles through the inner boundary, giving rise to the sustained long term inflow observed in Figures \ref{fig: all acc rate}, \ref{fig: 
comp prim disc} and \ref{fig: comp shells}.

\subsection{Inner evolution}

\begin{figure*}
\includegraphics[width=0.49\textwidth]{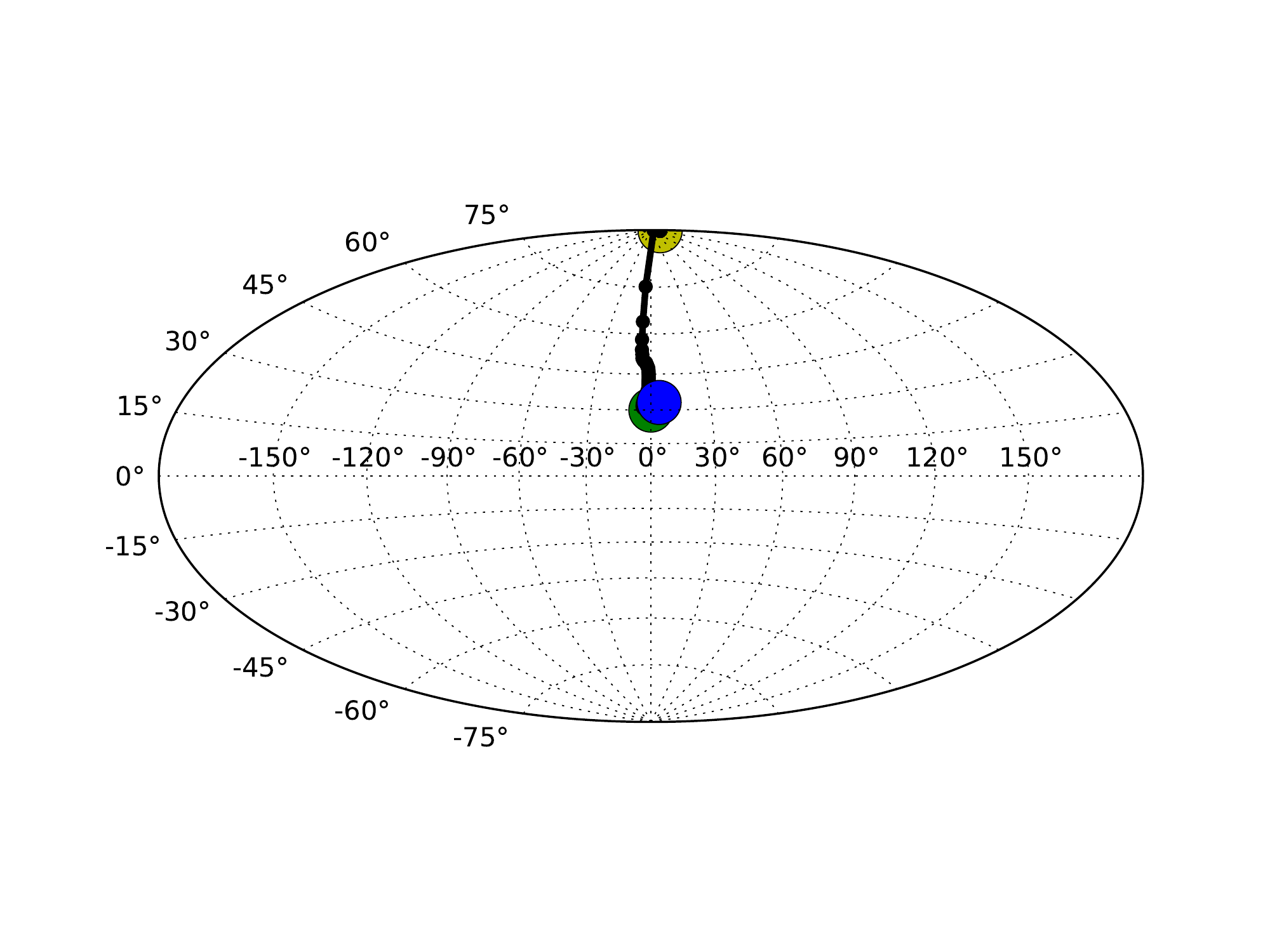}
\includegraphics[width=0.49\textwidth]{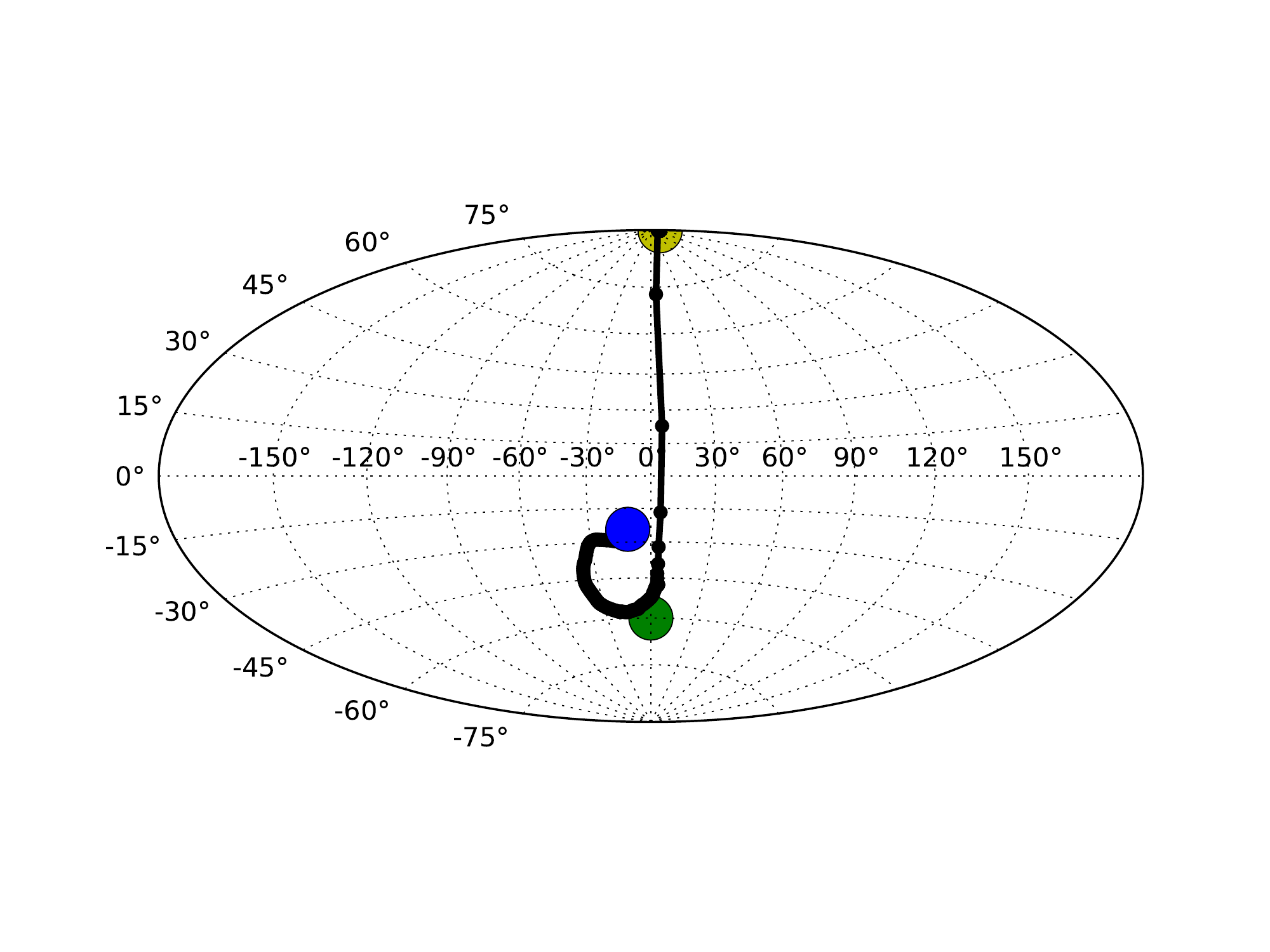}\\
\includegraphics[width=0.49\textwidth]{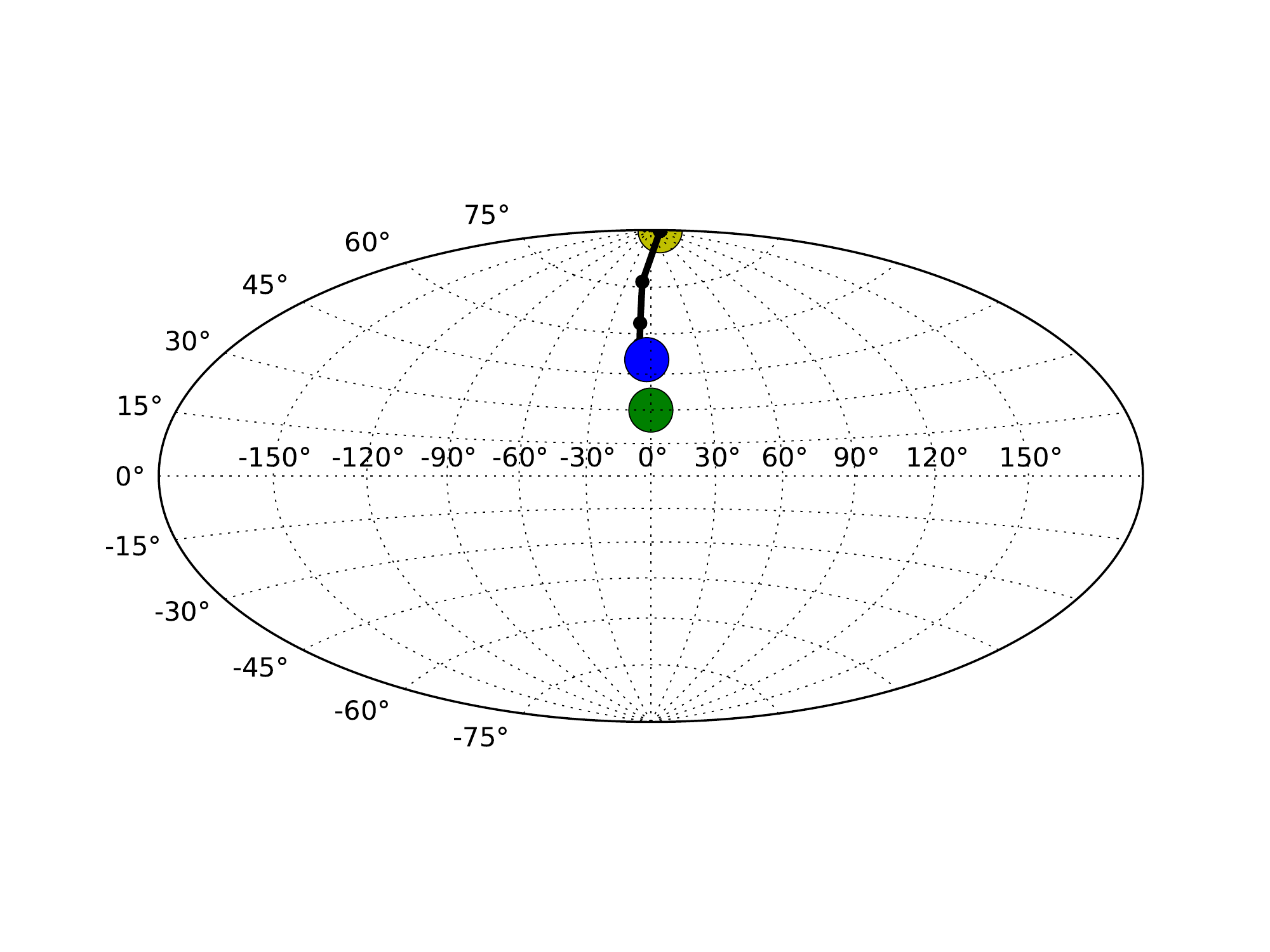}
\includegraphics[width=0.49\textwidth]{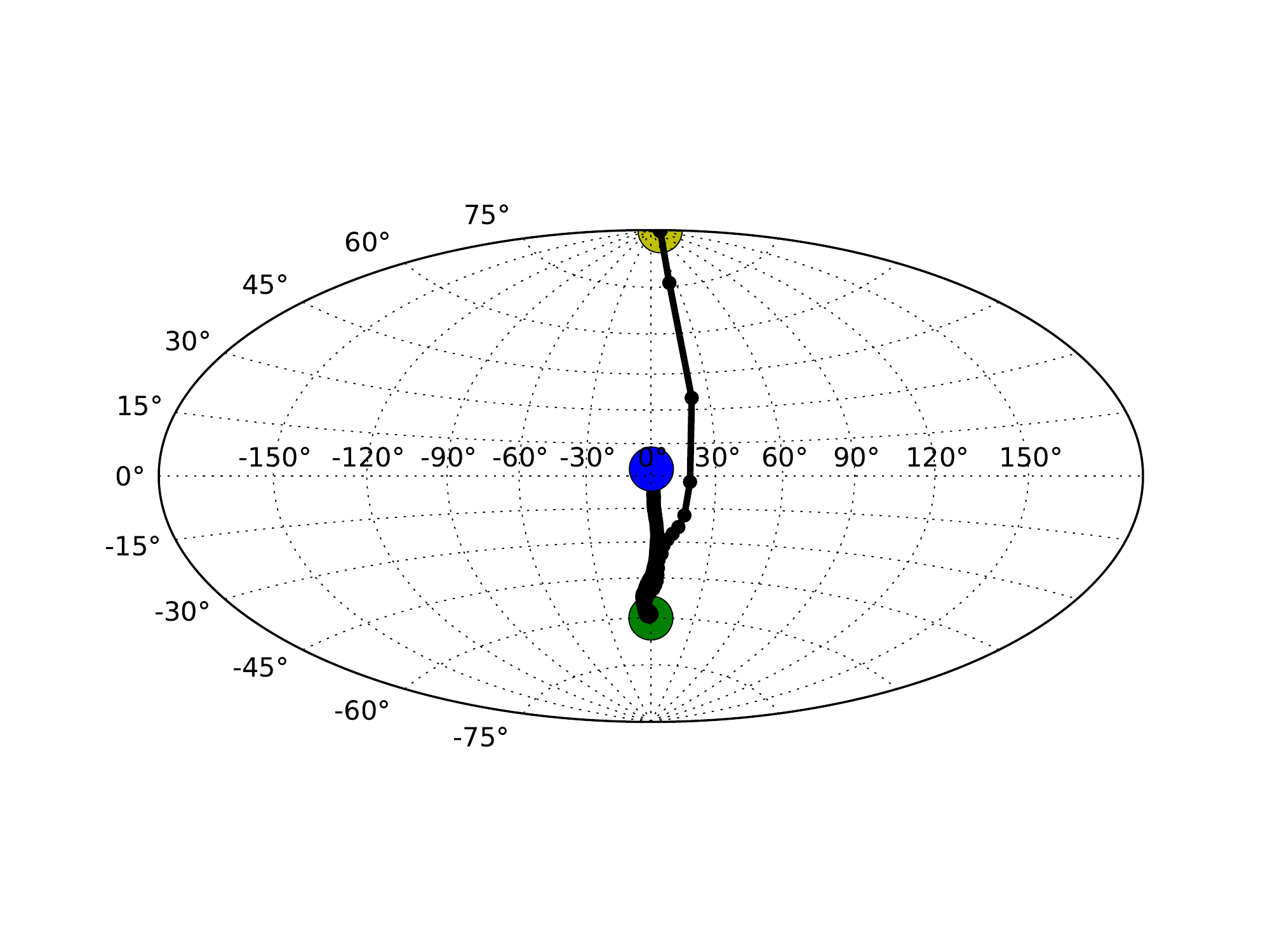}\\
\includegraphics[width=0.49\textwidth]{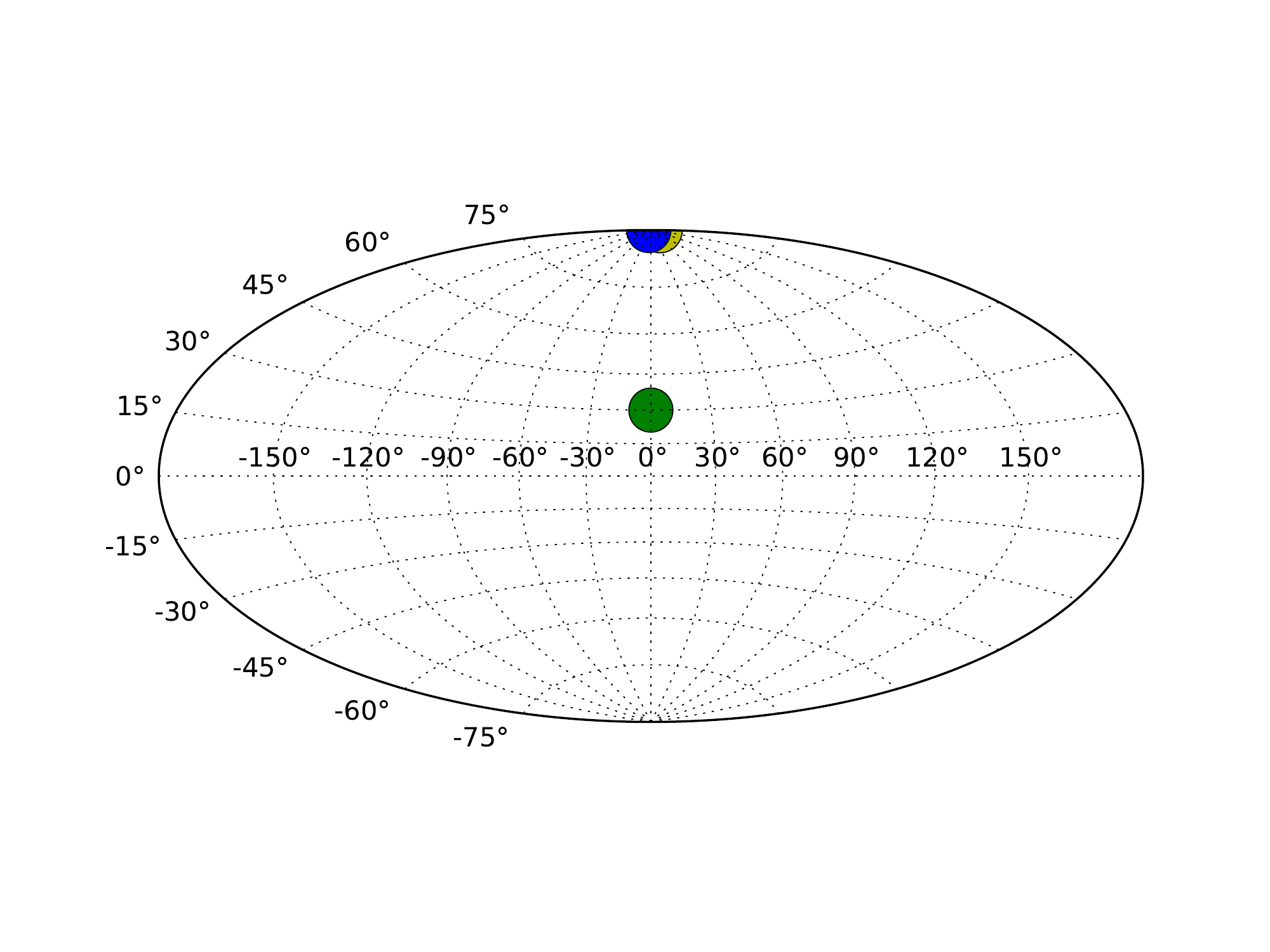}
\includegraphics[width=0.49\textwidth]{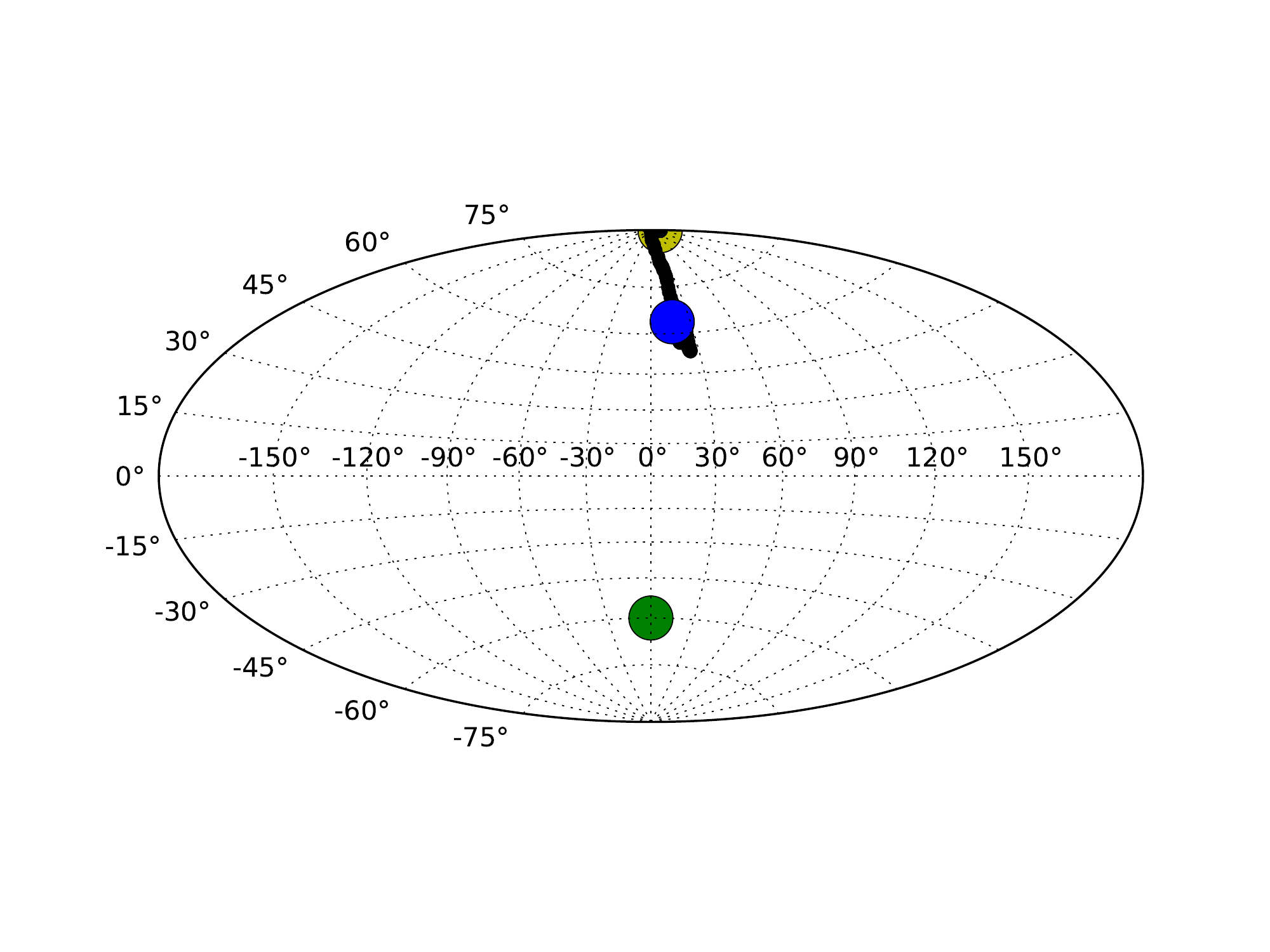}
\caption{Angular momentum direction in the region between $ r_{\rm acc}<r<0.005$ (corresponding to
the inner 5 pc). The yellow dot marks the start of the interaction when the angular momentum points in the $z$ direction, while the blue dot denotes the final state and the green dot the angular momentum direction of the shell, initially. From top to bottom, $\vrot=$ 0.2, 0.3, 0.7. The left and right columns correspond to $\tilt =60\degree$ and $150\degree$ respectively.}
\label{fig: inner evol}
\end{figure*}

Figure \ref{fig: inner evol} shows the orientation of the angular momentum of the gas in the inner 
region close to the accretion radius, i.e. within $r_{\rm acc}<r<r_{\rm inner} = 0.005,$ sampled at 
regular intervals of time ($\Delta t = 10^{-3}$).  The first column refers to co-rotating events,
the second to counter-rotating events.  Yellow dots refer to the initial orientation of the primitive disc;
the blue dots refer to the final, equilibrium state, and green dots mark the orientation of 
the shell at the onset of the dynamical interaction.
The angular momentum orientation which initially coincides with the 
orientation of the disc angular momentum soon shifts toward that of the shell, the final size of the shift 
being in proportion to the level of mixing that drove the inflow.
For  $J_{\rm shell}>J_{\rm disc}$, the weakness of the interaction leads to a final orientation which is closer to 
that of the disc, even in the case of counter-rotation.
For $J_{\rm shell}=J_{\rm disc}$, when the shell falls in co-rotation, the direction of the innermost flow is at an
angle $\theta_{\rm
acc}\sim 3 \, \theta_{\rm tilt} /4 $ while in the counter-rotating case, it passes through $\theta_{\rm tilt}$ 
but soon recedes to $\theta_{\rm acc}\sim \theta_{\rm tilt}/2$.
For $J_{\rm shell}<J_{\rm disc}$, in the co-rotating case the direction of the inflow is at an angle 
$\theta_{\rm acc}\sim \theta_{\rm tilt} $ while in the counter-rotating case $\theta_{\rm acc}$ at some moment 
reaches $\theta_{\rm tilt}$ for later converging to a value $\theta_{\rm acc}\sim 4 \,\tilt/5 \sim120 \degree$.

Figure \ref{fig: inner evol} indicates that if a boost in the black hole feeding occurs through the collision of gas clumps on the scale 
probed in this simulations, the inner accretion flow would have to be changing orientation 
on the Myr time scale. The small scale flow does
not preserve full memory of the larger scale orientation of the feeding events.

\section{Summary and Conclusions}\label{sec:conclusions}

We studied the hydro-dynamics of a rotating gas shell falling onto a primitive gaseous disc 
that rotates around a central black hole in a galactic bulge.  The shell mimics a large scale 
instability in the nuclear region of a galaxy where a relic 
disc is present from an earlier inflow episode.  
The simulations are idealised and represent a toy model that helps us to understand how warped discs or nested misaligned rings can form in the interaction of different flows that coexist to possibly feed a central black hole.  

The infalling shell and the primitive disc perturb each other, causing a re-distribution of their angular momentum 
content via shocks. Considerable mixing between the gas and shell particles occurs when the shell and disc have 
comparable angular momentum, forming a warped disc orthogonal to the primitive disc when some degree of counter-rotation is present. Co- and 
counter-rotating events from infalling shells with higher or lower angular momentum than the disc, produce 
nested ring like structures rotating on disjoint planes. 
A comparative analysis of the inflow rates onto the central black hole in the models explored shows that the infall of gas is boosted by more than two orders of magnitude relative to cases in absence of interaction. When the perturbation
has ceased to act, inflows continue due to wobbling of the disc induced by gas particles in eccentric motion.
This idealised representation shows that subsequent accretion episodes
provide a viable way to drive major gas inflows toward the central parsec, and make relic discs more susceptible to accretion. Thus, nested inflow episodes 
possibly triggered by larger scale perturbations act as {\it catalysts} for the fuelling and growth of the central black hole.
 
Accretion in the very vicinity of massive black holes may occur chaotically, i.e. 
through a sequence of accretion events that give rise to smaller-scale viscous discs, each with a different and random orientation of their angular momentum relative to the spin axis of the rotating central massive black hole (e.g. \citealt{King06, King07}). 
Our simulations explored the formation of larger scale discs and the misaligned, nested rings resulting from the disc-shell
interaction have no connection with the black hole spin, nor with the warping or breaking of the disc 
due to Lense-Thirring precession (\citealt{Lodato10, Nixon12oct}), and should not be confused. 
Instead, this study is in line and extends that  of 
\cite{Nixon12may},  who simulated the evolution of two misaligned (non-self gravitating) discs in the gravitational (Keplerian) potential of a massive black hole. They showed that
the accretion rate is enhanced in a similar way by cancellation of angular momentum at the junction of the
two pre-placed discs. We learned that non-correlated inflows at large scales generate nested, misaligned disc-like 
structures together with inflows across the inner pc without necessarily preserving the orientation of the parent 
flows --especially when the parent flows have some degree of counter-rotation.
Further steps should go in the direction of assessing the effects that self-gravity would have
on these converging flows starting from realistic initial conditions, as e.g.  those created in a merger of gas
rich galaxies, or from gravitational unstable modes as the ones observed by \cite{Hopkins10sim}. 

\acknowledgments 
\noindent The authors are very grateful to John Miller for his critical comments, careful reading and
enlightening  suggestions.

\vspace{1.0cm}

\appendix
\section{Test on numerical viscosity}
\label{app: visc}

\begin{figure}
\includegraphics[width=0.5\textwidth]{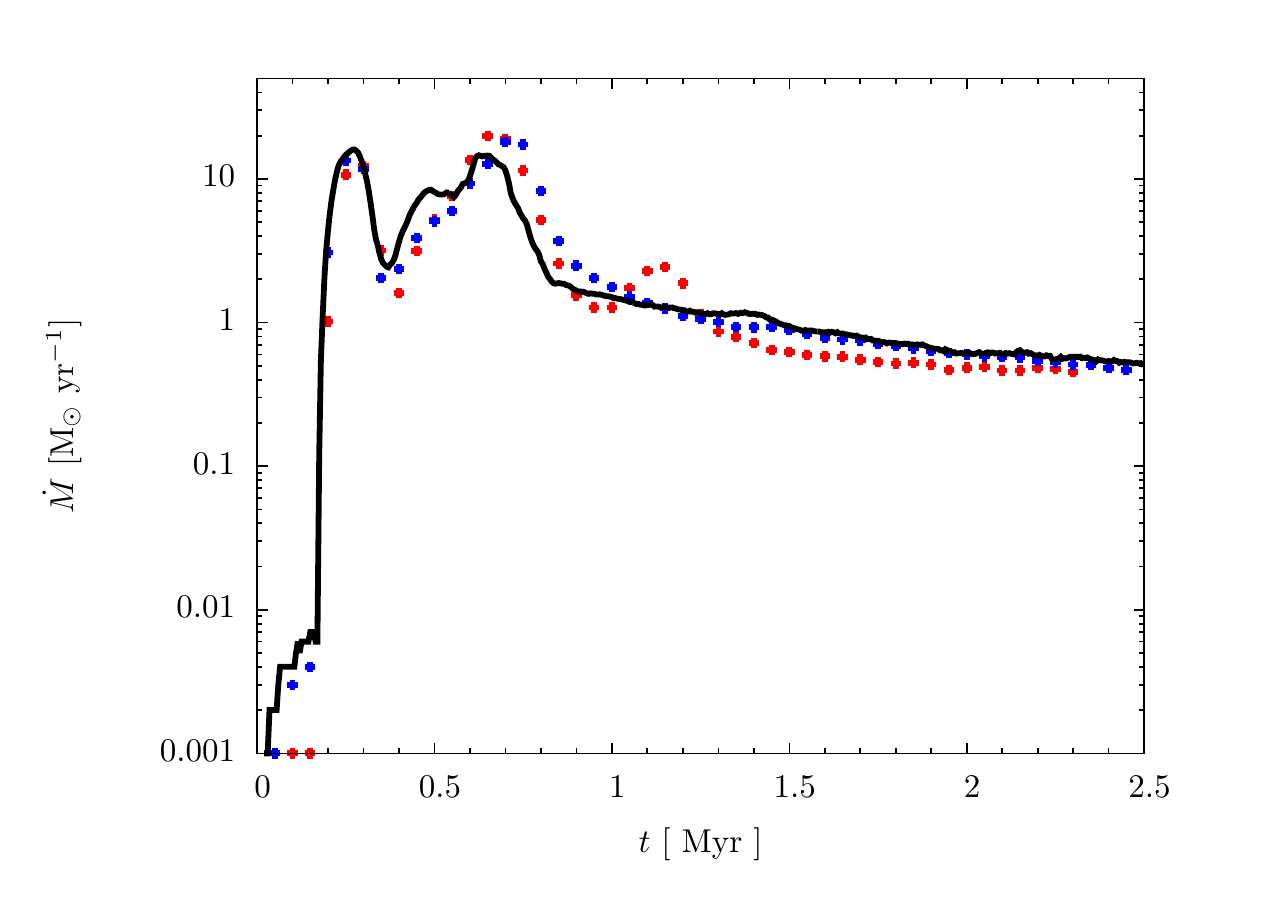}
\caption{Accretion rates for the case $J_{\rm shell}=J_{\rm disc}$ with $\vrot = 0.3$ and $\tilt = 150 \degree$ obtained by varying $\avisc$. The red and blue dots show the accretion rate obtained setting $\avisc = 0.2$ and 0.5 respectively, while the black continuous line shows the accretion rate obtained using $\avisc = 1.0$, the fiducial value.}
\label{fig: visc accretion}
\end{figure} 

\begin{figure}
\includegraphics[width=0.5\textwidth]{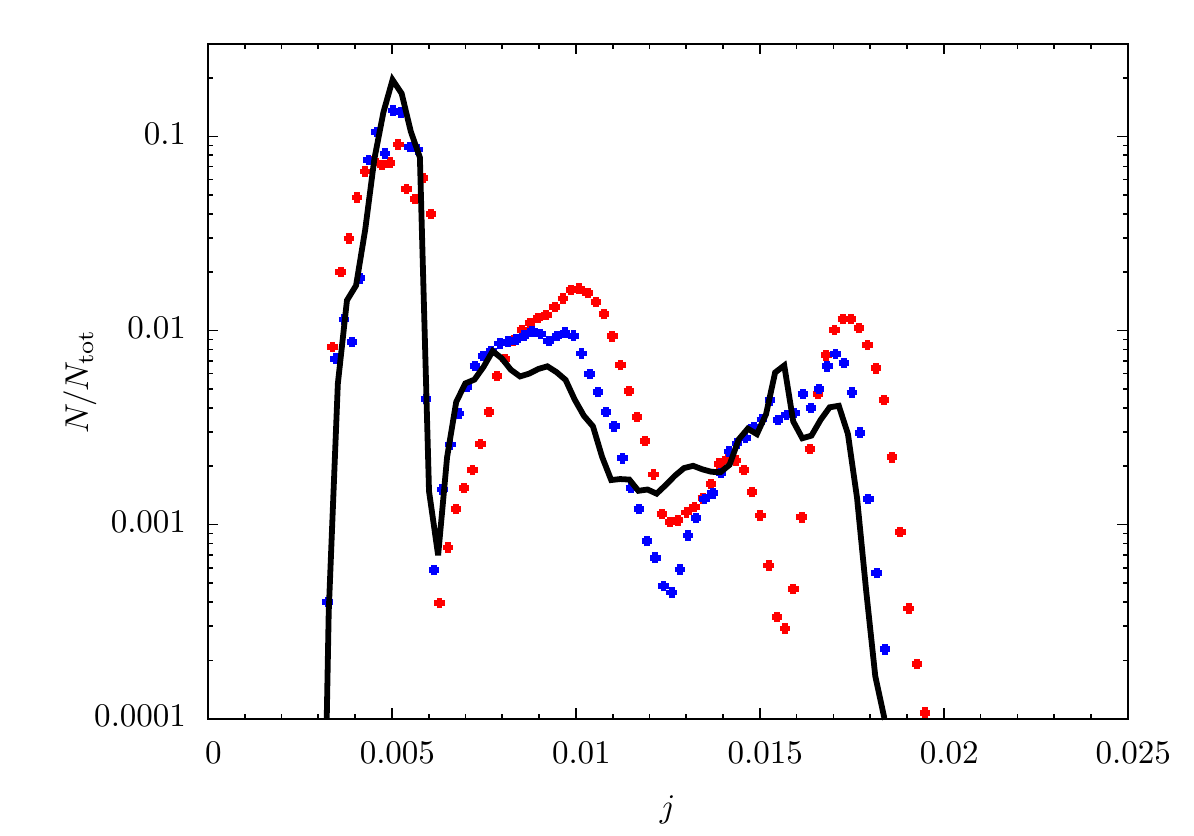}
\caption{Angular momentum distributions for the case  $J_{\rm shell}=J_{\rm disc}$ with $\vrot = 0.3$ and $\tilt = 150 \degree$ obtained by varying $\avisc$. The red and blue dots show the angular momentum distribution obtained setting $\avisc = 0.2$ and 0.5 respectively, while the black continuous line shows the angular momentum distribution obtained  using $\avisc = 1.0$, the fiducial value.}
\label{fig: visc Lhist}
\end{figure} 

\begin{figure}
\includegraphics[width=0.5\textwidth]{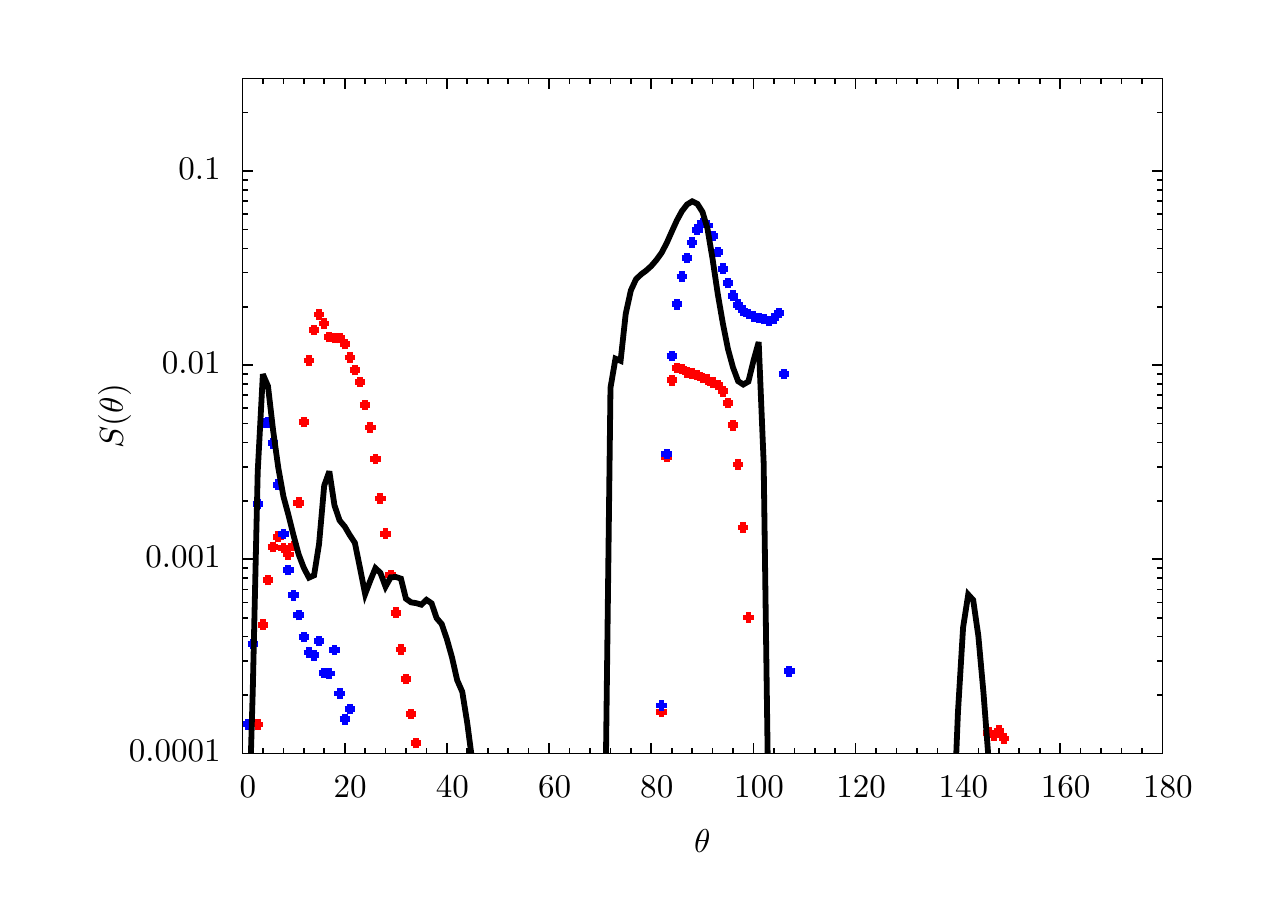}
\caption{Main planes of rotation for the case  $J_{\rm shell}=J_{\rm disc}$ with $\vrot = 0.3$ and $\tilt = 150 \degree$ obtained by varying $\avisc$. The red and blue dots show the main planes of rotation obtained setting $\avisc = 0.2$ and 0.5 respectively, while the black continuous line shows the main planes of rotation obtained using $\avisc = 1.0$, the fiducial value.}
\label{fig: visc mainplane}
\end{figure}

In this appendix we run tests to explore the dependence of the results on the adopted value 
of the artificial viscosity $\avisc$  in order to quantify how much accretion is numerical and 
how the cancellation and transport of angular momentum via shocks and shear, 
respectively, differ with varying $\avisc$.

Throughout all our simulations, the artificial viscosity coefficient was set, as implemented in \textsc{Gadget-2}, to $\avisc = 1.0$. In this appendix we run two simulations with  $\avisc = 0.2, 
0.5$  for the most violent case case $J_{\rm disc}=J_{\rm shell}$ when the shell is in counter rotation.
A comparison of the results is shown in Figures \ref{fig: visc accretion}, \ref{fig: visc Lhist} and \ref{fig: visc mainplane}, where we plot the evolution of the accretion rate and the end states of the angular momentum distribution and main planes of rotation.

As illustrated in Figure \ref{fig: visc accretion}, the effect of
of a change in $\avisc$ on the inflow rate across the
inner boundary of the simulated volume leads on average a change $\Delta
\dot{M}/\dot{M}\approx 0.4$, with a maximum variation of
$\approx 1$. A change of this magnitude does not alter our conclusions.

Figure \ref{fig: visc Lhist} shows the distribution of the angular momentum moduli of all the particles, 
when the system has reached an equilibrium state. The three distributions are 
qualitatively the same, indicating the robustness of our results (see also Appendix B of Hobbs11).
Artificial viscosity affects the angular momentum transport in two ways: In discs dominated by shear viscosity, a larger $\avisc$ implies faster angular momentum transport and faster broadening of an initial ring of gas. This may affect the long term evolution of the ring like structures that appear in our simulations, but these late stages are not explored in our study. By contrast, in our simulations, shell particles have initially a broad angular momentum distribution (a homogeneous shell with a cylindrical velocity profile; see e.g. Figure \ref{fig: Fig5Hobbs}) and move on  eccentric orbits crossing one with the other, as the initial configuration is out of equilibrium. With time, orbit crossing causes particles to shock and settle into a common ring. The larger the $\avisc$, the faster the circularization of the orbits, as shocks are better resolved, an effect observed in our simulations. 
We also compared the main planes of rotation obtained by using different values of $\avisc$ (Figure \ref{fig: visc mainplane}) to asses the impact of the artificial viscosity on the chaotic structure of the resulting flows. We find no significant differences.

\section{Test on resolution}
\label{app: resol}

\begin{figure}
\includegraphics[width=0.5\textwidth]{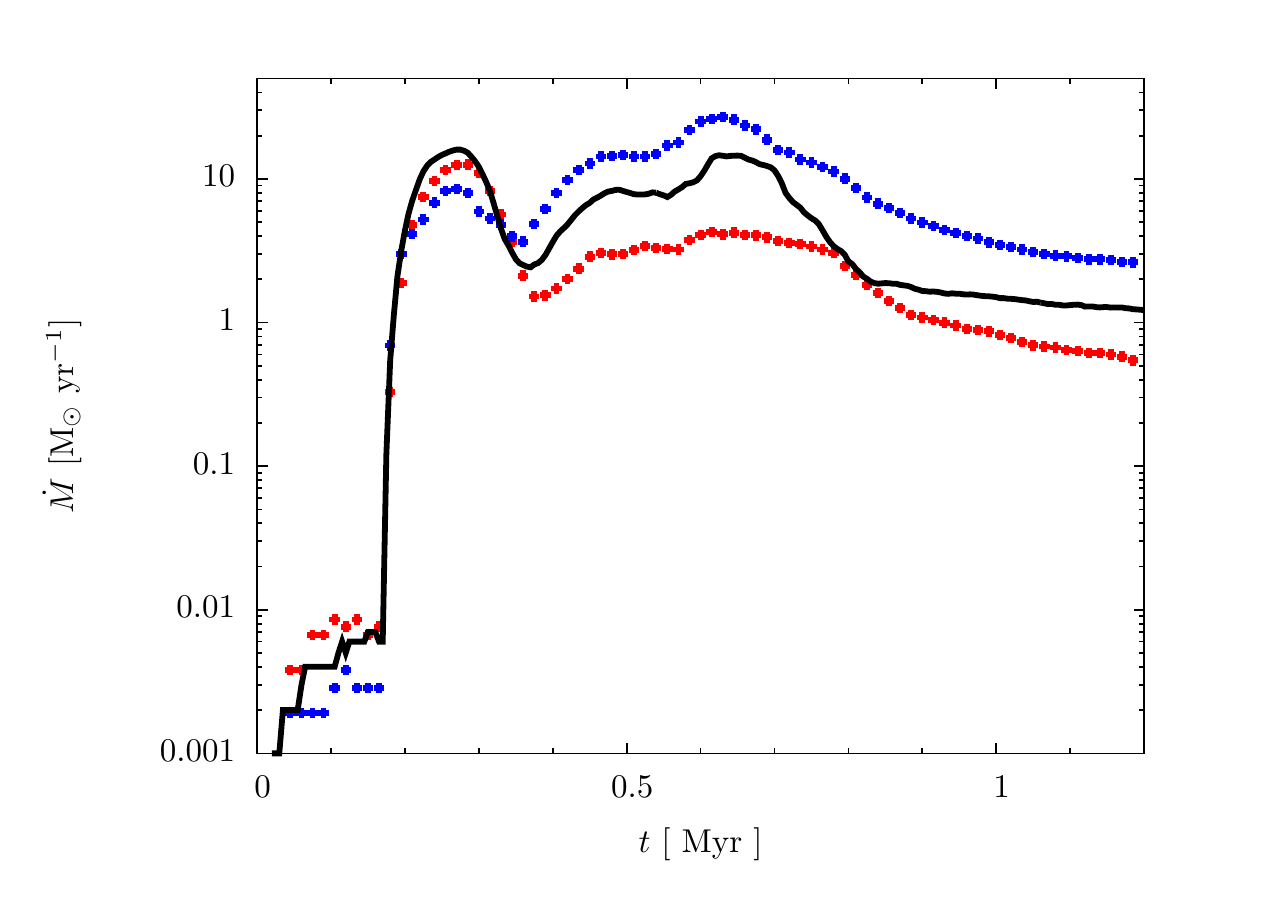}
\caption{Accretion rates for the case $J_{\rm shell}=J_{\rm disc}$ with $\vrot = 0.3$ and $\tilt = 150 \degree$ obtained by varying $N_{\rm part}$. The red and blue dots show the accretion rate obtained setting $N_{\rm part} = 0.5\times10^6$ and $N_{\rm part} = 2.0\times10^6$ respectively, while the black continuous line shows the accretion rate obtained using $N_{\rm part} = 1.0\times10^6$, the fiducial value.}
\label{fig: resol accretion}
\end{figure} 

\begin{figure}
\includegraphics[width=0.5\textwidth]{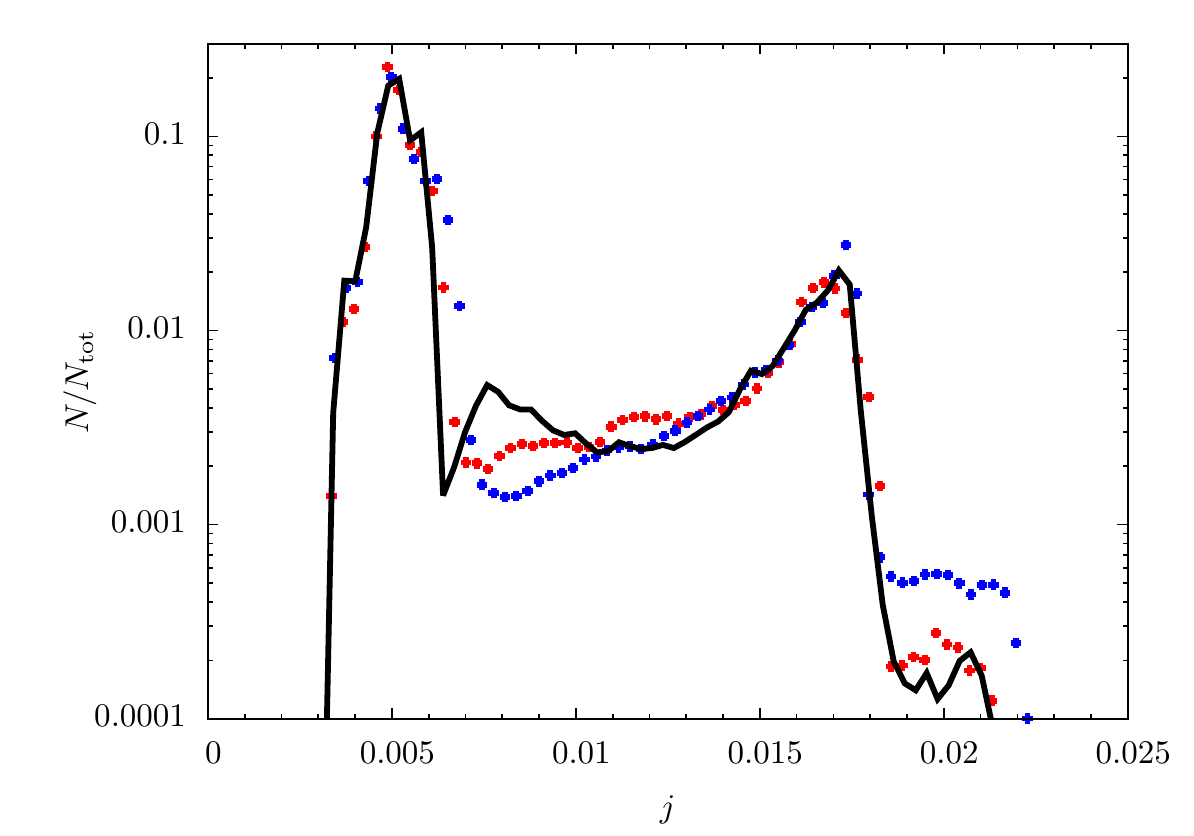}
\caption{Angular momentum distributions for the case  $J_{\rm shell}=J_{\rm disc}$ with $\vrot = 0.3$ and $\tilt = 150 \degree$ obtained by varying $N_{\rm part}$. The red and blue dots show the angular momentum distribution obtained setting  $N_{\rm part} = 0.5\times10^6$ and  $N_{\rm part} = 2.0\times10^6$ respectively, while the black continuous line shows the angular momentum distribution obtained  using  $N_{\rm part} = 1.0\times10^6$, the fiducial value.}
\label{fig: resol Lhist}
\end{figure} 

\begin{figure}
\includegraphics[width=0.5\textwidth]{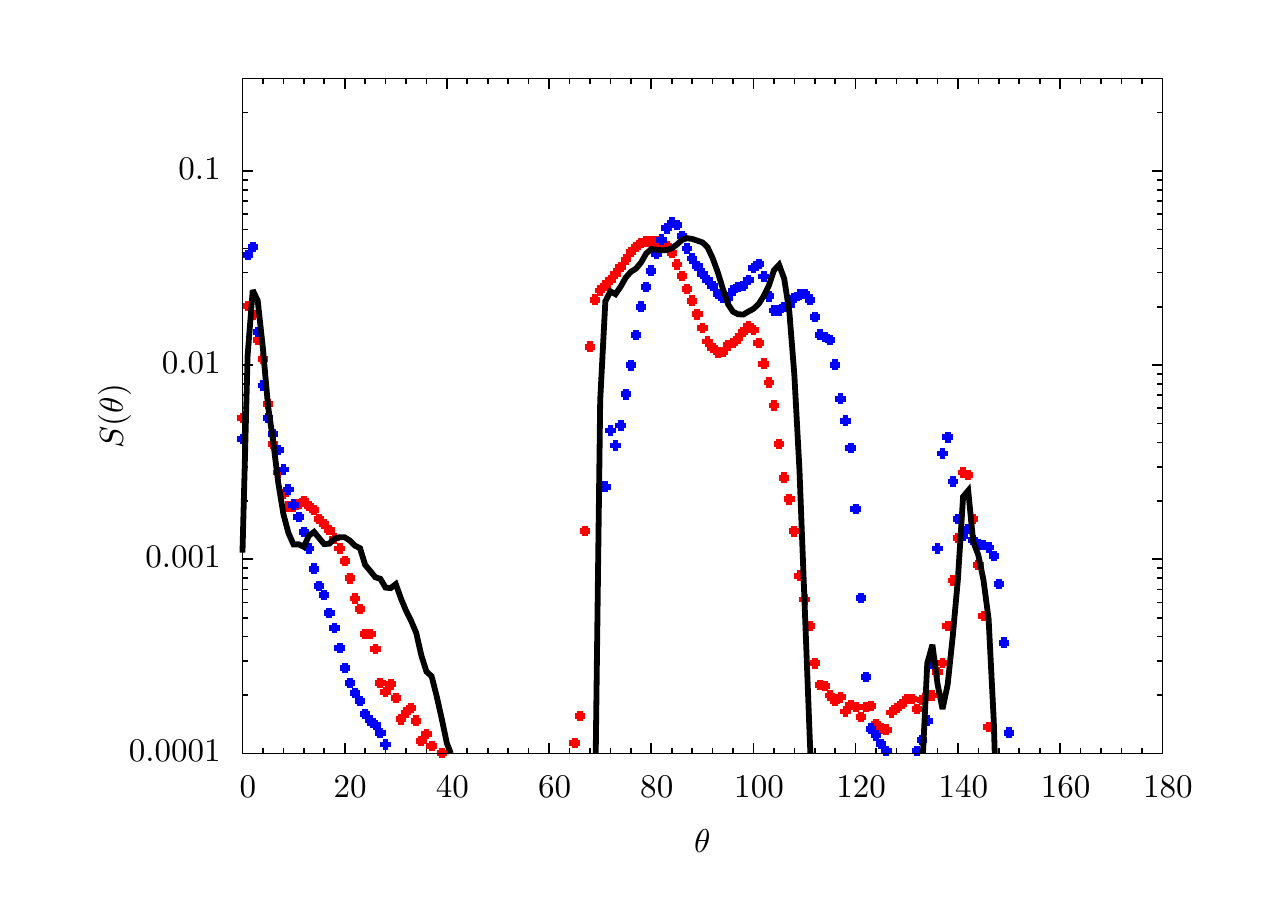}
\caption{Main planes of rotation for the case,  $J_{\rm shell}=J_{\rm disc}$ with $\vrot = 0.3$ and $\tilt = 150 \degree$ obtained by varying  $N_{\rm part}$. The red and blue dots show the main planes of rotation obtained setting  $N_{\rm part} = 0.5\times10^6$ and $N_{\rm part} = 2.0\times10^6$ respectively, while the black continuous line shows the main planes of rotation obtained using  $N_{\rm part} = 1.0\times10^6$, the fiducial value.}
\label{fig: resol mainplane}
\end{figure} 

We here  explore the dependence of our results on the number of particles $\npart$ used 
by running two extra simulations.
Throughout all our simulations, the number of particles was set to $\npart = 1.0\times10^6$ (with $\ndisc = \nshell = \npart/2$). The two extra simulations we discuss here have $\npart = 0.5\times 10^6$ and $\npart = 2.0 \times 10^6$. In the same fashion as in Appendix \ref{app: visc}, the simulations were run for the most violent case case, $J_{\rm disc}=J_{\rm shell}$ when the shell is in counter rotation.
A comparison of the results is shown in Figures \ref{fig: resol accretion}, \ref{fig: resol Lhist} and \ref{fig: resol mainplane}, where we plot the evolution of the accretion rate and the end states of the angular momentum distribution and main planes of rotation.

As illustrated in Figure \ref{fig: resol accretion}, the effect of a change in resolution on the inflow rate across the
inner boundary of the simulated volume is not dramatic. The  ratio $\Delta
\dot{M}/\dot{M}\lsim 1.0$ always. We emphasise again that the inflow rates
cannot be taken at face value, in this study. Thus, variations on the accretion rate of this magnitude do not alter our conclusions.

Figure \ref{fig: resol Lhist} shows the distribution of the angular momentum moduli of all the particles at $t = 0.25$, when the system is still actively evolving. The three distributions are 
qualitatively the same, indicating the robustness of our results (see also Appendix B of Hobbs11). We also compared the main planes of rotation obtained by varying the number of particles used (Figure \ref{fig: visc mainplane}) to asses the impact that improving the resolution has on the chaotic structure of the resulting flows. Once again, we find no significant differences.

\label{lastpage}
\bibliographystyle{apj}
\bibliography{Overlapping_Inflows}

\end{document}